\documentclass[aps,prb, onecolumn, superscriptaddress,notitlepage]{revtex4-1}

\usepackage{amsmath,amsfonts, amssymb, amsthm, dsfont}
\usepackage{yfonts}
\usepackage{bm,bbm}
\usepackage{mathrsfs}
\usepackage{graphicx}
\usepackage{hyperref}
\usepackage{multirow}
\usepackage[normalem]{ulem}
\usepackage{tikz}
\usetikzlibrary{calc}
\usetikzlibrary{shapes.multipart}

\usepackage{hyperref}
\hypersetup{hypertex=true,
colorlinks=true,
linktocpage=true,
linkcolor=blue,
citecolor=teal,
urlcolor=purple}

\newcommand{\dd}{\mathrm{d}}

\newcommand{\sq}{\mathrm{Sq}}
\newcommand{\Sq}{\mathbbm{Sq}}

\newcommand{\sqb}[1]{\left[#1\right]}

\newcommand{\ket}[1]{|#1\rangle}
\newcommand{\bra}[1]{\langle#1|}

\newcommand{\comments}[1]{}
\newcommand{\om}{\omega}
\newcommand{\mb}[1]{{\mathbf{#1}}}

\renewcommand{\=}[1]{\overset{#1}{=}}
\newcommand{\eq}[1]{Eq.~(\ref{#1})}
\newcommand{\eqs}[2]{Eqs.~(\ref{#1}) and (\ref{#2})}
\newcommand{\eqss}[2]{Eqs.~(\ref{#1})--(\ref{#2})}
\newcommand{\Zf}{\mathbb Z_2^f}
\newcommand{\refn}[1]{Ref.~\onlinecite{#1}}
\newcommand{\refns}[1]{Refs.~\onlinecite{#1}}
\newcommand{\dtilde}[1]{\tilde{\raisebox{0pt}[0.9\height]{$\tilde{#1}$}}}

\newcommand{\ag}[2]{{#1}_\mb{#2}}

\def\U{\mathrm{U}(1)}

\def\T{{{\mathcal{T}}}}
\def\R{\mathbb{R}}
\def\Z{\mathbb{Z}}

\makeatletter
\def\l@subsubsection#1#2{}
\makeatother

\usetikzlibrary{decorations.pathreplacing,decorations.markings}
\tikzset{middlearrow/.style={
        decoration={markings,
            mark= at position 0.55 with {\arrow{#1}} ,
        },
        postaction={decorate}
    }
}

\begin{document}

\title{Systematic Construction of Interfaces and Anomalous Boundaries\\ for Fermionic Symmetry-Protected Topological Phases}

\date{May 20, 2025}

\author{Kevin Loo}
\email{lu-jy20@mails.tsinghua.edu.cn}
\affiliation{Yau Mathematical Sciences Center, Tsinghua University, Haidian, Beijing 100084, China}
\author{Qing-Rui Wang}
\email{wangqr@mail.tsinghua.edu.cn}
\affiliation{Yau Mathematical Sciences Center, Tsinghua University, Haidian, Beijing 100084, China}

\begin{abstract}
We use the pullback trivialization technique to systematically construct gapped interfaces and anomalous boundaries for fermionic symmetry-protected topological (FSPT) states by extending their symmetry group $G_f = \mathbb{Z}_2^f \times_{\omega_2} G$ to larger groups. These FSPT states may involve decoration layers of both Majorana chains and complex fermions. We derive general consistency formulas explicitly for (2+1)D and (3+1)D systems, where nontrivial twists arise from fermionic symmetric local unitaries or ``gauge transformations'' that ensure coboundaries vanish at the cochain level. Additionally, we present explicit example for a (3+1)D FSPT of symmetry group $G_f=\mathbb{Z}_2^f \times \mathbb{Z}_4 \times \mathbb{Z}_4$ with Majorana chain decorations.
\end{abstract}

\maketitle
\tableofcontents

\section{Introduction}

Topological phases of quantum matter fundamentally differ from traditional phases in that they cannot be described by the Landau paradigm of spontaneous symmetry breaking. Through the lens of quantum information theory \cite{wen1989,Wen1990to,Levin2006to,Chen2010lu}, certain systems can exhibit distinct phases even in the absence of symmetry, arising from different patterns of long-range entanglement (LRE), commonly referred to as topological orders. These phases with topological order are typically associated with anyonic excitations in the bulk and nontrivial edge states at their boundaries. The concept of finite-depth local unitary (LU) transformations has since become central to constructing, understanding, and classifying these unconventional quantum phases. By definition, the ground states of systems with intrinsic topological order cannot be mapped to a trivial product state by any finite-depth LU transformation. In other words, a phase transition must occur between the topologically ordered states and the trivial states under continuous deformations.

%In contrast, there are systems belonging to the short-range entangled (SRE) class which do not contain intrinsic topological order. All the phases in this class can be adiabatically connected to a trivial phase under certain LU transformations. Moreover, we can allow SRE phases with global symmetry, which are called symmetry-protected topological phases.  

%A refinement of the notion of LU transformation to symmetric local unitary (SLU) transformation suggested an equivalence class of states without intrinsic topological order -- short-range entangled (SRE) state -- which has states belonging to different phases even if they do not break the global symmetry. These phases are dubbed the symmetry-protected topological (SPT) phases.
%It was realized in \refns{gu2009spt,chen2011spt} that in a class of short-range entangled (SRE) phases with global symmetry, the SRE states can belong to many different phases without breaking the symmetry (without a phase transition which closes the gap). SPT phases are gapped phases that are robust against any local perturbations preserving its global symmetry. 

In contrast, there are quantum systems that can be connected to a trivial phase under finite-depth LU transformations. These systems are known as short-range entangled (SRE) states. However, when considering the symmetry of the system, they are nontrivial in the sense that they cannot be connected to a trivial state via symmetric local unitary (SLU) transformations, which are a refinement of LU transformations that take symmetry into account. These states are referred to as symmetry-protected topological (SPT) phases \cite{Gu2009spt,CTSR,Ludwig,Senthil2015,wen2017review}. While SPT phases do not feature anyonic excitations, they can possess nontrivial symmetric edge states, which are often characterized by quantum field theories with 't Hooft anomalies.

Tremendous progress has been made in the construction, classification, and characterization of SPT phases over the past few decades. One of the simplest and most prominent examples of SPT phases, which have been observed experimentally, are topological insulators and superconductors \cite{Hasan_2010, Qi_2011, Hasan_2011}. These systems are SPT states in the non-interacting limit. A complete classification of SPT phases in non-interacting fermion systems, across any symmetry group and dimension, has been achieved through the K-theory \cite{Kitaev:2009mg, Ryu_2010, Wen2011free,KdvKS2017}.
In systems with interactions, the group cohomology theory has shown to be a systematic and fruitful framework for classifying interacting bosonic SPT (BSPT) states \cite{Gu2009spt,chen2013bspt,Chen:2011pg,Wen2014nlsm}. Additionally, SPT phases have been constructed using a variety of other approaches, including field theory and cobordism theories \cite{Kapustin:2014tfa, KTTW2015, Yonekura2018, WGW2014, PWY2016, GK2016, FH2016, BGK2016,TCSR2017, WCWG2018, WOPZ2018, NWWG2021, GJ2017}, domain wall decoration \cite{Chen2014dw, WNC2018, Yang_LSMSPT, HuangPRB2017}, matrix product state constructions \cite{Kou:2009vea,Chen:2010zpc,CGW2011, SPC2011, kitaev2011, Lam2023}, and category theory \cite{Lan:2016rcq,Lan:2016seg,Kong:2020cie,Kong:2021equ}.
Although the bulk of SPT phases are trivial in the sense of no anyons, after gauging, these phases can be characterized by the braiding statistics of anyonic or loop excitations \cite{Levin2012spt,WL2014,JMR2014,LL2015,wang2016bs,CTW2018,WCWG2018,ZWWG2019,Non-Abelian,fbraiding,Hai:2023osv}, topological response theory \cite{YW2013, cg2014}, and various topological invariants \cite{kitaev2010free, Hung:2012dx, Wen:2013ue, Hung:2013cda, Qi2013, Santos:2013uda, Freed2014, Gu:2015lfa, WangLevin2015, Tantivasadakarn_2017, SSGR2017, CKTY2018, Inamura:2023ldn,Kobayashi:2024bts}. On the boundary of SPT phases, there are anomalous boundary states \cite{Wen2013a, WSW2014, KT2014anomaly, Witten2016parity, Witten2016f, Bulmash2021anomaly,Lu_LSMSPT, ChoPRB2017, Cheng2018f, JianPRB2018, JCQL2019, Kobayashi:2021jsc}.
The notion of SPT phases has also been generalized to subsystem symmetries \cite{You:2018oai,Devakul:2018fhz,Devakul:2019duj,Burnell:2021reh,Zhou:2022eig,You:2024syf}, higher-form symmetries \cite{Thorngren2015, Tachikawa2017, WW2018, Tsui:2019ykk, Kong:2020cie, KLWZZ2020class, Inamura2021, IO2024}, and even non-invertible symmetries \cite{Lan:2016rcq, Lan:2016seg, SS2024, MF2024, LOZ2023, Meng:2024nxx,Aksoy:2025rmg}.

In fermionic interacting systems, the construction and classification of fermionic SPT (FSPT) phases are even more intricate. A physical construction of FSPT phases is provided by the general group supercohomology theory \cite{GuWen2012, wang2017towards, wang2018construction}, which builds on the concepts of fermionic SLU (FSLU) and domain wall decorations \cite{Chen2014dw}. Mathematically, the construction of domain wall decorations can be described using spectral sequences \cite{wang2021domain}, and the computations can be performed with the algorithm proposed in Ref.~\onlinecite{Ouyang2020}.
Other approaches to constructing FSPT phases include spin bordisms and others \cite{ryuzhang2012, yaoryu2013, wps2014, gulevin2014, youxu2014, mfm2015, TF2016, wlg2017, KT2017, GPW2017, cbyg2018, CW2018, ThorngrenPRX2018, WangASPT, PW2020letter, PW2020, Barkeshli:2021ypb, ZNQG2022}.

The nontrivial nature of an SPT phase is manifested in the presence of gapless or degenerate edge states, provided that the system has a symmetric boundary \cite{haldane1983, AKLT1988, Hagiwara1990, glarum1991, NG1994, qhe2005z2, qhe2005, qhe2006, topinv2007, ti2007, ti2008, Chen:2012hc,Liu:2012taf}. The low-energy effective theory of a nontrivial SPT phase is described by an invertible topological quantum field theory that incorporates some global symmetry. This theory contributes a U(1) phase factor to the partition function of a closed spacetime manifold, which depends on the configuration of background gauge fields associated with the global symmetry.
At the boundary, however, the system is usually described by a symmetric gapless quantum field theory characterized by a 't Hooft anomaly \cite{tHooft1979}, which can be canceled by the bulk SPT phase through the process of anomaly inflow \cite{chen2011spt, Chen:2011pg, Levin2012spt,Witten2016f}. 
The presence of this 't Hooft anomaly means that we cannot independently gauge the global symmetry on the boundary alone. However, the situation is different when we gauge the entire system, including both the bulk and the boundary.

In addition to symmetric gapless states, the boundary of an SPT phase can also be gapped, provided it still possesses the same 't Hooft anomaly, as long as the boundary is in (2+1)D or higher dimensions \cite{fcv2013, BNQ2013, wps2013, VS2013, MKF2013, BCFV2014, cfv2014, mfcv2014, ws2014, mkf2015, mkf2015, MEA2015, WLL2016, FVM2018, NMLW2021, TKBB2021, YC2023, CWY2024}. In (2+1)D, a gapped state can be described by a modular tensor category, which characterizes its anyonic excitations. The global symmetry action on the boundary may permute the anyons and even fractionalize the symmetry on them, a feature known as symmetry-enriched topological (SET) states \cite{etingof_fusion_2009, Heinrich_2016, Barkeshli:2016mew, Wen:2016cij, Cheng_2017, BBCW2019, Barkeshli:2019vtb, Aasen2021, Bulmash:2020flp, Bulmash2021frac, Bulmash2021anomaly, Lan:2023uuq, ZG2024}. In higher dimensions, however, the complete mathematical description of topological orders remains unknown, and thus a full generalization of SET states in higher dimensions is still lacking.
Interestingly, a less general but notable perspective was proposed in Ref.~\onlinecite{WWW2017}, where the authors presented a systematic construction of gapped boundary SPT phases in bosonic systems by extending the symmetry at the boundary. Upon gauging the new boundary symmetry, a symmetry-enriched gauge theory emerges at the boundary. This construction has been supported by various recent works on fully symmetric gapped surfaces with topological orders.

In this paper, we aim to generalize this pullback construction to fermionic systems. We propose a systematic framework for constructing gapped boundaries for arbitrary FSPT phases.

\subsection{Overview of setup}\label{sec:introsetup}

To systematically construct symmetry-preserving gapped boundaries for FSPT phases, the main idea is to extend the global symmetry in the bulk to a larger symmetry on the boundary. 
More precisely, we mean that the global symmetry group $G$ acting on both the bulk and boundary is extended to a larger group, say, $G'$, via a short exact sequence, $1 \rightarrow A \rightarrow G' \overset{\pi}{\rightarrow} G \rightarrow 1$. However, the group $A$  acts trivially in the bulk as $\pi(a)=1\in G$, for $a\in A\subset G'$. This renders the quotient group $G=G'/A$ as the true symmetry acting in the bulk rather than $G'$.

A well-known example where this symmetry-lifting mechanism is applicable is the spin-1 Haldane phase, as demonstrated in \refns{WWW2017, PWW2018}. Specifically, the (1+1)D spin-1 Haldane chain \cite{haldane1983,haldanechain}, protected by SO(3) symmetry, represents a nontrivial BSPT phase. In the case of an \emph{open} spin chain, the ground state exhibits a fourfold degenerate edge state, formed by dangling spin-1/2 qubits at its two ends. These spin-1/2 states transform as projective representations of the global SO(3) symmetry. However, we can alternatively interpret the spin-1/2 as a genuine irreducible representation of the larger group SU(2). In other words, the global SO(3) symmetry can be extended to SU(2) via the short exact sequence:
\begin{align}
1 \rightarrow \Z_2 \rightarrow \mathrm{SU}(2) \rightarrow \mathrm{SO}(3) \rightarrow 1.
\end{align}
By extending the bulk SO(3) symmetry to SU(2) on the boundary, we can introduce an additional spin-1/2 degree of freedom on the boundary. This new spin-1/2 can pair with the dangling spin-1/2 edge state to form a spin-0 singlet. This pairing results in a gapped edge, effectively removing the edge degeneracy. In other words, the symmetry-extended boundary achieves a unique ground state, characterized by the spin-0 singlet formed between the edge state of the Haldane chain and the newly introduced spin-1/2.

We aim to generalize the pullback construction to fermionic systems in order to systematically construct symmetry-preserving gapped boundaries. The symmetry of a fermionic system is given by $G_f=\Z_2^f\times_{\omega_2}G$, where $\Z_2^f$  represents the fermion parity group, $G=G_f/\Z_2^f$  is the bosonic symmetry group, and $\omega_2\in H^2(G,\Z_2)$ is the 2-cocycle specifying the central extension of the sequence 
\begin{align}\label{Gf}
1\rightarrow \Z_2^f \rightarrow G_f \rightarrow G\rightarrow 1.
\end{align}
In the general supercohomology framework for constructing FSPTs, a $G_f$-FSPT in, for example, (3+1)D is constructed from a tuple of cocycles or cochains $(n_2,n_3,\nu_4)$ that satisfy certain consistency conditions. To also specify the symmetry extension, we will use the tuple $(\omega_2;n_2,n_3,\nu_4)$ to denote a particular FSPT state.

In parallel to the bosonic case, we can also consider a fermionic symmetry group extension
\begin{align}
1\rightarrow A \rightarrow G_f' \rightarrow G_f \rightarrow 1,
\end{align}
such that a fermionic state with symmetry $G_f$, labeled by $(\omega_2;n_2,n_3,\nu_4)$ as an element in the generalized cohomology for FSPT phases, is pulled back to a trivial state with the extended symmetry $G_f'$. There should also exist an Atiyah-Hirzebruch spectral sequence associated with this generalized cohomology to address the trivialization problem. However, computing the differentials in this spectral sequence is challenging, if not impossible, and they often lack a physical derivation or clear interpretation.

\begin{widetext}
\begin{figure*}[ht]
    \centering
    \includegraphics[scale=.4]{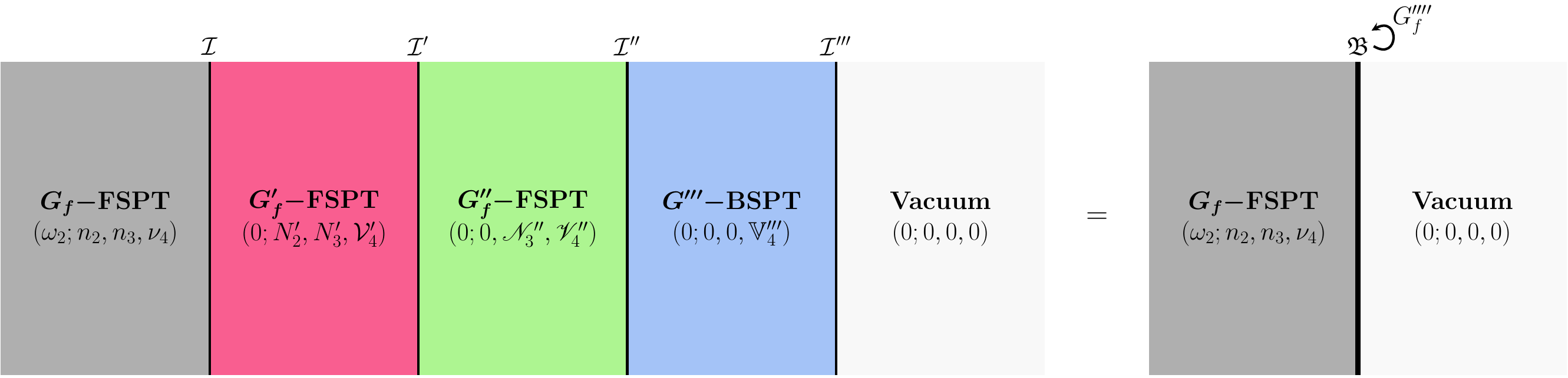}
    \caption{
    Illustration of the pullback trivialization procedure to construct symmetric gapped domain walls or boundaries for (3+1)D FSPT states. Each region corresponds to an FSPT phase with a different symmetry groups and decoration data, with gapped interfaces separating them, according to the group extensions Eqs.~(\ref{I0}), (\ref{I1}), (\ref{I2}) and (\ref{I3}). By shrinking all the intermediate FSPT layers, we ultimately obtain the symmetric gapped boundary $\mathfrak B$ with $G_f''''$ symmetry for the original FSPT state.}
    \label{fig:ext}
\end{figure*}
\end{widetext}

The procedure we adopt in this paper to construct the symmetric gapped boundary of an FSPT $(\omega_2;n_2,n_3,\nu_4)$ involves a sequence of pullback trivializations of the FSPT layers, one by one. The physical picture is illustrated in Fig.~\ref{fig:ext}. The basic idea is to pullback and trivialize the four data components $(\omega_2;n_2,n_3,\nu_4)$ step by step by extending the symmetry group to larger symmetry groups. Each pullback trivialization generates a gapped interface between FSPT layers with different decoration structures. After trivializing all the fermionic data, we are left with a BSPT, which can then be trivialized again to produce a gapped interface with the vacuum. By shrinking all the intermediate steps, we ultimately obtain a symmetric gapped boundary between the original FSPT and the vacuum (see the right-hand side of Fig.~\ref{fig:ext}). This process yields the desired symmetric gapped boundary.

To be more specific, the first gapped interface $\mathcal I$ is between the original FSPT with symmetry group $G_f = \Z_2^f \times_{\omega_2} G$ and nontrivial extension $\omega_2$, and an FSPT with symmetry group $G_f' = \Z_2^f \times G'$ and a trivial extension. The pullback trivialization process involves a short exact sequence of groups:
\begin{align}\label{I0}
&1 \rightarrow A \rightarrow G' \overset{\pi}{\rightarrow} G \rightarrow 1\qquad
(\omega_2\text{-interface\ }\mathcal{I}),
\end{align}
where the extension 2-cocycle $\omega_2$ is pulled back to $\omega_2' = \pi^\ast \omega_2$, which becomes a coboundary in $H^2(G', \Z_2)$. In this way, the FSPT $(\omega_2; n_2, n_3, \nu_4)$ of symmetry $G_f$ can be identified with an FSPT $(0; N_2', N_3', \mathcal{V}_4')$ of symmetry $G_f'$. The new decoration data (denoted with a prime) are obtained by twisting the original decoration data (without primes) due to the trivialization of $\omega_2'$. As a result, there is a gapped interface between these two FSPTs.

Similarly, the Majorana chain decoration data $n_2$ and complex fermion decoration data $n_3$ can be pullback-trivialized by further extending the symmetry groups
\begin{align}\label{I1}
&1 \rightarrow A' \rightarrow {\ } G'' \overset{\pi'}{\rightarrow} G' \rightarrow 1\qquad (\text{Majorana chain interface\ }\mathcal{I}'),\\\label{I2}
&1 \rightarrow A'' \rightarrow {\ } G''' \overset{\pi''}{\rightarrow} G'' \rightarrow 1\qquad (\text{complex fermion interface\ }\mathcal{I}''),
\end{align}
resulting in a BSPT with symmetry group $G'''$, labeled by a 4-cocycle $\mathbb{V}_4'''$. There are two gapped interfaces, $\mathcal{I}'$ and $\mathcal{I}''$, corresponding to the two extensions. The construction of the gapped boundary $\mathcal{I}'''$ of the BSPT $\mathbb{V}_4'''$ can be done again using the pullback construction \cite{WWW2017, PWW2018}:
\begin{align}\label{I3}
&1 \rightarrow A''' \rightarrow {\ } G'''' \overset{\pi'''}{\rightarrow} G''' \rightarrow 1\qquad (\text{bosonic interface\ }\mathcal{I}''').
\end{align}
When shrinking all the intermediate FSPT layers and treating them as effectively a (2+1)D system, we obtain a gapped boundary for the original FSPT $(\omega_2; n_2, n_3, \nu_4)$ with symmetry group $G_f$. However, the symmetry of the boundary is enlarged to a larger group $G_f''''$, where a subgroup acts only on the boundary, and a quotient group acts on both the bulk and the boundary.

Another way to understand the symmetry extension is by gauging the new subgroup that acts solely on the boundary, leaving only the quotient group to act on both the bulk and the boundary. In this framework, the (2+1)D boundary becomes a gauge theory with the original $G_f$ symmetry, effectively forming a symmetry-enriched gauge theory.

\subsection{Summary of main results}

Let us specify the extension of the bosonic symmetry $G$ to the fermionic symmetry via the 2-cocycle $\omega_2 \in H^2(G, \mathbb{Z}_2)$. The cochains $n_{d-1} \in C^{d-1}(G, \mathbb{Z}_2)$ and $n_d \in C^d(G, \mathbb{Z}_2)$ describe the decorations of (1+1)D Majorana chains and (0+1)D spinless complex fermions in the $(d+1)$D model, respectively. The cochain $\nu_{d+1} \in C^{d+1}[G, \U]$ is the bosonic $\U$ phase factor that appears in the wavefunction of SPT phases. Similarly, we define lower-degree cochains $\tau_1 \in C^1(G', \mathbb{Z}_2)$, $m_{d-2} \in C^{d-2}(G', \mathbb{Z}_2)$, $m_{d-1} \in C^{d-1}(G', \mathbb{Z}_2)$, and $\mu_d \in C^{d}[G', \U]$ for the largest extended symmetry group [which may be, for example, $G''''$ as in \eq{I3}]. These cochains represent the trivialization of the decoration data for the pullback constructions. Physically, they represent the relevant decoration data on the boundary with the larger symmetry group $G'$ in the last pullback step.

Our construction of symmetric gapped boundaries for $(\Zf \times_{\omega_2} G)$-FSPT phases is achieved by extending all relevant cocycles or cochains to a larger group $G'$, such that these cocycles, possibly twisted by some cochains (because of FSLU or ``gauge transformations''), become coboundaries. These cochains must satisfy the following consistency equations when pulled back to the cochains of $G'$.

\begin{widetext}
(2+1)D $(\Zf\times_{\om_2'}G')$-FSPT phase
$(\om_2';n_1',n_2',\nu_3')$:
\begin{align}
    \om_2'&\={2}\dd\tau_1',\\
    n_1'&\={2}\dd m_0',\\
    n_2'&\={2}\dd m_1'+\tau_1'\smile \dd m_0',\\\nonumber
    \nu_3'&\={1} \dd \mu'_2+ \frac12 m'_1\smile \dd m'_1-\frac14 \tau'_1\smile(\dd m'_0)^2
    +\frac12\dd m'_1\smile_1(\tau'_1 \smile\dd m'_0) \\&\quad
    +\frac12\tau'_1\smile(\dd m'_1+\tau'_1 \smile\dd m'_0)
    +\frac12 \tau'_1\smile(\tau'_1\smile_1\dd m'_0)\smile\dd m'_0.
\end{align}

(3+1)D $(\Zf\times_{\om_2'}G')$-FSPT phase $(\om_2';n_2',n_3',\nu_4')$:
\begin{align}
    \om_2' &\={2} \dd\tau'_1,\\
    n_2' &\={2} \dd m'_1,\\
    n_3' &\={2} \dd m'_2+(\tau'_1+m'_1)\smile \dd m'_1,\\\nonumber
    \nu_4' &\={1} \dd \mu'_3 -\frac14\dd m'_1\smile\Sq^1(m_1')
    +\frac12 [\Sq^2(m_2') + \dd m'_2\smile_2 (m_1' \smile\dd m'_1)]\\\nonumber
    &\quad+\frac12\dd m'_1\smile\sqb{(m_1')^2\smile_2 \dd m'_1}
    +\frac12 (m_1')^2 \smile(\dd m'_1\smile_1 m'_1)
    +\frac12 m_1'(01)m_1'(12)m_1'(14)\dd m'_1(234)\\\nonumber
    &\quad +\frac12\tau'_1\smile\sqb{\dd m'_2+(\tau'_1+m_1')\smile \dd m'_1}
    +\frac12\sqb{\dd m'_2+(\tau'_1+m_1')\smile \dd m'_1}\smile_2(\tau'_1\smile\dd m'_1)
    +\frac12\Sq^1(\tau'_1)\smile\dd m'_1\\
    &\quad
    +\frac12 (\tau'_1\smile\dd m'_1)\smile_3 (\dd m'_1)^2
    + \frac12 \dd m'_1(134)\dd m'_1(123)\tau'_1(01)
    +\frac12[1+\dd m'_1(124)]\dd m'_1(234)\tau'_1(01)\tau'_1(02).
\end{align}
\end{widetext}

In addition to the general results presented above, we also provide nontrivial examples based on the pullback construction. In (2+1)D, a nontrivial FSPT with symmetry $\mathbb{Z}_2^f \times \mathbb{Z}_4 \times \mathbb{Z}_4$ is pulled back to a nontrivial $\mathtt{SmallGroup}(32,2)$-BSPT, and further to a trivial $\mathtt{SmallGroup}(64,23)$-BSPT. This means that all three phases have symmetric gapped interfaces. By shrinking the last two SPTs, we obtain a gapped boundary for the nontrivial (2+1)D $(\mathbb{Z}_2^f \times \mathbb{Z}_4 \times \mathbb{Z}_4)$-FSPT. A similar result holds in (3+1)D: a nontrivial $(\mathbb{Z}_2^f \times \mathbb{Z}_4 \times \mathbb{Z}_4)$-FSPT with Majorana chain decorations is pulled back to a nontrivial $(\mathbb{Z}_2^f \times \mathtt{SmallGroup}(32,2))$-FSPT with only complex fermion decorations and further pulled back to a $\mathtt{SmallGroup}(64,23)$-BSPT. The latter BSPT can be trivialized using known results in bosonic system, thus completing the construction of the symmetric gapped boundary for the (3+1)D $(\mathbb{Z}_2^f \times \mathbb{Z}_4 \times \mathbb{Z}_4)$-FSPT with Majorana chain decorations after shrinking the other layers.

\subsection{Organization of the paper}

The remainder of the paper is organized as follows. In Sec.~\ref{sec:FSPTconstruction}, we provide a brief review of the construction of $(\mathbb{Z}_2^f \times_{\omega_2} G)$-FSPT phases within the framework of general supercohomology theory. In Sec.~\ref{sec:bspt}, we revisit the gapped boundary construction for BSPT phases, demonstrating the pullback trivialization procedure for a (3+1)D BSPT phase as a foundational example for generalizing to FSPT cases.

In Sec.~\ref{sec:2Dbdy}, we extend the pullback construction approach to (2+1)D FSPT phases with $\mathbb{Z}_2^f \times_{\omega_2} G$ symmetry. Sections \ref{sec:3Dbdy1} and \ref{sec:3Dbdy2} focus on gapped boundary constructions for (3+1)D $(\mathbb{Z}_2^f \times_{\omega_2} G)$-FSPT phases in two specific settings: Sec.~\ref{sec:3Dbdy1} addresses the case of $\omega_2 = 0$ with only complex fermion decorations, while Sec.~\ref{sec:3Dbdy2} explores $\omega_2 = 0$ with Majorana chain decorations. Each procedure is illustrated with concrete examples. Sec.~\ref{sec:3Dbdy3} generalizes the pullback trivialization method to handle (3+1)D FSPT phases with nontrivial $\omega_2$ extensions. Finally, we summarize our findings and discuss potential future directions in Sec.~\ref{sec:conclusion}.

Appendix~\ref{sec:1Dbdy} discusses the boundary construction of a (1+1)D FSPT phase, which is in some sense a degenerate case. Appendix~\ref{sec:O5om} demonstrates how the obstruction function $\mathcal{O}_5$, as defined in Ref.~\onlinecite{wang2018construction}, can be simplified using consistency equations and the Bockstein homomorphism. Appendix~\ref{sec:triveqz} provides the trivialization equation for $n_2 \in H^2(G, \mathbb{Z}_2 = \{0, 1\})$ viewed as a $\mathbb{Z}$-valued 2-cochain. Detailed calculations for the pullback trivialization of the obstruction function $\mathcal{O}_5$ in the $\omega_2 = 0$ case are given in Appendix~\ref{sec:O5pb}.

\noindent \textbf{Notations and conventions}: Throughout the text, the cyclic group is denoted as $\mathbb{Z}_N = \{0, 1, \cdots, N-1\}$, with group multiplication defined as addition modulo $N$. The phase factor $\mathrm{U}(1)$ is identified with the additive group $\mathbb{R}/\mathbb{Z}$ of real numbers modulo integers. We use the notations $\={n}$ to denote equality modulo $n$ and $\={\mathbb{Z}}$ to indicate equality within the ring of integers.
%$\={\dd}$ to represent equality up to coboundaries.

\section{Review of FSPT construction and pullback construction}

In this section, we review the two fundamental cornerstones for the systematic construction of symmetric gapped boundaries for FSPT states. The first is the construction of fixed-point states for FSPT phases using the generalized supercohomology theory \cite{GuWen2012,wang2017towards,wang2018construction}. The second is the pullback construction, which has been employed in constructing gapped boundaries for bosonic SPT phases \cite{WWW2017, PWW2018}. In this paper, we integrate these two approaches to develop a unified and systematic framework for constructing symmetric gapped boundaries for FSPT states by pullback trivialization.

\subsection{Construction of FSPT states using the generalized supercohomology theory}
\label{sec:FSPTconstruction}

In a fermionic system, the symmetry group  $G_f=\Zf\times_{\om_2}G$ is a central extension of the bosonic symmetry group $G$ by the fermion parity symmetry group $\Z_2^f=\{1,P_f\}$, where $P_f=(-1)^F$ denotes the fermion parity operator. All physical symmetries should not change the fermion parity of the FSPT state, i.e., commute with the fermion parity operator $P_f=(-1)^F$, where $F$ counts the total fermion number. In other words, the subgroup $\Z_2^f$ is in the center of $G_f$. For a given bosonic symmetry group $G$, there are many different plausible extensions to $G_f$. We can specify the extension by a 2-cocycle $\om_2\in H^2(G,\Z_2=\{0,1\})$, which satisfies the cocycle equation $\dd\om_2(g,h,k):=\om_2(h,k)+\om_2(gh,k)+\om_2(g,hk)+\om_2(g,h)=0\ (\text{mod 2})$.

With an adequate understanding of fermionic invertible topological orders with only $\Zf$ symmetry, we can construct FSPT fixed-point states with symmetry group $G_f=\Zf\times_{\om_2}G$ using domain wall decorations \cite{chen2014symmetry,wang2021domain}. The key idea is to decorate fermionic invertible topological states at the intersection points, lines, and surfaces (and even hypersurfaces) of the domain walls of the bosonic $G$ symmetry. After proliferating the $G$ symmetry domain walls, the full $G_f$ symmetry can be restored, allowing the system to be an FSPT state.

To better understand the construction of gapped boundaries, it is essential to first grasp the process of constructing the bulk FSPT state. In this section, we will outline the basic idea behind the construction of a (3+1)-dimensional FSPT. For a more detailed treatment of the construction, we refer to \refns{wang2017towards, wang2018construction}.

As the first step, we triangulate a 3D spatial spin manifold (incorporating branching structures) and assign a label to each vertex $i$ using elements $g_i$ from the bosonic symmetry group $G$ \footnote{The manifold should be spin with a chosen spin structure, as this is necessary to define fermions on it. However, since we are constructing gapped boundaries for a state on a region with the topology of an open disk, the issue of spin structure does not arise.}. The label $g_i \in G$ represents a $G$-symmetry domain associated with vertex $i$. For simplicity, we assume that $G$ is a \emph{unitary} finite symmetry group in this paper.

The decorated fermionic invertible topological order resides on different subsimplices of the tetrahedra within the triangulation of the 3D spatial manifold. The degrees of freedom for a single 3-simplex (tetrahedron) are described as follows.
\begin{align}\label{decoration3d}
    \vcenter{\hbox{\includegraphics[scale=1]{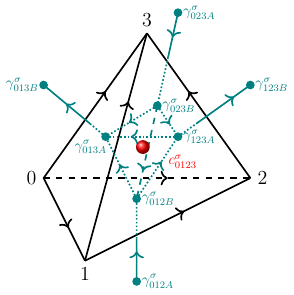}}}
\end{align}
Here, the group elements $g_0, g_1, g_2, g_3$ on the vertices are occasionally abbreviated as $0, 1, 2, 3$. The center of a tetrahedron $\langle ijkl \rangle$, located at the intersection of six $G$-symmetry domain walls perpendicular to its edges, is decorated with $|G|$ species of complex fermion modes, $c_{ijkl}^\sigma$, indexed by $\sigma \in G$ [as represented by the red ball in \eq{decoration3d}]. The decoration is controlled by a $\mathbb{Z}_2$-valued function $n_3(g_i, g_j, g_k, g_l)$, which depends on the tuple $(g_i, g_j, g_k, g_l) \in G^{\times 4}$. If $n_3 = 0$, all complex fermion modes $c_{ijkl}^\sigma$ at the center of the tetrahedron are unoccupied. Conversely, if $n_3 = 1$, exactly one fermion mode, $c_{ijkl}^{g_i}$, is occupied at the center of the tetrahedron $\langle ijkl \rangle$, while all other modes $c_{ijkl}^\sigma$ (with $\sigma \neq g_i$) remain in their vacuum states.

An additional layer of decoration involves placing (1+1)D Kitaev's Majorana chains at the intersection lines of the $G$-domain walls. This is achieved by assigning $|G|$ species of Majorana fermions, $\gamma^\sigma_{ijk,A}$ and $\gamma^\sigma_{ijk,B}$, to the (green) vertices on the two sides of each (black) triangle $\langle ijk \rangle$, where $\sigma$ corresponds to the group element in $G$. The decoration of the Kitaev chain is determined by the pairing of Majorana fermions, which is specified by a function $n_2(g_i, g_j, g_k) \in \mathbb{Z}_2$, with $g_i, g_j, g_k \in G$. If $n_2 = 0$, the Majorana fermions $\gamma^\sigma_{ijk,A}$ and $\gamma^\sigma_{ijk,B}$ on the two sides of the triangle $\langle ijk \rangle$ are paired in their vacuum state. If $n_2 = 1$, a Kitaev chain is formed by pairing the Majorana fermions along the link passing through the triangle $\langle ijk \rangle$. For each of the $|G|$ species of Majorana fermions, we choose to pair $\gamma^{g_i}_{ijk,A}$ and $\gamma^{g_i}_{ijk,B}$ nontrivially. All other $|G| - 1$ species of Majorana fermions, $\gamma^\sigma_{ijk,A}$ and $\gamma^\sigma_{ijk,B}$ for $\sigma \neq g_i$, remain in their vacuum pairings.

The decoration procedure outlined above involves three layers of degrees of freedom on the 3D triangulated lattice, which can be captured by the data $(\omega_2; n_2, n_3, \nu_4)$, where $\nu_4$ represents the bosonic phase factor. To ensure that the decoration corresponds to a fixed-point state of a symmetric gapped phase of matter, the set of decoration data must satisfy several consistency conditions. For instance, the decorated Kitaev chains must form closed loops without any dangling Majorana fermions within any tetrahedron. This requires that the total number of Kitaev chains passing through the boundary of a tetrahedron be even, which leads to the consistency condition $\dd n_2 = 0$. The wavefunctions of the FSPT state on two different triangulated lattices are related by wavefunction renormalization, which can be induced by Pachner moves or $F$-moves. These moves are interpreted as fermionic symmetric local unitary (FSLU) transformations between the fermionic Fock spaces of the two triangulations. Under such an $F$-move, the fermion parity of both the Majorana and complex fermions should be conserved. Therefore, we have the consistency condition $\dd n_3 = (\omega_2 + n_2) \smile n_2$. The FSLU transformations must also satisfy a coherence condition, known as the fermionic superhexagon equation. The reordering of fermionic operators in this equation introduces a twist in the cocycle condition for the bosonic $\U$-valued phase factor $\nu_4 \in C^4[G, \U]$, influenced by an obstruction function involving $\omega_2$ and lower-dimensional cochains $n_2 \in C^2(G, \mathbb{Z}_2)$ and $n_3 \in C^3(G, \mathbb{Z}_2)$. This results in the twisted cocycle equation $\dd \nu_4 = \mathcal{O}_5[\omega_2; n_2, n_3]$. The explicit form of $\mathcal{O}_5[\omega_2; n_2, n_3]$ is given in \eq{dnu4om}. In summary, the decoration data must satisfy the following consistency equations:
\begin{align}
\dd n_2 &= 0,\\
\dd n_3 &= (\om_2+n_2)\smile n_2,\\
\dd \nu_4 &= \mathcal O_5[\om_2;n_2,n_3].
\end{align}

In addition to the (3+1)D FSPT, we can also apply the general framework of supercohomology theory to construct FSPTs in other dimensions. For the (1+1)D case, we assign each vertex with $|G|$ degrees of freedom and decorate complex fermions on the oriented links connecting the vertices. The decoration data is described by $(\omega_2; n_1, \nu_2)$, where $n_1$ governs the occupation of complex fermions on each link, and $\nu_2$ corresponds to the standard BSPT classification data (see Appendix \ref{sec:1Dbdy} for the associated consistency conditions). For (2+1)D $G_f$-FSPT phases, we similarly assign the vertices with $|G|$ degrees of freedom, and we decorate complex fermions at the center of each triangle and Majorana fermions on the two sides of each edge of a triangle. The decorations of the complex fermions and Kitaev chains are specified by the cochains $n_1$ and $n_2$, with $\nu_3$ encoding the usual BSPT phase factor. Together with $\omega_2$, the algebraic data characterizing an FSPT phase in (2+1)D are encoded in the set $(\omega_2; n_1, n_2, \nu_3)$, which should satisfy various consistency conditions (see Sec.~\ref{sec:2Dbdy}).

For unitary group $G$, the symmetry transformation of the bosonic states under $G$ is $U(g)\ket{g_i}=\ket{gg_i}$ on arbitrary vertex $i$, while the symmetry transformation of complex fermions $c^\sigma_{ij\cdots k}$ and Majorana fermions $\gamma^\sigma_{ij\cdots k, A/B}$ under $G$ are given by~\cite{wang2018construction}
\begin{align}\label{symmc}
    U(g)c^\sigma_{ij\cdots k}U(g)^\dagger &=(-1)^{\om_2(g,\sigma)}c^{g\sigma}_{ij\cdots k},\\\label{symmgamma}
    U(g)\gamma^\sigma_{ij\cdots k, A/B}U(g)^\dagger &=(-1)^{\om_2(g,\sigma)}\gamma^{g\sigma}_{ij\cdots k,A/B},
\end{align}
where the subscript $ij\cdots k$ denotes the vertices on a simplex $\langle ij\cdots k\rangle$ of the corresponding spatial dimension.
These transformation rules guarantee that the FSPT states constructed through the above decorations are $G$-symmetric.

\subsection{Pullback construction of symmetric gapped boundaries for BSPT phases}
\label{sec:bspt}

Before discussing the pullback construction of symmetric gapped boundaries for FSPTs, it is instructive to first review the framework of symmetric gapped boundaries for bosonic SPT phases. In particular, we will outline the pullback trivialization framework used in constructing gapped boundaries, as presented in \refn{WWW2017}. We will focus on a (3+1)D BSPT phase with a finite unitary symmetry group $G$ and a group cocycle $\nu_4^G \in H^4[G, \U]$. This review will help build the necessary foundation for understanding how the pullback construction can be systematically used to construct gapped boundaries.

Assuming that a closed 3D spatial manifold $\mathcal M_{3}$ is triangulated, we assign to each (locally ordered) vertex $i$ a bosonic degree of freedom labeled by a group element $g_i\in G$. Equivalently, this means assigning a local Hilbert space of dimension $|G|$ to each vertex, where $|G|$ is the order of the group $G$. The symmetry $g\in G$ acts on $g_i$ by $gg_i$. The fixed-point ground state of the BSPT phase on a closed manifold is symmetric and is given by \cite{Chen:2011pg}:
\begin{align}
    \ket{\Psi}=\sum_{\{g_i\}} \Psi(\{g_i\}) \ket{\{g_i\}},
\end{align}
where the wavefunction $\Psi(\{g_i\})$ is a product of $\U$ complex phases determined by the 4-cocycle $\nu_4^G$ within each tetrahedron $\T$:
\begin{align}
\Psi(\{g_i\}) =\prod_{\T=\langle 1234\rangle\in \mathcal M_{3}} e^{2\pi i s_{\T} \nu_4^G (e,g_1,g_2,g_3,g_4)}.
\end{align}
Here, the factor $s_{\T}=\pm 1$ depends on the orientation of a tetrahedron $\T$. The 4-cocycle $\nu_4^G$ is homogeneous, satisfying $\nu_4^G (\{g_i\})=\nu_4^G (\{gg_i\})$ for all $g\in G$. We have assigned $g_0=e$ (the identity element of $G$) to a fixed point outside $\mathcal M_3$. For a closed manifold $\mathcal M_3$, the BSPT wavefunction is independent of the choice of $g_0$, as will be demonstrated below.

In the BSPT fixed-point construction, the global $G$-symmetry is represented by the operator
\begin{align}
    V(g)\equiv\sum_{\{g_i\}}\ket{\{gg_i\}}\bra{\{g_i\}},
\end{align}
for $g\in G$. It acts on the ground state $\ket{\Psi}$, yielding a shifted ground state $\ket{\Tilde{\Psi}}=\sum_{\{g_i\}}\Tilde{\Psi}(\{g_i\})\ket{\{g_i\}}$ with wavefunction
\begin{align}\label{shiftwavefn}
    \Tilde{\Psi}(\{g_i\})
    =\prod_{\T \in \mathcal M_{3}} e^{2\pi i s_{\T} \nu_4^G (e,g^{-1}g_1,g^{-1}g_2,g^{-1}g_3,g^{-1}g_4)} 
    =\prod_{\T \in \mathcal M_{3}} e^{2\pi i s_{\T} \nu_4^G (g,g_1,g_2,g_3,g_4)},
\end{align}
where we assigned $g$ to a virtual site, and the homogeneity property of $\nu_4^G$ is used in the second equality. Using the cocycle condition for $\nu_4^G$, i.e., $\dd\nu_4^G (e,g,g_1,g_2,g_3,g_4)=0$, the change in the wavefunction can be simplified using $\nu_4^G (g,g_1,g_2,g_3,g_4)-\nu_4^G (e,g_1,g_2,g_3,g_4) =
    -\nu_4^G (e,g,g_2,g_3,g_4)+\nu_4^G (e,g,g_1,g_3,g_4)
     -\nu_4^G (e,g,g_1,g_2,g_4)+\nu_4^G (e,g,g_1,g_2,g_3)$.
When substituted into the expression \eq{shiftwavefn}, the right-hand side cancels out, because each face of $\T\in\mathcal M_3$ is shared by two neighboring tetrahedra with opposite orientations, and these contributions cancel out. Each tetrahedron contributes a factor of $\nu_4^G(e,g,g_i,g_j,g_k)$ with opposite signs. Therefore, we have $\Tilde{\Psi}(\{g_i\})=\Psi(\{g_i\})$, and the BSPT state $\ket{\Psi}$ is $G$-symmetric if the spatial manifold $\mathcal M_3$ has no boundary.

On a manifold $\mathcal M_3$ with boundary $\partial \mathcal M_3$, the BSPT wavefunction is not invariant under $G$-symmetry transformation. In this case, the symmetry action $V(g)$ on the fixed-point wavefunction at the boundary may introduce a residual phase factor $\vartheta(\{g_i\})$. Specifically, the wavefunction in the transformed state becomes:
\begin{align}\label{residual}
    \Tilde{\Psi}(\{g_i\})&=\vartheta(\{g_i\}) \Psi(\{g_i\}),\\
    \vartheta(\{g_i\})&=\prod_{F \in \partial \mathcal M_3} e^{2\pi i s_F \nu_4^G (e,g,g_1,g_2,g_3)}.
\end{align}
The product here runs over all faces (triangles) $F$ on the boundary $\partial \mathcal M_3$. The factor $s_F=\pm 1$ indicates the orientation of the triangles. The presence of this residual phase factor indicates the existence of a 't Hooft anomaly — an obstruction to promoting the global symmetry to a local gauge symmetry at the boundary. This anomaly arises because the boundary is not gauge-invariant, with the appearance of a nontrivial phase factor. However, this phase factor can be canceled by the bulk through spectral inflow, where the bulk compensates for the gauge non-invariance at the boundary, ensuring the consistency of the overall system.

The pullback trivialization construction is a method to cancel the residual phase factor by enlarging the symmetry such that the cocycle becomes trivial in the larger symmetry group, as discussed in \refn{WWW2017}. Specifically, consider a group extension via the short exact sequence:
\begin{align}
1 \rightarrow A \rightarrow G' \overset{\pi}{\rightarrow} G \rightarrow 1,
\end{align}
where $\pi$ is a surjective group homomorphism from the bigger group $G'$ to the quotient group $G=G'/A$ and $A$ is a normal subgroup of $G'$.
In this framework, the gapped boundary $\partial \mathcal M_3$ can be viewed as a gapped interface (or domain wall) between two different BSPT phases: a $G$-BSPT phase with cocycle $\nu_4^G\in H^4[G,\U]$ and a ${G'}$-BSPT phase with the pulled-back cocycle $\nu_4^{G'}=\pi^\ast(\nu_4)\in H^4[G',\U]$. 
We note that any nontrivial cocycle of $G$ can be pulled back to a trivial one in $G'$ through an extension of $G$ by a finite group $A$~\cite{WWW2017,Tachikawa2017}.
Now, let us assume that we choose the extension such that the pulled-back cocycle $\nu_4^{G'}$ is a coboundary in $H^4[G', \mathrm{U}(1)]$, i.e.,
\begin{align}\label{cobdy}
\nu_4^{G'}\={1}\dd \mu_3^{G'},
\end{align}
where $\mu_3^{G'}$ is a 3-cochain of $G'$. In this case, the ${G'}$-BSPT state is trivial. The boundary then represents a $G'$-symmetric gapped interface between the trivial product state (or vacuum state) and the bulk state. After shrinking the $G'$-BSPT state, the interface effectively becomes a gapped boundary with an emergent $G'$ symmetry of the bulk $G$-BSPT. The pullback construction ensures that the boundary behaves in a $G$-symmetric manner, since $G$ is the quotient group of the bigger boundary symmetry $G'$.

More explicitly, we can show that the composite system consisting of the $G$-symmetric bulk and $G'$-symmetric boundary is anomaly-free. For brevity, we will henceforth denote the cochains $\mu_d^{G}$ and $\mu_d^{G'}$ of the two symmetries simply as $\mu_d$ and $\mu_d'$, respectively. The global $G'$-symmetry operator is given by 
\begin{align}
    V(g')\equiv \sum_{\{b_i\in G',g_j\in G\}}\ket{\{g'b_i,\pi(g')g_j\}}\bra{\{b_i,g_j\}}, \quad g'\in G',
\end{align}
where $b_i\in G'$ labels the boundary vertex $i\in\partial\mathcal M_3$, and $g_j\in G$ labels the bulk vertex $j\in\mathcal M_3\backslash\partial\mathcal M_3$. The total BSPT ground state is given by
\begin{align}\label{largegs}
    \ket{\Psi'}=\sum_{\{b_i,g_j\}} 
    \Psi'_b(\{b_i\})\Psi'(\{\pi(b_i),g_j\}) \ket{\{b_i,g_j\}},
\end{align}
where the wavefunction $\Psi'_b(\{b_i\})$ is a product of $\mu'_3$-dependent phase factors corresponding to all the faces $F$ in $\partial \mathcal M_3$:
\begin{align}
    \Psi'_b(\{b_i\})= \prod_{F\in\partial \mathcal M_3} e^{2\pi i s_F \mu_3' (e,b_1,b_2,b_3)}.
\end{align}
Acting $V(g')$ on the ground state in \eq{largegs} yields the shifted wavefunctions,
\begin{widetext}
\begin{align}
    \Psi'(\{\pi(b_i),g_j\})&\rightarrow \Tilde\Psi'(\{\pi(b_i),g_j\})=
    \vartheta(\{b_i,g_j\})\Psi'(\{\pi(b_i),g_j\}),\\\label{shiftwavefn1}
    \Psi'_b(\{b_i\})&\rightarrow \Tilde\Psi'_b(\{b_i\})=\prod_{F\in \partial\mathcal M_3} e^{2\pi i s_F \mu_3' (g',b_1,b_2,b_3)},
\end{align}
\end{widetext}
where $\vartheta(\{b_i,g_j\})\equiv \prod_{F\in \partial \mathcal M_3} e^{2\pi i s_F \nu'_4 (e,g',b_1,b_2,b_3)}$ is the residual phase factor.
The coboundary relation in \eq{cobdy} tells us that
\begin{align} 
    \mu_3' (g',b_1,b_2,b_3)=\nu'_4(e,g',b_1,b_2,b_3)+\mu_3' (e,b_1,b_2,b_3)
    -\mu_3' (e,g',b_2,b_3)+\mu_3' (e,g',b_1,b_3)
    -\mu_3' (e,g',b_1,b_2).
\end{align}
Substituting this relation into \eq{shiftwavefn1} not only cancels the last three terms pairwise, but also leaves an additional $\nu'_4$-dependent phase factor that precisely cancels the residual phase factor $\vartheta(\{b_i,g_j\})$. This way, the total fixed-point wavefunction of $G'$-BSPT state with both the bulk and the boundary is symmetric.

In the above construction, we enlarged the symmetry from $G$ in the bulk to $G'$ on the boundary. Alternatively, we can preserve the bulk symmetry $G$ by gauging the normal subgroup $A$ of $G'$ on the boundary. In this way, the boundary effectively supports a gauge theory corresponding to $A$, enriched by the quotient symmetry $G$. This setup is known as a symmetry-enriched topological (SET) phase. In a boundary SET phase, it hosts an anyonic model where the anyons are additionally acted upon by the symmetry $G$. The anyonic system exhibits a $G$-anomaly, indicating that the boundary cannot be realized independently without a bulk state. This anomaly reflects the interplay between the $A$ gauge structure and the $G$ symmetry action \cite{chen2014symmetry, Chen_2017}.

%In the gauging procedure, the group extension structure determines how gauge charges transform under the quotient symmetry. When a gauge charge cannot be described by $\operatorname{Rep}(A\times G)$, then there is a ``fractionalization'' of the quotient symmetry, i.e., the gauge charge carries fractionalized quantum number of the symmetry $G$.

\section{(2+1)D $(\mathbb Z_2^f \times_{\om_2} G)$-FSPT states with nontrivial extension symmetry group}
\label{sec:2Dbdy}

Before the discussion of the more interesting (3+1)D case, let us first focus on the construction of symmetric gapped boundaries for FSPT phases in (2+1)D.

In a (2+1)D FSPT phase with fermionic symmetry $G_f=\mathbb Z_2^f \times_{\om_2} G$, the fixed-point wavefunction can be described by a set of cochain data $(\om_2;n_1,n_2,\nu_3)$ in the context of the general supercohomology theory. They should satisfy the following consistency equations \cite{wang2018construction}:
% \begin{align}\label{dn1om2d}
%     \dd n_1 &\={2} 0,\\\label{dn2om2d}
%     \dd n_2 &\={2} \om_2\smile n_1,\\\nonumber\label{dnu3om2d}
%     \dd \nu_3 
%     &\={1} \frac12 [\om_2 \smile n_2+\Sq^2(n_2)]
%     -\frac14 \om_2\smile n_1^2\\
%     &\quad+\frac12\sqb{\om_2(012)\om_2(023)n_1(24)+\om_2(013)\om_2(123)}n_1(34).
% \end{align}
\begin{align}
    \dd \om_2 &\={2} 0,\\\label{dn1om2d}
    \dd n_1 &\={2} 0,\\\label{dn2om2d}
    \dd n_2 &\={2} \om_2\smile n_1,\\\label{dnu3om2d}
    \dd \nu_3 
    &\={1} \frac12 [\om_2 \smile n_2+\Sq^2(n_2)]
    +\frac14 \om_2 \smile n_1^2 
    +\frac12\om_2(012)\om_2(023)n_1(23)n_1(34).
\end{align}
Here, the right-hand side of the last equation \eq{dnu3om2d} matches the expression given in Ref.~\onlinecite{wang2018construction}, with the subtraction of a coboundary term $ \dd\!\left(\frac{1}{4}\omega_2n_1\right)$. This can be derived by setting $\dd\omega_2\={2}0$ in Ref.~\onlinecite{wang2018construction} and applying the consistency conditions $\dd n_1\={2}0$ and $\dd n_2\={2}\omega_2 n_1$.
We emphasize that $\omega_2$ in the term $\frac{1}{4} \omega_2 \smile n_1^2$ only takes values in $\{0,1\}$, rather than arbitrary values modulo 2.

The notation $\Sq^2$ in the twisted cocycle equation \eq{dnu3om2d} refers to the Steenrod square acting on cochains. It is defined as \cite{steenrod1947,wen2019fspt}:
\begin{align}\label{sqdef}
\Sq^i(n_d) := n_d\smile_{d-i}n_d + n_d\smile_{d-i+1}\dd n_d,
\end{align}
for a $d$-cochain $n_d$, using the higher cup product $\smile_i$. When $n_d$ is a cocycle (i.e., it satisfies $\dd n_d=0$), the cochain-level Steenrod square $\Sq^i$ reduces to the usual cocycle-level Steenrod square $\sq^i$, as:
\begin{align}
\sq^i(n_d) = n_d\smile_{d-i}n_d,
\end{align}
which is more commonly used in the literature.

With the known construction data $(\om_2;n_1,n_2,\nu_3)$ of (2+1)D FSPT, we can outline the pullback trivialization procedure for the symmetric gapped boundary of this phase as
\begin{align}\nonumber
(\om_2;n_1,n_2,\nu_3)
&\,\overset{\text{  pullback  }}{\sim} (\om'_2;n'_1,n'_2,\nu'_3) = (\dd \tau'_1;n'_1,n'_2,\nu'_3)\\\nonumber
&\overset{\text{coboundary}}{\sim} (0;N'_1,N'_2,\mathcal V'_3)
= (0;\dd m'_0,N'_2,\mathcal V'_3)
\\\nonumber
&\overset{\text{coboundary}}{\sim} (0;0,\mathscr N'_2,\mathscr V'_3)\\\nonumber
&\,\overset{\text{  pullback  }}{\sim} (0;0,\mathscr N''_2,\mathscr V''_3) = (0;0,\dd m''_1,\mathscr V''_3)\\\nonumber
&\overset{\text{coboundary}}{\sim} (0;0,0,\mathbb V''_3)\\\nonumber
&\,\overset{\text{  pullback  }}{\sim} (0;0,0,\mathbb V'''_3)=(0;0,0,\dd\mu'''_2)\\
&\overset{\text{coboundary}}{\sim} (0;0,0,0),
\end{align}
where ``$\sim$'' denotes the presence of a gapped interface between the two SPT phases.
More precisely, there are two kinds of ``$\sim$'' in the above procedure. The first one $\overset{\text{pullback}}{\sim}$ indicates the process of pulling back the symmetry of interest to a larger symmetry, where one of the decoration data is trivialized (i.e., mapped to a coboundary) in the extended symmetry. On the other hand, $\overset{\text{coboundary}}{\sim}$ means that we perform appropriate ``gauge" transformations to convert a trivial decoration cocycle into a truly trivial one at the cochain level.

\subsection{Lattice model realizations for gapped interfaces}

\begin{widetext}
\begin{figure*}[ht]
    \centering
    \includegraphics[width=.8\linewidth]{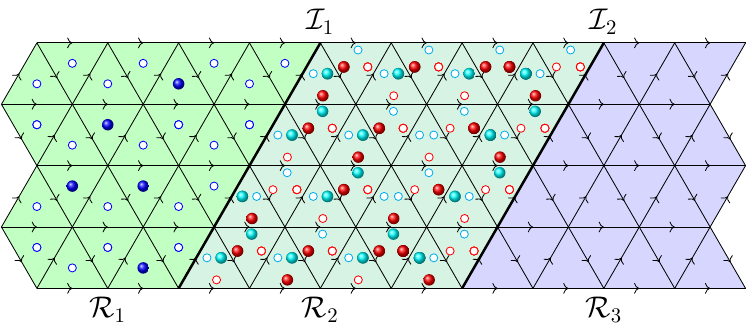}
    \caption{Lattice interface construction for pullback trivialization of complex fermion decorations in (2+1)D FSPT phases. The first region $\mathcal R_1$ describes an FSPT phase with symmetry $\Z_2^f\times G$ with complex fermion decorations $n_2$ at the center of each triangle. The second region $\mathcal R_2$ is obtained from region $\mathcal R_1$ by applying a FSLU $\mathcal U_1^f$, which enlarges the symmetry from $G$ to $G'$. This transformation results in a trivial $(\Z_2^f\times G')$-FSPT phase, where the decoration data $n_2'=\dd m_1'$ becomes a 2-coboundary. The third region $\mathcal R_3$ is obtained by further applying another FSLU transformation $\mathcal U_2^f$. It becomes a $G'$-BSPT phase.}
    \label{fig:2dlattice}
\end{figure*}
\end{widetext}

The formal pullback trivialization procedure can be explicitly implemented on lattice models, as illustrated in Fig.~\ref{fig:2dlattice}. For simplicity, we focus on the lattice model corresponding to the trivialization of complex fermion decoration data $n_2$. However, the ideas presented here can be generalized to other cases. On the 2D spatial manifold shown in Fig.~\ref{fig:2dlattice}, there are three distinct regions, $\mathcal R_1$, $\mathcal R_2$, and $\mathcal R_3$, each hosting a different SPT phase. These regions are separated by two gapped interfaces, denoted as $\mathcal I_1$ and $\mathcal I_2$. The pullback trivialization procedure for these regions can be succinctly summarized as:
\begin{align}\nonumber
(n_2,\nu_3)
&\,\overset{\text{  pullback  }}{\sim} (n'_2,\nu'_3) = (\dd m'_1,\nu'_3)\\
&\overset{\text{coboundary}}{\sim} (0,\mathcal V'_3).
\end{align}
Next, we will discuss the construction of the two gapped interfaces in details.

Initially, the entire 2D lattice (for all three regions $\mathcal R_1, \mathcal R_2$ and $\mathcal R_3$) is constructed as a $(\mathbb{Z}_2^f \times G)$-FSPT with nontrivial complex fermion decoration $n_2 \in H^2(G, \mathbb{Z}_2)$. This means that a complex fermion mode (depicted as a blue ball in the figure) is placed at the center of a triangle whenever $n_2 = 1$ for that triangle. 
The complex fermion mode always remains in the empty state for the regions $\mathcal{R}_2$ and $\mathcal{R}_3$ after the transformations discussed later. Therefore, we omit the blue balls in these regions.
The symmetry action on these complex fermions follows the standard rule given in \eq{symmc}.

To trivialize this 2-cocycle, we pull it back by extending the symmetry $G$ to a larger symmetry group $G'$ via the short exact sequence:
\begin{align}
1 \rightarrow A \rightarrow G' \overset{\pi}{\rightarrow} G \rightarrow 1,
\end{align}
where $A$ is a normal subgroup of $G'$ and $\pi$ is the projection map. The extension is chosen such that the 2-cocycle $n_2$ is pulled back to a trivial 2-cocycle $n_2' = \pi^*(n_2) = \dd m_1'$ in $H^2(G', \mathbb{Z}_2)$. If this condition holds, we can define a fermionic symmetry local unitary (FSLU) transformation as follows:
\begin{align}
\mathcal U_1^f =\prod_{\langle 1'2'3'\rangle\in \mathcal R_2,\mathcal R_3}
(c_{23A}^\dagger)^{m_1'(2'3')}(c_{12A}^\dagger)^{m_1'(1'2')} 
(c_{13B}^\dagger)^{m_1'(1'3')}
(c_{123})^{n_2(123)},
\end{align}
where the product runs over all triangles $\langle 1'2'3' \rangle$ in regions $\mathcal{R}_2$ and $\mathcal{R}_3$. This FSLU is similar to the construction of anomalous SPT states \cite{WangASPT}. It annihilates the fermions (blue balls) at the center of each triangle and create fermions (red or cyan balls in region $\mathcal{R}_2$) on the two sides of each link, depending on the vertex labels $g_i' \in G'$. This FSLU is $G$-symmetric, and it also preserves fermion parity due to the trivialization $n_2' = \dd m_1'$ (mod 2). As a result, all fermions at the centers of the triangles are pushed to the boundary near each link in region $\mathcal{R}_2$. In regions $\mathcal{R}_2$ and $\mathcal{R}_3$, there is no complex fermion mode (or it is in the vacuum empty state) at the center of each triangle. It is actually a trivial FSPT with symmetry $\Z_2^f\times G'$.

The third region $\mathcal{R}_3$ is obtained further by a FSLU or ``gauge transformation'' that makes the 2-cocycle $n_2' = \dd m_1'$ truly trivial at the cochain level. This process does not require extending the symmetry but only applying a second FSLU:
\begin{align}
\mathcal U_2^f=\prod_{\langle 1'2'\rangle\in\mathcal R_3}
(c_{12A})^{m_1'(1'2')}
(c_{12B})^{m_1'(1'2')}.
\end{align}
This FSLU annihilates the possible two fermions on the two sides of each link in region $\mathcal{R}_3$. It also preserves fermion parity because the presence or absence of fermions on each side of a link depends on the 1-cochain $m_1'(1'2')$ of the same link. After the action of this FSLU in region $\mathcal{R}_3$, we obtain a purely bosonic state in this region, with no fermionic modes remaining. Therefore, region $\mathcal{R}_3$ is, in general, a BSPT phase protected by symmetry $G'$. As a consequence of the fermion reordering induced by FSLU $\mathcal{U}_2^f$, we added some cochains to modify the pullback $\nu_3'$. This leads to the creation of a new 3-cocycle $\mathcal V_3':=\nu_3'+f[m_1']$, which is the shifted version of $\nu_3'$ by a functional of $m_1'$
(the explicit form of $f[m_1']$ is defined in \eq{nu3pbn2}, with both $\tau_1''$ and $m_0''$ being set to zero). This 3-cocycle $\mathcal V_3'$ in $H^3[G', \U]$ describes the $G'$-BSPT phase in region $\mathcal R_3$ after the pullback trivialization of the $(\mathbb{Z}_2^f \times G)$-FSPT phase. This BSPT phase can also be further pullback-trivialized using established methods.

In the following subsections, we provide explicit calculations of the full pullback trivializations on $(\omega_2; n_1, n_2, \nu_3)$. This process is achieved by performing the pullback trivialization of each of the components $\omega_2$, $n_1$, $n_2$, and $\nu_3$ one by one. Note that $n_1$ cannot be trivialized by group extension, the reason for which will be discussed later. However, for completeness, we still include it in the calculation later.

\subsection{Pullback trivialization of $\om_2$}

The first step in the pullback trivialization construction is to create a gapped interface between an FSPT phase with an extended symmetry group $ G_f=\Z_2^f \times_{\omega_2} G $ and an FSPT phase with a direct product symmetry group $ G_f'=\Z_2^f \times G'$.

The symmetry extension is given by the short exact sequence:
\begin{align}
1 \rightarrow A \rightarrow G' \overset{\pi}{\rightarrow} G \rightarrow 1.
\end{align}
This extension is chosen such that the 2-cocycle $\omega_2'$ of $G'$, related to the 2-cocycle $\omega_2$ of $G$ via the pullback $\om'_2 :=\pi^\ast (\om_2)$, belongs to the trivial cohomology class $[\om'_2]\={2}[0]\in H^2(G',\Z_2)$. Consequently, we can write $\omega_2'$ as a coboundary
\begin{align}
    \om'_2 \={2} \dd \tau'_1.
\end{align}
However, we note that in the $\frac{1}{4}$-term of \eq{dnu3om2d}, $\omega_2'$ can only take values in ${0,1}$, while $\dd \tau_1'$ can take values in ${-1,0,1,2}$. Thus, we refine the trivialization equation as a mod 4 equation:
\begin{align}\label{om2triv2d}
\om'_2 \={4} \dd \tau'_1-2\Sq^1(\tau_1')
\={4} \dd \tau'_1-2\tau_1'\tau'_1-2\tau'_1 \smile_1\dd \tau'_1.
\end{align}
A detailed derivation of this result can be found near Eq.~(\ref{App_dm1}) in Appendix~\ref{sec:triveqz}.

With the trivialization of $\omega_2'$, the consistency equation for $n_2'$ becomes
\begin{align}
    \dd n_2' \={2} \dd\tau_1'\smile n_1'
    \={2} \dd\!\left(\tau_1' n_1'\right).
\end{align}
By shifting $n_2'$ with certain cochain terms, we can define new decoration data:
\begin{align}
    N_1' &\={2}n_1',\\
    N_2' &\={2} n_2'+\tau_1' n_1',
\end{align}
such that their consistency equations become
\begin{align}
\dd N_1'\={2}0,\\
\dd N_2'\={2}0.
\end{align}
These new decoration data correspond to an FSPT phase protected by the direct product symmetry group $G_f' = \mathbb{Z}_2^f \times G'$.

\begin{widetext}
The cochain shift of the phase factor $\nu_3$ is more involved. With the trivialization of $\omega_2'$, the obstruction function on the right-hand side of the consistency equation \ref{dnu3om2d} for $\nu_3$ becomes:
\begin{align}
    \mathcal O_4[\dd\tau_1';n_1',n_2']
    \={1} \frac14 \dd\tau_1'\smile n_1'^2+\frac12 \dd\tau_1'\smile n_2'+\frac12 \Sq^2(n_2')
    +\frac12 \Sq^1(\tau_1')\smile n_1'^2
    +\frac12\dd\tau_1'(012)\dd\tau_1'(023)n_1'(23)n_1'(34).
\end{align}
On the other hand, for the phase factor $\mathcal{V}_3'$ of the new FSPT with decoration data $(N_1', N_2', \mathcal{V}_3')$ and symmetry $G_f'=\Z_2^f\times G'$, it should satisfy the consistency equation $\dd \mathcal V_3'=\Tilde{\mathcal O}_4[0;N_1',N_2']$ where the right-hand side obstruction function is given by
\begin{align}\nonumber
    \Tilde{\mathcal O}_4[0;N_1',N_2']
    &\={1} \frac12 \Sq^2(N_2')
    \={1} \frac12 \Sq^2(n_2'+\tau_1' n_1')\\
    &\={1} \frac12 \Sq^2(n_2')
    +\frac12 \Sq^2(\tau_1' n_1' )
    +\frac12 \dd\!\sqb{(n_2'+\tau_1'n_1')\smile_1 (\tau_1'n_1')}.
\end{align}
Here, we simplified the Steenrod operation acting on the sum of two cochains in the last line using the following formula \cite{wen2019fspt}:
\begin{align}\label{sqk2c}
    \Sq^k(a_i+b_i)\={2} \Sq^k(a_i)+\Sq^k(b_i)+\dd b_i\smile_{i-k+2}\dd a_i+\dd(b_i\smile_{i-k+1}a_i) +
    \dd(\dd b_i\smile_{i-k+2} a_i),
\end{align}
where $a_i$ and $b_i$ are two cochains of degree $i$.

We can then calculate the difference between the two obstruction functions as follows:
\begin{align}\nonumber%\label{O4cobdy1}
    \mathcal O_4[\dd\tau_1';n_1',n_2']-\Tilde{\mathcal O}_4[0;N_1',N_2']
    &\={1} \dd\!\sqb{\frac12 \tau_1'n_2'+\frac12(n_2'+\tau_1'n_1')\smile_1 (\tau_1'n_1')+\frac14 \tau_1'n_1'^2}\\\nonumber
    &\quad
    +\frac12 \Sq^2(\tau_1' n_1' )
    +\frac12 \Sq^1(\tau_1')n_1'^2
    +\frac12 \tau_1'\dd\tau_1'n_1'
    +\frac12\dd\tau_1'(012)\dd\tau_1'(023)n_1'(23)n_1'(34)\\%\label{O4cobdy2}
    &\={1} \dd\!\sqb{\frac12 \tau_1'n_2'+\frac12(n_2'+\tau_1'n_1')\smile_1 (\tau_1'n_1')+\frac14 \tau_1'n_1'^2+\frac12\tau_1'(\tau_1'\smile_1 n_1')n_1'}.
\end{align}
\end{widetext}
The final equality can be verified through direct calculation numerically. Therefore, by defining a 3-cochain $f_3$ as 
\begin{align}
    f_3 \={1}\frac12 \tau_1'n_2'+\frac12(n_2'+\tau_1'n_1')\smile_1 (\tau_1'n_1')
    +\frac14 \tau_1'n_1'^2
    +\frac12\tau_1'(\tau_1'\smile_1 n_1')n_1',
\end{align}
we obtain the phase factor for the new FSPT
\begin{align}
    \mathcal V_3':\={1} \nu_3'-f_3[\tau_1';n_1',n_2'].
\end{align}
It corresponds to the phase factor of the FSPT with symmetry $G_f' = \mathbb{Z}_2^f \times G'$, as it satisfies the required consistency condition
\begin{align}
    \dd \mathcal V_3'\={1} \Tilde{\mathcal O}_4[0;N_1',N_2'] \={1} \frac12 \Sq^2(N_2'),
\end{align}
for the Majorana chain and complex fermion decoration data $(N_1',N_2')$.

In this way, we have constructed a gapped interface between two distinct FSPT states: the state $(\omega_2; n_1, n_2, \nu_3)$ with an extended symmetry group $G_f = \mathbb{Z}_2^f \times_{\omega_2} G$, and the state $(0; N_1', N_2', \mathcal{V}_3')$ with a direct product symmetry group $G_f' = \mathbb{Z}_2^f \times G'$. The interface  between these two states, denoted as $\mathcal{I}$, is shown in Fig.~\ref{fig:ext} and represents the first of such interfaces in our construction.

\subsection{Majorana chain decoration $n_1$}

Now that we have an FSPT phase with a direct product symmetry group $G_f' = \mathbb{Z}_2^f \times G'$ and decoration data $(0; N_1', N_2', \mathcal{V}_3')$, our next task is to pullback and trivialize the cochain $N_1'$ in this subsection.

There is, however, a significant difference in handling the decoration data $N_1'$ compared to other cocycles of higher degree. In the pullback construction, the goal is to trivialize a cocycle by embedding it in a larger group, effectively making the cocycle part of the image of some differential in the Lyndon–Hochschild–Serre spectral sequence associated with the short exact sequence of the symmetry extension \cite{wang2021domain}. However, for a 1-cocycle, the degree is too low, and no nontrivial differentials exist to trivialize it. In other words, if $N_1'$ is a nontrivial cocycle, it cannot be pulled back into a trivial 1-cocycle for any larger group. This result can also be demonstrated more directly: if $N_1'' = \dd m_0''$ is trivial in some larger group $G''$, then by the definition of the coboundary operator, we have $N_1''(g) = \dd m_0''(g) = m_0''$ as a constant for any $g \in G''$. Consequently, for the original group $G'$, $N_1'(g) = m_0''$ must also be constant for all $g \in G'$, implying that $N_1'$ itself is a coboundary in $G'$. In summary, a 1-cocycle $N_1'$ can only be pullback-trivialized if and only if it is already a trivial 1-cocycle in the original group $G'$.

Although the degree of $N_1'$ is too low for pullback trivialization, we still include it in the discussion for completeness. This means, initially, we assume that $N_1' \={2} \dd m_0'$ is a 1-coboundary for $G'$. Therefore, no symmetry extension is needed. 

Now, let us focus on the FSLU or ``gauge transformation'' that can be performed on $N_1' \={2} \dd m_0'$ such that we obtain an FSPT phase with symmetry group $G_f' = \Z_2^f \times G'$, and trivial Majorana chain decoration at the cochain level. The decoration data for this phase are given by a redefinition of the 2- and 3-cochains
\begin{align}
\mathscr N'_2 &:\={2} N'_2\={2} n_2'-\tau_1' n_1',\\
\mathscr V'_3 &:\={1} \mathcal V'_3\={1} \nu'_3 - f'_3[\tau_1';m_0',n_2'],
\end{align}
where the 3-cochain $f'_3$ is chosen to be
\begin{align}
    f'_3[\tau_1';m_0',n_2'] \={1}
    \frac14 \tau_1'(\dd m_0')^2+\frac12 \tau_1'n_2'
    +\frac12(n_2'+\tau_1'\dd m_0')\smile_1 (\tau_1'\dd m_0')
    +\frac12\tau_1'(\tau_1'\smile_1 \dd m_0')\dd m_0'.
\end{align}
With these definitions, one can verify directly that the new decoration data $(\mathscr{N}'_2, \mathscr{V}'_3)$ satisfy the usual consistency conditions
\begin{align}
\dd \mathscr N'_2 &\={2} 0,\\\label{nu3pbn1}
\dd \mathscr V'_3 &\={1} \frac12 \Sq^2(N_2'),
\end{align}
which corresponds to an FSPT phase with only complex fermion decorations and no Majorana chain decorations. This phase is labelled as $(0; 0, \mathscr{N}'_2, \mathscr{V}'_3)$, representing a $(\Z_2^f \times G')$-FSPT phase.

The gauge transformation ensures that the Majorana chain is trivialized in the cochain level, and the remaining decorations now correspond purely to complex fermions. The interface $\mathcal{I}'$ between these two $(\Z_2^f\times G')$-FSPT states $(0; N_1', N_2', \mathcal{V}_3')$ and $(0; 0, \mathscr N'_2, \mathscr V'_3)$ is shown in Fig.~\ref{fig:ext} and represents the second of such interfaces in our construction.

\subsection{Complex fermion decoration $n_2$}

The next task is to trivialize the complex fermion decoration 2-cocycle $\mathscr{N}'_2$. This can be done by extending the symmetry to a larger group $G''$, as follows:
\begin{align}
    1 \rightarrow A' \rightarrow G'' \overset{\pi'}{\rightarrow} G' \rightarrow 1.
\end{align}
This extension is chosen such that we obtain a trivial 2-cocycle in $G''$. Specifically, we define the extended 2-cocycle as
\begin{align}
    \mathscr N_2'':=\pi'^\ast(\mathscr N_2')\={2} \dd m_1'',
\end{align}
where the pullback operation $\pi'^\ast$ maps the 2-cocycle from $G'$ to $G''$, ensuring the trivialization of the cocycle in the larger group.
After this extension, the twisted 3-cocycle consistency equation, as given in \eq{nu3pbn1}, becomes
\begin{align}
    \dd\mathscr V_3''
    \={1} \frac12 \Sq^2(\dd m_1'')
    \={1} \frac12 \dd m_1''\dd m_1''
    \={1} \frac12 \dd (m_1'' \dd m_1'').
\end{align}
Thus, the decoration data for the 3-cocycle can be redefined by an appropriate FSLU or ``gauge transformation'' to ensure the trivialization of the complex fermion sector, as follows:
\begin{align}
    \mathbb V_3''
    &\={1} \mathscr V_3''+\frac12 m_1'' \dd m_1''\\\label{nu3pbn2}
    &\={1} \nu''_3 -\frac14 \tau_1''(\dd m_0'')^2
    +\frac12 m_1'' \dd m_1''
    +\frac12 \tau_1''(\dd m_1''+\tau_1''\dd m_0'') +\frac12(\dd m_1'')\smile_1 (\tau_1''\dd m_0'')+\frac12\tau_1''(\tau_1''\smile_1 \dd m_0'')\dd m_0''.
\end{align}
It describes a $G''$-BSPT phase with the untwisted cocycle consistency equation
\begin{align}
\dd\mathbb V_3'' \={1}0.
\end{align}

This construction successfully trivializes the complex fermion decoration data, resulting in an interface $\mathcal{I}''$ between the $(\Z_2^f \times G')$-FSPT phase with decoration data $(0; 0, \mathscr{N}_2', \mathscr{V}_3')$ and the $(\Z_2^f \times G'')$-FSPT phase with decoration data $(0; 0, 0, \mathbb{V}_3'')$. This interface is the third one in our construction, as depicted in Fig.~\ref{fig:ext}. In the 2D lattice model shown in Fig.~\ref{fig:2dlattice}, the pullback trivialization corresponds to interface $\mathcal{I}_1$, and the ``gauge transformation'' corresponds to interface $\mathcal{I}_2$. Together, these two interfaces combine to form the interface $\mathcal{I}''$, as shown in Fig.~\ref{fig:ext}.

\subsection{Bosonic SPT $\nu_3$}

The final step of our construction involves the pullback trivialization of the bosonic phase factor $\mathbb{V}_3''$. This can be achieved by performing an extension of the global symmetry group $G''$ to a larger group $G'''$, via the sequence
\begin{align}\label{ext:G'''}
1 \rightarrow A'' \rightarrow G''' \overset{\pi''}{\rightarrow} G'' \rightarrow 1.
\end{align}
By pulling back the 3-cocycle, we define the new 3-cocycle $\mathbb{V}_3''' := \pi''^\ast(\mathbb{V}_3'')$ as a coboundary, which is expressed in terms of a 2-cochain $\mu'''_2$:
\begin{align}
\mathbb V'''_3 \={1} \dd \mu'''_2.
\end{align}
This is the standard pullback trivialization construction for a BSPT phase.

The last interface we construct is between the $G''$-BSPT phase, with 3-cocycle $\mathbb{V}_3''$, and a trivial $G'''$-BSPT phase. This trivial phase can be viewed as having a gapped boundary with the vacuum. This interface  $\mathcal{I}'''$ after shrinking the trivial $G'''$-BSPT is shown in Fig.~\ref{fig:ext}. By shrinking all the previously constructed interfaces corresponding to the different enhanced symmetry groups $G'$, $G''$, and $G'''$, we ultimately obtain the gapped boundary of the original $(\Z_2^f \times_{\omega_2} G)$-FSPT phase, labeled by $(\omega_2; n_1, n_2, \nu_3)$.

\subsection{Example: $(\Z_2^f\times\Z_4\times\Z_4)$-FSPT, $\mathtt{SmallGroup}(32,2)$-BSPT and trivial $\mathtt{SmallGroup}(64,23)$-BSPT}
\label{sec:2dexample}

In this subsection, we will demonstrate our previous pullback construction of an FSPT state with a gapped boundary using an explicit, concrete example. As discussed earlier, a nontrivial Majorana chain decoration data $n_1$ cannot be pullback-trivialized, so we will primarily focus on an FSPT with complex fermion decoration data $n_2$. For simplicity, we will assume that the symmetry group is of direct product type, i.e., $\omega_2 = 0$, and that the bosonic symmetry is a finite unitary Abelian group. In this case, the full classification is well-established \cite{wlg2017,wang2018construction,RNQWG2023}.

\subsubsection{The nontrivial $(\Z_2^f\times\Z_4\times\Z_4)$-FSPT with complex fermion decorations}

To be more explicit, we consider an example with the global symmetry $G_f = \mathbb{Z}_2^f \times G = \mathbb{Z}_2^f \times (\mathbb{Z}_4 \times \mathbb{Z}_4)$. There are nontrivial (2+1)D $G_f$-FSPT phases with complex fermion decoration. The decoration data for this phase are given as:
\begin{align}
n_2 &\={2} n_1^{(1)} n_1^{(2)},\\\label{nu3example}
\nu_3 &\={1} \frac12 n_1^{(1)}\sqb{n_1^{(1)}\smile_1n_1^{(2)}}n_1^{(2)} +\frac12 x_1^{(1)}\sqb{n_1^{(2)}}^2.
\end{align}
Here, the notation $n_1^{(i)} \in H^1(\mathbb{Z}_4^{(i)}, \mathbb{Z}_2)$ is used, where $\mathbb{Z}_4^{(i)}$ ($i=1, 2$) are the two subgroups $\mathbb{Z}_4^{(1)}$ and $\mathbb{Z}_4^{(2)}$ of $G = \mathbb{Z}_4 \times \mathbb{Z}_4$. The explicit expression for $n_1^{(i)}$ is
\begin{align}
n_1^{(i)}(\mb{g})=[\mb{g}_i]_2, \quad \forall \mb{g}=(\mb{g}_1,\mb{g}_2)\in \mathbb{Z}_4 \times \mathbb{Z}_4,
\end{align}
where $[x]_N$ represents the mod-$N$ value of an integer $x$. We also define a $\mathbb{Z}_2$-valued 1-cochain as:
\begin{align}
    x_1^{(i)}(\mb{g}) :\={2} \frac{[\mb{g}_i]_4-[\mb{g}_i]_2}{2}, \quad \forall \mb{g}=(\mb g_1,\mb g_2)\in \Z_4\times \Z_4,
\end{align}
which satisfies the relation
\begin{align}
\dd x_1^{(i)} \={2}\left[n_1^{(i)}\right]^2,
\end{align}
for the $\mathbb{Z}_4^{(i)}$ group. Using this relation, we can verify that the bosonic phase factor $\nu_3$ in Eq.~\eqref{nu3example} satisfies the required obstruction equation $\dd \nu_3 \={1} \frac12 n_2\smile n_2$.

\subsubsection{Trivialization of $n_2$ to $\mathtt{SmallGroup}(32,2)$-BSPT}

To trivialize $n_2\in H^2(G,\Z_2)$, we choose the normal subgroup to be $A=\Z_2$ tautologically and consider the central extension
\begin{align}\label{ses2d}
0 \rightarrow (A=\Z_2) \rightarrow \Tilde{G} \rightarrow (G=\Z_4\times\Z_4) \rightarrow 0.
\end{align}
Here, we use the notation $\Tilde{G}$ to represent $G''$ from the earlier discussion for convenience, as we omit the trivialization procedure of $\omega_2$ and $n_1$ in this specific example.
The extension is defined by a 2-cocycle $\phi_2 \in H^2(G, A) = H^2(\mathbb{Z}_4 \times \mathbb{Z}_4, \mathbb{Z}_2) = (\mathbb{Z}_2)^3$. For this example, we set $\phi_2$ to be equal to $n_2$:
\begin{align}
\phi_2 (\mb{g,h}) = n_2 (\mb{g,h}) = \left(n_1^{(1)} n_1^{(2)}\right)(\mb{g,h}) = \mb g_1 \mb h_2\ (\mathrm{mod}\ 2),
\end{align}
for $\mb g=(\mb g_1,\mb g_2)$ and $\mb h=(\mb h_1,\mb h_2)\in G=\Z_4\times\Z_4$.
Let us denote the elements in $\Tilde{G}=\Z_2\times_{\phi_2}(\Z_4\times\Z_4)$ by $\mb g=(\mb g_0,\mb g_1,\mb g_2)$, where $\mb g_0\in A=\Z_2$ and $(\mb g_1,\mb g_2)\in G=\Z_4\times\Z_4$. Based on the expression for $\phi_2$, the group multiplication rule in $\Tilde{G}$ is given by
\begin{align}\nonumber
\mb g\times_{\Tilde{G}} \mb h 
&= (\mb g_0,\mb g_1,\mb g_2)\times_{\Tilde{G}} (\mb h_0,\mb h_1,\mb h_2)\\ &
= \left([\mb g_0+\mb h_0+\mb g_1\mb h_2]_2,[\mb g_1+\mb h_1]_4,[\mb g_2+\mb h_2]_4\right),
\end{align}
where $[x]_N$ represents $x$ modulo $N$, as previously defined.

In this construction, the larger group $\Tilde{G}=\Z_2\times_{\phi_2}(\Z_4)^2$ can be expressed with the following group presentation:
\begin{align}
\langle a,b,c\, |\, a^2=b^4=c^4=1, ab=ba, ac=ca, bc=acb \rangle,
\end{align}
where $a$, $b$, and $c$ are the generators of $A = \mathbb{Z}_2$ and $G = \mathbb{Z}_4 \times \mathbb{Z}_4$, respectively.
The group $\Tilde{G}$ can also be identified as
\begin{align}
\Tilde{G}=\Z_2\times_{\phi_2}(\Z_4\times\Z_4)=\mathtt{SmallGroup}(32,2),
\end{align}
which corresponds to the second group of order 32 in the GAP Small Group Library \cite{SmallGrp}.

When $n_2$ is pulled back to $\tilde{n}_2 \in H^2(\Tilde{G}, \mathbb{Z}_2)$, it can be shown to be a coboundary. This result follows because $\tilde{n}_2$ lies in the image of the differential $d_2$ in the second page of the Lyndon-Hochschild-Serre spectral sequence associated with the central extension \eq{ses2d}. More explicitly, we can define a $\mathbb{Z}_2$-valued 1-cochain of $\Tilde{G}$ as
\begin{align}
\Tilde m_1(\mb g)=[\mb g_0]_2,\quad \forall \mb g\in \Tilde{G},
\end{align}
where $[\mathbf{g}_0]_2$ denotes the modulo 2 value of $\mathbf{g}_0$. The differential of this 1-cochain is
\begin{align}\nonumber
\dd \Tilde m_1(\mb{g},\mb{h})
&= \Tilde m_1(\mb {h}) - \Tilde m_1(\mb{g}\times_{\Tilde{G}}\mb{h}) + \Tilde m_1(\mb{g}) \\\nonumber &
= [\mb h_0]_2 - [\mb g_0+\mb h_0+\mb g_1\mb h_2]_2+ [\mb g_0]_2 \\ &
\={2} \mb g_1 \mb h_2.
\end{align}
This result matches precisely with $\tilde{n}_2(\mathbf{g}, \mathbf{h}) = \mathbf{g}_1 \mathbf{h}_2$, which is the pullback of $n_2 = n_1^{(1)} n_1^{(2)}$ from $G$ to $\Tilde{G}$.
Thus, we find that $\tilde{n}_2 \in H^2(\Tilde{G}, \mathbb{Z}_2)$, as the pullback 2-cocycle of $n_2 \in H^2(G, \mathbb{Z}_2)$ along the short exact sequence \eq{ses2d},  is indeed trivial because it can be expressed as a coboundary:
\begin{align}
\Tilde n_2 \={2} \dd \Tilde m_1.
\end{align}

For the group $\Tilde{G}=\mathtt{SmallGroup}(32,2)=\Z_2\times_{\phi_2}(\Z_4\times\Z_4)$, the 3-cocycle of \eq{nu3pbn2} in $H^3[\Tilde{G},\U]$ becomes (after setting $\omega_1=\tau_1=n_1=m_0=0$)
\begin{align}\label{V3'}
\Tilde{\mathbb V}_3(\mb g,\mb h,\mb k)
&\={1} \Tilde \nu_3(\mb g,\mb h,\mb k)+\frac12
\Tilde m_1(\mb g)\Tilde n_2(\mb h,\mb k)\\\nonumber
&\={1} \nu_3[(\mb g_1,\mb g_2),(\mb h_1,\mb h_2),(\mb k_1,\mb k_2)] + \frac12 \Tilde m_1 (\mb g)\, n_2[(\mb h_1,\mb h_2),(\mb k_1,\mb k_2)] \\
&\={1}\frac12 \mb g_1 \mb h_1 \mb h_2 \mb k_2+\frac14\!\left([\mb g_1]_4-[\mb g_1]_2\right)\mb h_2 \mb k_2+\frac12\mb g_0 \mb h_1 \mb k_2
\end{align}
for $\mb g,\mb h,\mb k\in \Tilde{G}$. Here, we used the expression for $\nu_3 \in H^3[G, \U]$ from Eq.~\eqref{nu3example}. It can be shown that $\Tilde{\mathbb V}_3$ is a nontrivial element in $H^3[\Tilde{G},\U]=(\Z_4)^4$.

In summary, we demonstrated that for the $(\mathbb{Z}_2^f \times \mathbb{Z}_4 \times \mathbb{Z}_4)$-FSPT phase in (2+1)D, characterized by the previously defined decoration data $(n_2, \nu_3)$, a gapped interface exists between this phase and a BSPT phase described by $(0, \Tilde{\mathbb{V}}_3)$. The global symmetry of the BSPT phase is given by $\Tilde{G} = \mathtt{SmallGroup}(32, 2)$.

\begin{widetext}
\subsubsection{Trivialization of $\nu_3$ to trivial $\mathtt{SmallGroup}(64,23)$-BSPT}

Now, in order to further trivialize $\Tilde{\mathbb V}_3$, we introduce a second central extension
\begin{align}\label{sesGtt}
1 \rightarrow (\Tilde{A}=\Z_2) \rightarrow \dtilde{G} \overset{\Tilde\pi}{\rightarrow} [\Tilde G=\Z_2\times_{\phi_2}(\Z_4)^2] \rightarrow 1.
\end{align} 
Here $\dtilde{G}$ represents $G'''$ in \eq{ext:G'''} in earlier discussions.
We can choose the extension 2-cocycle $\Tilde \phi_2\in H^2(\Tilde{G},\Tilde{A})$ to be
\begin{align}
\Tilde \phi_2(\mb g,\mb h)
=
\mb g_1\mb h_1\mb h_2+\frac{[\mb g_1]_4-[\mb g_1]_2}{2} \mb h_2+\mb g_0\mb h_1\ (\mathrm{mod}\ 2), \quad \forall \mb g,\mb h\in \Tilde{G}.
\end{align}
% \begin{align}
% \Tilde \phi_2(\mb g,\mb h)=\left[\mb g_1\mb h_1\mb h_2+\frac{[\mb g_1]_4-[\mb g_1]_2}{2} \mb h_2+\mb g_0\mb h_1\right]_2.
% \end{align}
It can be checked by definition that $\Tilde\phi_2$ is a 2-cocycle in $H^2(\Tilde{G},\Z_2)$. With this choice, the multiplication rule in $\dtilde{G}$ is given by
\begin{align}\label{xGtt}\nonumber
\ag{a}{g}\times_{\dtilde{G}}\ag{b}{h}
&=
\left(a+b+\Tilde \phi_2(\mb g,\mb h)\right)_{\mb g\times_{\Tilde{G}}\mb h}\\
&=
\left(\left[a+b+\mb g_1\mb h_1\mb h_2+\frac{[\mb g_1]_4-[\mb g_1]_2}{2} \mb h_2+\mb g_0\mb h_1\right]_2\right)_{([\mb g_0+\mb h_0+\mb g_1\mb h_2]_2,[\mb g_1+\mb h_1]_4,[\mb g_2+\mb h_2]_4)},
\end{align}
where $\ag{a}{g}$ denotes an element in $\dtilde{G}$ with $a\in \Tilde{A}=\Z_2$ and $\mb g=(\mb g_0,\mb g_1,\mb g_2)\in \Tilde{G}=\Z_2\times_{\phi_2}(\Z_4)^2$.

Using the multiplication rule \eq{xGtt}, the group $\dtilde{G}$ can be presented as
\begin{align}
\langle a,b,c,d\, |\, a^2=b^2=c^4=d^4=1, ab=ba, ac=ca, ad=da,
bc=acb, bd=db,  
cd=dcb\rangle,
\end{align}
where $a$ and $b$ are the generators of $\Tilde{A} = \mathbb{Z}_2$ and $A = \mathbb{Z}_2$, respectively, and $c$ and $d$ are the generators of the two $\Z_4$ subgroups of $G = \mathbb{Z}_4 \times \mathbb{Z}_4$. The group $\dtilde{G}$ can be identified as
\begin{align}
\dtilde{G}=\Z_2\times_{\Tilde \phi_2}\left[\Z_2\times_{\phi_2}(\Z_4)^2\right] = \mathtt{SmallGroup}(64,23),
\end{align}
which is the 23rd group of order 64 in the GAP small group library \cite{SmallGrp}. Additional information about $\dtilde{G}=\mathtt{SmallGroup}(64,23)$ can be found, for example, in \refn{matydGroupNames}.

Our next goal is to show that $\dtilde{\mathbb V}_3 = \Tilde\pi^\ast(\Tilde{\mathbb V}_3)$ is trivialized, i.e., that it is a 3-coboundary in $H^3[\dtilde{G},\U]$. This can be demonstrated in two different ways.

The first approach to demonstrate that $\dtilde{\mathbb V}_3$ is a coboundary is to use the Lyndon–Hochschild–Serre spectral sequence (LHSSS) 
$E_2^{p,q}=H^p[\Tilde{G},H^q(\Tilde{A},\U)] \Rightarrow H^{p+q}[\dtilde{G},\U]$
associated with the short exact sequence \eqref{sesGtt}. The cocycle $\dtilde{\mathbb V}_3$ in \eq{V3'}, as the pullback of $\Tilde{\mathbb V}_3$ for $\Tilde{G}$, is located at 
\begin{align}
    E_2^{3,0}=H^3[\Tilde{G},H^0[\Tilde{A},\U]]=H^3[\Tilde{G},\U]
\end{align}
of the second page of the LHSSS. This cocycle becomes a coboundary if it lies in the image of the differential $d_2: E_2^{1,1}\rightarrow E_2^{3,0}$. Here, one can show that 
\begin{align}
    E_2^{1,1}=H^1[\Tilde{G},H^1[\Tilde{A},\U]]=H^1[\Tilde{G},\Z_2]=(\Z_2)^2.
\end{align} 
We choose the term located at $E_2^{1,1}$ as
\begin{align}
F_{1,1}(a,\mb h) = \frac12 a\mb h_2\ (\mathrm{mod}\ 1),\quad \forall a\in \Tilde{A},\; \mb h\in \Tilde{G}.
\end{align}
Using the results from Ref.~\onlinecite{wang2021domain}, the differential $d_2$ maps this $F_{1,1}\in E_2^{1,1}$ to a term in $E_2^{3,0}$ given by
\begin{align}\nonumber
d_2(F_{1,1})(\mb g,\mb h,\mb k) 
&= F_{1,1}(\Tilde \phi_2(\mb{g},\mb{h}),\mb{k})
=\frac12 \Tilde \phi_2(\mb{g},\mb{h})\mb{k}_2\\
&=\frac12 \!\left(\mb g_1\mb h_1\mb h_2+\frac{[\mb g_1]_4-[\mb g_1]_2}{2} \mb h_2+\mb g_0\mb h_1\right)\!\mb k_2,\quad \forall \mb{g,h,k}\in \Tilde{G}.
\end{align}
This expression matches exactly with $\dtilde{\mathbb V}_3 \in E_3^{3, 0} = H^3[\Tilde{G}, \U]$ in Eq.~\eqref{V3'}. Therefore, we see that $\dtilde{\mathbb V}$ becomes trivial on the third page of the LHSSS, confirming that it is a coboundary.

The second approach to show that $\dtilde{\mathbb V}_3$ is a coboundary is more explicit. Motivated by the definition of $F_{1,1}$ above, we define a 2-cochain on $\dtilde{G}$ as follows:
\begin{align}
\dtilde{\mu}_2(\ag{a}{g},\ag{b}{h})=\frac12 a\mb h_2\ (\mathrm{mod}\ 1),\quad \forall a\in \Tilde{A},\;\mb h\in \Tilde{G}.
\end{align}
We can then calculate the differential of $\dtilde{\mu}_2$, given by
\begin{align}\nonumber
\dd \dtilde{\mu}_2(\ag{a}{g},\ag{b}{h},\ag{c}{k}) 
&:= \dtilde{\mu}_2(\ag{b}{h},\ag{c}{k}) - \dtilde{\mu}_2(\ag{a}{g}\times_{\dtilde{G}}\ag{b}{h},\ag{c}{k}) + \dtilde{\mu}_2(\ag{a}{g},\ag{b}{h}\times_{\dtilde{G}}\ag{c}{k}) - \dtilde{\mu}_2(\ag{a}{g},\ag{b}{h}) \\\nonumber
&\={1}
\frac12 b\mb k_2
-\frac12\!\left(a+b+\mb g_1\mb h_1\mb h_2+\frac{[\mb g_1]_4-[\mb g_1]_2}{2} \mb h_2+\mb g_0\mb h_1\right)\!\mb k_2
+\frac12 a(\mb h_2+\mb k_2)
-\frac12 a\mb h_2\\
&\={1}
\frac12\!\left(\mb g_1\mb h_1\mb h_2+\frac{[\mb g_1]_4-[\mb g_1]_2}{2} \mb h_2+\mb g_0\mb h_1\right)\!\mb k_2, \quad\forall \ag{a}{g},\ag{b}{h},\ag{c}{k}\in \dtilde{G},
\end{align}
which is exactly $\dtilde{\mathbb V}_3 \in H^3[\dtilde{G}, \U]$ as the pullback of $\Tilde{\mathbb V}_3\in H^3[\Tilde{G}, \U]$ in Eq.~\eqref{V3'}. Therefore, we conclude that \begin{align} \dtilde{\mathbb V}_3 = \dd \dtilde{\mu}_2, \end{align} confirming that $\dtilde{\mathbb V}_3$ is a coboundary for $\dtilde{G}$ explicitly.
In this way, we obtain a trivial BSPT phase with symmetry $\dtilde{G} = \mathtt{SmallGroup}(64, 23)$. Shrinking this trivial $\mathtt{SmallGroup}(64,23)$-BSPT state and the nontrivial $\mathtt{SmallGroup}(32,2)$-BSPT state effectively leads to a gapped boundary for the nontrivial $(\mathbb{Z}_2^f \times \mathbb{Z}_4 \times \mathbb{Z}_4)$-FSPT state with complex fermion decorations.
\end{widetext}

\section{(3+1)D $(\mathbb Z_2^f \times G)$-FSPT states with complex fermion decorations}
\label{sec:3Dbdy1}

In this section, we will discuss the construction of gapped boundaries for FSPT states in (3+1) dimensions, decorated with complex fermions. This analysis serves as a foundation for extending to the more complex case of Majorana chain decorations in the next section. For related discussions, please refer to \refn{chen2020}. An alternative approach, utilizing the framework of spacetime partition functions, is presented in \refn{KOT2019}.

Since we restrict to FSPT phases decorated with only complex fermions and trivial $\om_2$ extension, the decoration data here is simply $(n_3,\nu_4)$ for FSPT with the direct product symmetry $G_f=\Zf\times G$. Here, the data $n_3\in H^3(G,\Z_2)$ is a 3-cocycle specifying the complex fermion decoration. And $\nu_4$ is the bosonic phase factor appears in the ground state wavefunction. 
In the special supercohomology classification, there are only two consistency equations:
\begin{align}
\dd n_3 &\={2} 0,\\ \label{dnu4}
\dd \nu_4 &\={1} \frac12 \sq^2(n_3). 
\end{align}
Here, $\sq^2(n_3) :=n_3\smile_1 n_3$ is the cocycle-level Steenrod square acting on the 3-cocycle $n_3\in H^3(G,\Z_2)$.

For a given $(\Z_2^f\times G)$-FSPT with decoration data $(n_3,\nu_4)$, the pullback trivialization
procedure for the symmetric gapped boundary can be summarized as
\begin{align}\nonumber
(n_3,\nu_4)
&\,\overset{\text{  pullback  }}{\sim} (n'_3,\nu'_4) = (\dd m'_2,\nu'_4)\\\nonumber
&\overset{\text{coboundary}}{\sim} (0,\mathcal V'_4)\\\nonumber
&\,\overset{\text{  pullback  }}{\sim} (0,\mathcal V''_4) = (0,\dd\mu''_3)\\
&\overset{\text{coboundary}}{\sim} (0,0),
\end{align}
where there exists a gapped interface indicated by ``$\sim$'' between the two SPT phases on the two sides in each step. They are obtained by either pullback to a bigger symmetry group or a coboundary FSLU (``gauge transformation'').

\subsection{Lattice model realizations for gapped interfaces}

\begin{figure*}[ht]
\centering\includegraphics[width=.8\linewidth]{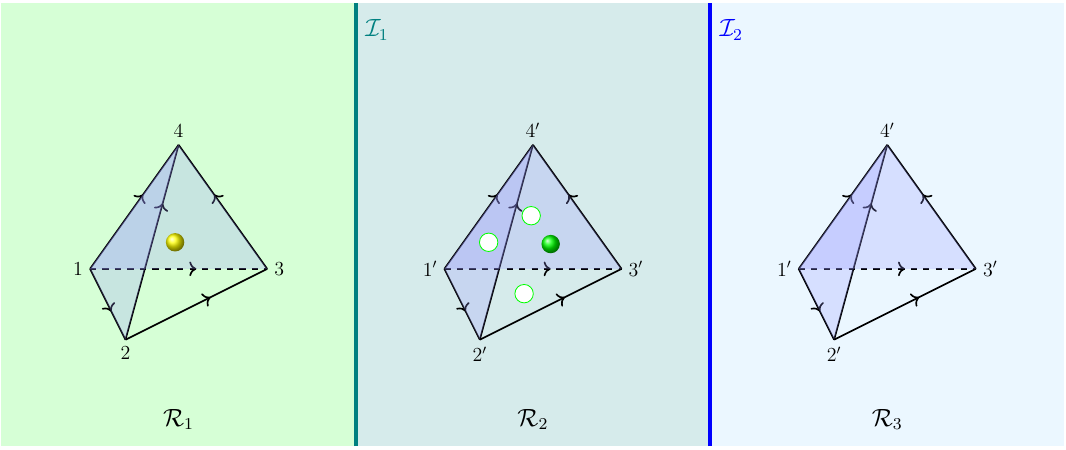}
\caption{Lattice interface construction for pullback trivialization of complex fermion decorations in (3+1)D FSPT phases. 
The first region, $\mathcal R_1$, represents an FSPT state with symmetry $\Z_2^f \times G$, featuring complex fermion decorations $n_3$ at the center of each tetrahedron. The second region, $\mathcal R_2$, is obtained from $\mathcal R_1$ by applying an FSLU $\mathcal U_1^f$, which enlarges the symmetry from $G$ to $G'$. The decoration data in this region is given by a 3-coboundary $n_3' = \dd m_2'$. Finally, the third region, $\mathcal R_3$, is obtained by applying another FSLU transformation, $\mathcal U_2^f$, and represents a $G'$-BSPT state.}
\label{fig:lattice3d}
\end{figure*}

The pullback construction and coboundary transformations can be explicitly described on the lattice, as illustrated schematically in Fig.~\ref{fig:lattice3d}. The diagram depicts three regions ($\mathcal R_1$, $\mathcal R_2$, and $\mathcal R_3$), with two gapped interfaces ($\mathcal I_1$ and $\mathcal I_2$) between them. For simplicity, we represent each region of 3D spatial triangulation with a single tetrahedron, corresponding to an FSPT or BSPT state.

% \begin{widetext}
% \end{widetext}

Initially, the entire 3D lattice is constructed as a $(\mathbb{Z}_2^f \times G)$-FSPT, featuring a nontrivial complex fermion decoration $n_3 \in H^3(G, \mathbb{Z}_2)$. This means that, in addition to assigning $|G|$ degrees of freedom at each vertex, the center of each tetrahedron may or may not be decorated with a spinless complex fermion (depicted as a yellow ball in Fig.~\ref{fig:lattice3d}), with the decoration specified by $n_3 \in H^3(G, \mathbb{Z}_2)$. The ground state wavefunction is a superposition of all possible configurations ${g_i}$, each with its corresponding complex fermion decoration.

In the second region, $\mathcal{R}_2$, shown in Fig.~\ref{fig:lattice3d}, we have a coboundary of the complex fermion decoration data, which is obtained by extending the symmetry from $G$ to a larger group $G'$, while trivializing the 3-cocycle $n_3$ via the short exact sequence:
\begin{align}
1 \rightarrow A \rightarrow G' \overset{\pi}{\rightarrow} G \rightarrow 1.
\end{align}
Here, $A$ is a normal subgroup of $G'$, and $\pi$ is the projection map. The extension is chosen such that the 3-cocycle $n_3$ is pulled back to a trivial 3-cocycle, $n_3' = \pi^\ast(n_3) = \dd m_2'$, in $H^3(G', \mathbb{Z}_2)$. This allows us to define an FSLU transformation as follows:
\begin{align}
\mathcal U_1^f =\prod_{\langle 1'2'3'4'\rangle\in \mathcal R_2,\mathcal R_3}
(c_{234A}^\dagger)^{m_2'(2'3'4')}(c_{124A}^\dagger)^{m_2'(1'2'4')}
(c_{134B}^\dagger)^{m_2'(1'3'4')}  
(c_{123B}^\dagger)^{m_2'(1'2'3')}  
(c_{1234})^{n_2(1234)},
\end{align}
where the product runs over all tetrahedra $\langle 1'2'3'4' \rangle$ in regions $\mathcal{R}_2$ and $\mathcal{R}_3$. This FSLU transformation annihilates the fermions (yellow balls) at the center of each tetrahedron and creates fermions (green circles or balls) at the centers of the boundary triangles, depending on the vertex labels $g_i' \in G'$. The 2-cochain $m_2'$ specifies the decoration of complex fermions (green balls) at the center of each triangle. This FSLU is $G$-symmetric and preserves fermion parity, due to the trivialization $n_3' = \dd m_2'$ (mod 2). As a result, all fermions at the centers of the tetrahedra are pushed to the boundaries near each triangle in region $\mathcal{R}_2$. This region thus represents an FSPT with symmetry $\mathbb{Z}_2^f \times G'$ and trivial complex fermion decoration $n_3' = \dd m_2$.
This pullback construction creates a gapped interface, $\mathcal{I}_1$, between the nontrivial $(\mathbb{Z}_2^f \times G)$-FSPT, labeled by $(n_3, \nu_4)$, in region $\mathcal{R}_1$ and a $(\mathbb{Z}_2^f \times G')$-FSPT, labeled by $(\dd m_2', \nu_4')$, in region $\mathcal{R}_2$.

Since the decoration data $n_3' = \dd m_2'$ is a coboundary, we can perform a FSLU or ``gauge transformation'' to remove this coboundary and set it to zero at the cochain level. This process is achieved by applying another FSLU transformation:
\begin{align}
\mathcal U_2^f=\prod_{\langle 1'2'3'\rangle\in\mathcal R_3}
(c_{123A})^{m_2'(1'2'3')}
(c_{123B})^{m_2'(1'2'3')},
\end{align}
which acts on the third region $\mathcal{R}_3$. This transformation annihilates the two fermions on the two sides of each triangle $\langle 1'2'3' \rangle$ in region $\mathcal{R}_3$. It preserves fermion parity, as the presence or absence of fermions on each side of a triangle is determined by the 2-cochain $m_2'(1'2'3')$ associated with that triangle. After applying this FSLU, the region becomes a purely bosonic state, with no remaining fermionic modes. Thus, region $\mathcal{R}_3$ is, in general, a BSPT phase protected by symmetry $G'$, with a 3-cocycle given by $\mathcal{V}_4' := \nu_4' + f[m_2']$. This phase has a gapped interface, $\mathcal{I}_2$, with the $(\mathbb{Z}_2^f \times G')$-FSPT, labeled by $(\dd m_2', \nu_4')$, in region $\mathcal{R}_2$, as shown in Fig.~\ref{fig:lattice3d}. Furthermore, this BSPT phase $\mathcal V_3'$ can be further trivialized using established pullback methods in bosonic systems.

By shrinking all intermediate SPT phases and interfaces, we ultimately obtain a symmetric gapped boundary for the original $(\mathbb{Z}_2^f \times G)$-FSPT $(n_3, \nu_4)$ in region $\mathcal{R}_1$. In the following, we will provide a step-by-step construction and the corresponding formulas for the pullback trivializations.

\subsection{Complex fermion decoration $n_3$}

Let us start with a $(\mathbb{Z}_2^f \times G)$-FSPT phase with complex decoration data $n_3 \in H^3(G, \mathbb{Z}_2)$ and a bosonic phase factor $\nu_4$. We can use a symmetry extension
\begin{align}
1 \rightarrow A \rightarrow {G'} \overset{\pi}{\rightarrow} G \rightarrow 1,
\end{align}
to trivialize the 3-cocycle $n_3$. Specifically, we want its pullback $n_3' = \pi^\ast(n_3)$ to be a 3-coboundary:
\begin{align}
n'_3 \={2} \dd m'_2
\end{align}
with a 2-cochain $m_2' \in C^2(G', \mathbb{Z}_2)$. In this way, we obtain a gapped interface $\mathcal{I}_1$ between the $(\mathbb{Z}_2^f \times G)$-FSPT phase $(n_3, \nu_4)$ in region $\mathcal{R}_1$ and the $(\mathbb{Z}_2^f \times G')$-FSPT phase $(\dd m_2', \nu_4')$ with complex fermion decoration data $n_3'=\dd m_2'$, which is a 3-coboundary, in region $\mathcal{R}_2$, as shown in Fig.~\ref{fig:lattice3d}.

The next task is to perform a FSLU or ``gauge transformation'' to make $n'_3 = \dd m'_2$ vanish at the cochain level. By setting $n'_3 = \dd m'_2$, the consistency condition in \eq{dnu4} becomes:
\begin{align}
\dd \nu'_4
\={1} \mathcal O_5[n_3']
\={1} \frac12 \Sq^2(n'_3)
\={1} \frac12 \Sq^2(\dd m'_2)
\={1} \frac12 \dd [\Sq^2(m'_2)],
\end{align}
where $\Sq^2(m'_2) \={2} m'_2 \smile m'_2 + m'_2 \smile_1 \dd m'_2$ represents the Steenrod square acting on 2-cochains. Using this, we can define
\begin{align}\label{V4}
\mathcal V'_4 :\={1} \nu'_4-\frac12\Sq^2(m'_2).
\end{align}
This $\mathcal{V}'_4$ becomes a 4-cocycle of $G'$, satisfying the condition
\begin{align}
    \dd \mathcal V'_4\={1} \mathcal O_5[0] \={1}0.
\end{align}
In this way, we obtain a $G'$-BSPT phase characterized by the 4-cocycle $\mathcal{V}'_4$. A gapped interface, $\mathcal{I}_2$, exists between the $(\mathbb{Z}_2^f \times G')$-FSPT state labeled by $(\dd m_2', \nu_4')$ in region $\mathcal{R}_2$ and the $G'$-BSPT state labeled by the 4-cocycle $\mathcal{V}'_4$ in region $\mathcal{R}_3$, as shown in Fig.~\ref{fig:lattice3d}.

\subsection{Bosonic SPT $\nu_4$}

The final task is to trivialize the bosonic cocycle $\mathcal{V}'_4$ describing the $G'$-BSPT. To achieve this, we use another central extension:
\begin{align}
1 \rightarrow A' \rightarrow G'' \overset{\pi'}{\rightarrow} G' \rightarrow 1,
\end{align}
to trivialize $\mathcal{V}''_4 := \pi'^\ast(\mathcal{V}'_4)$ as
\begin{align}
\mathcal V''_4 \={1} \dd \mu''_3.
\end{align}
Here, $\mu''_3$ is a 3-cochain in $C^3[G'', \U]$.

The BSPT phase characterized by the cocycle $\mathcal{V}''_4$ in the trivial class has an interface with the vacuum (or a direct product state). By shrinking all intermediate SPT states and the interfaces between them, the final result is a symmetric gapped boundary for the original $(\mathbb{Z}_2^f \times G)$-FSPT phase with decoration data $(n_3, \nu_4)$.

\subsection{Example: $(\Z_2^f\times\Z_2\times\Z_4)$-FSPT, $(\Z_4\times\Z_4)$-BSPT and trivial $(\Z_8\times\Z_4)$-BSPT}

In this subsection, we will provide an explicit example of the pullback construction for gapped coboundaries of (3+1)D FSPT phases with complex fermion decorations.

For simplicity, let's consider an FSPT associated with an Abelian symmetry group \cite{ZWWG2019}. It is known that the smallest Abelian symmetry supporting an FSPT with complex fermion decoration is $G_f = \mathbb{Z}_2^f \times G = \mathbb{Z}_2^f \times (\mathbb{Z}_2 \times \mathbb{Z}_4)$. The decoration data are as follows:
\begin{align}
n_3 &\={2} \left[n_1^{(1)}\right]^2 n_1^{(2)},\\
\nu_4 &\={1} \frac14 \left[n_1^{(1)}\right]^3 n_1^{(2)}.
\end{align}
Again, we use the notation $n_1^{(i)}(\mb{g}) = [\mb{g}_i]_2 := \mb{g}_i \ (\text{mod} \ 2)$ for an arbitrary $\mb{g} = (\mb{g}_1, \mb{g}_2) \in \mathbb{Z}_2 \times \mathbb{Z}_4$. One can verify that these decoration data satisfy the consistency conditions $\dd n_3=0$ and $\dd \nu_4=\frac{1}{2}\sq^2(n_3)$.

From the previous constructions, we need to trivialize $n_3$ and $\nu_4$ using two symmetry extensions. Let us choose $A = A' = \mathbb{Z}_2$, and we can construct two consecutive central extensions that extend the $\mathbb{Z}_2$ subgroup of $G$:
\begin{align}
0 \rightarrow (A=\Z_2) \rightarrow ( G'=\Z_4\times\Z_4) \rightarrow(G=\Z_2\times\Z_4) \rightarrow 0,\\
0 \rightarrow (A'=\Z_2) \rightarrow ( G''=\Z_8\times\Z_4) \rightarrow(G'=\Z_4\times\Z_4) \rightarrow 0,
\end{align}
which can be shown to trivialize the cocycles $n_3$ and $\nu_4$, respectively.
Alternatively, the composition of these two central extensions gives us a new symmetry extension, with $\tilde{A} = \Z_4$ being an extension of both $A$ and $A'$:
\begin{align}\label{Z8Z4}
0 \rightarrow (\Tilde A=\Z_4) \rightarrow (\Tilde G=\Z_8\times\Z_4) \overset{\Tilde\pi}{\rightarrow} (G=\Z_2\times\Z_4) \rightarrow 0,
\end{align}
The extension 2-cocycle $\phi_2 \in H^2(G, \tilde{A})$ is given by $\phi_2 = \sqb{n_1^{(1)}}^2 \in H^2(\Z_2,\Z_4)=\Z_2 \subset H^2(G,\Z_4)$ which extends the first $\Z_2$ subgroup of $G$ to $\Z_8$ in $\tilde G$.

In the second approach, given by \eq{Z8Z4}, the complex fermion decoration data $\tilde{n}_3 := \tilde{\pi}^*(n_3) \in H^3(\tilde{G}, \mathbb{Z}_2)$ is trivialized as follows:
\begin{align}
\Tilde n_3 \={2} \dd \Tilde m_2,
\end{align}
where the 2-cochain $\tilde{m}_2$ is defined by
\begin{align}
\Tilde m_2 :&\={2} \Tilde x_1^{(1)} \smile \Tilde n_1^{(2)},\\\label{x1}
\Tilde x_1^{(1)}(\Tilde{\mb{g}}) :&\={2} \frac{[\Tilde{\mb{g}}_1]_4-[\Tilde{\mb{g}}_1]_2}{2}, \quad\forall \Tilde{\mb{g}}=(\Tilde{\mb{g}}_1,\Tilde{\mb{g}}_2)\in \Tilde G.
\end{align}
This is because we have $\dd \tilde{x}_1 \={2} (\tilde{n}_1)^2$ for $\tilde G$.

From \eq{V4}, the explicit expression for the (3+1)-dimensional BSPT cocycle $\tilde{\mathcal{V}}_4$ is given by
\begin{align}
\Tilde{\mathcal V}_4 &\={1} \Tilde \nu_4-\frac12\Sq^2(\Tilde m_2)\\
&\={1} \frac14 \left[\Tilde n_1^{(1)} \right]^3 \Tilde n_1^{(2)} 
+ \frac12 \Tilde x_1^{(1)}\Tilde n_1^{(2)}\Tilde x_1^{(1)}\Tilde n_1^{(2)} 
+ \frac12 (\Tilde x_1^{(1)}\Tilde n_1^{(2)})\smile_1(\Tilde n_1^{(1)}\Tilde n_1^{(1)}\Tilde n_1^{(2)}).
\end{align}
Using topological invariants of cocycles for Abelian groups from loop braiding statistics \cite{WangLevin2015, Tantivasadakarn_2017, ZWWG2019}, one can verify that the above 4-cocycle $\tilde{\mathcal{V}}_4$ is indeed a 4-coboundary for $\tilde{G} = \mathbb{Z}_8 \times \mathbb{Z}_4$.

Therefore, we conclude that there exists a gapped boundary for the $(\mathbb{Z}_2^f \times \mathbb{Z}_2 \times \mathbb{Z}_4)$-FSPT with complex fermion decorations. This boundary exhibits a larger symmetry group $\mathbb{Z}_2^f \times \mathbb{Z}_8 \times \mathbb{Z}_4$. Furthermore, one can also gauge the normal subgroup $\tilde{A} = \mathbb{Z}_4$ in \eq{Z8Z4} on the (2+1)-dimensional boundary, resulting in a $(\mathbb{Z}_2^f \times \mathbb{Z}_2 \times \mathbb{Z}_4)$-symmetry enriched $\tilde{A} = \mathbb{Z}_4$ gauge theory, which supports anyonic excitations with nontrivial symmetry actions.

\section{(3+1)D $(\mathbb Z_2^f \times G)$-FSPT states with Majorana chain decorations}
\label{sec:3Dbdy2}

After constructing the pullback trivialization for the gapped boundary of the (3+1)D FSPT with complex fermion decorations in the previous section, we now turn to a more exotic construction for FSPT phases with Majorana chain decorations. For simplicity, we will focus on direct product symmetry groups $G_f = \mathbb{Z}_2^f \times G$. The construction for symmetry groups with nontrivial extensions will be discussed in the next section.

For a (3+1)D $(G_f = \mathbb{Z}_2^f \times G)$-FSPT state with a trivial $\omega_2$ extension, the decoration data in this case is $(n_2, n_3, \nu_4)$, which satisfy the following consistency equations:
\begin{align}\label{dn2}
\dd n_2 &\={2} 0,\\\label{dn3}
\dd n_3 &\={2} n_2\smile n_2,\\\label{dnu4_}
\dd \nu_4 &\={1} \frac12 \Sq^2(n_3)-\frac14 n_2\smile \beta_2  n_2 
+\frac12(n_2)^2(02345)(n_2)^2(01235). 
\end{align}
Here, the term $\beta_2 n_2$ on the right-hand side of \eq{dnu4_} represents the Bockstein homomorphism, given by $\beta_2  n_2(ijkl)\={\Z} n_2(jkl)n_2(ijl)-n_2(ikl)n_2(ijk)$. 
(This expression is identical to the standard Bockstein homomorphism, which is explicitly verified in Table \ref{tab:bockstein} in Appendix \ref{sec:O5om}.)
It is important to note that the last term in \eq{dnu4_} is not in the form of a (higher) cup product. Additionally, the right-hand side of the $\dd \nu_4$ equation differs from the result in Ref.~\onlinecite{wang2017towards} by a coboundary term $\dd\!\left(\frac{1}{8} n_2 n_2\right)=\frac14\beta_2 n_2\smile n_2+\frac14 n_2\smile\beta_2 n_2$, which can be absorbed as a redefinition of the phase factor $\nu_4$ for the fixed-point ground-state wavefunction.

The pullback construction now involves multiple steps, starting with the trivialization of the Majorana chain decoration data $n_2$, followed by the complex fermion decoration and the bosonic phase factor. This process can be summarized as follows:
\begin{align}\nonumber
(n_2,n_3,\nu_4)
&\,\overset{\text{  pullback  }}{\sim} (n'_2,n'_3,\nu'_4) = (\dd m'_1,n'_3,\nu'_4)\\\nonumber
&\overset{\text{coboundary}}{\sim} (0,N'_3,\mathcal V'_4)\\\nonumber
&\,\overset{\text{  pullback  }}{\sim} (0,N''_3,\mathcal V''_4) = (0,\dd m''_2,\mathcal V''_4)\\\nonumber
&\overset{\text{coboundary}}{\sim} (0,0,\mathscr V''_4)\\\nonumber
&\,\overset{\text{  pullback  }}{\sim} (0,0,\mathscr V'''_4)=(0,0,\dd\mu'''_3)\\
&\overset{\text{coboundary}}{\sim} (0,0,0),
\end{align}
where ``$\sim$'' denotes a gapped interface between two SPT phases on either side. In the following, we will first discuss the lattice model realizations of the pullback construction. Then, we will explain each step of the construction in detail, following the sequence outlined above.

\subsection{Lattice model realizations for gapped interfaces}
\begin{widetext}
\begin{figure*}[ht]
    \centering\includegraphics[width=.8\linewidth]{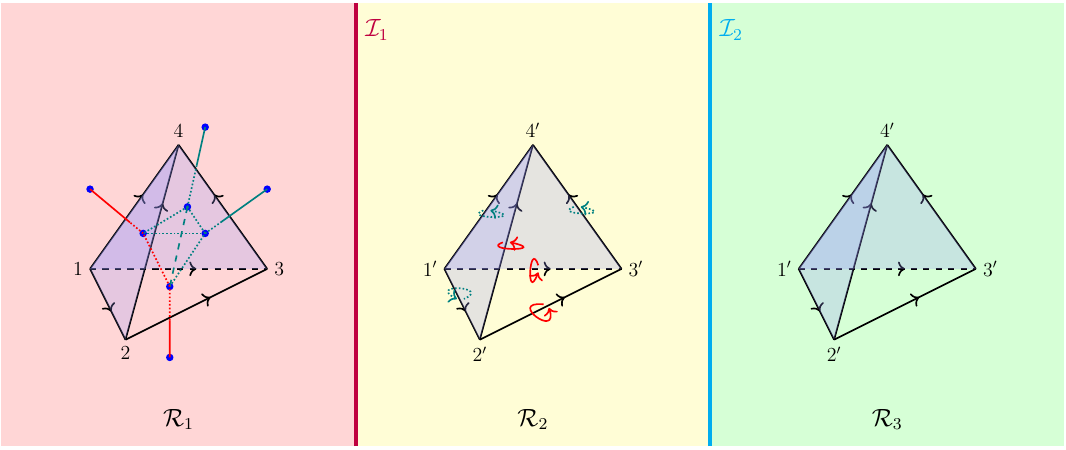}
    \caption{Lattice interface construction for pullback trivialization of Majorana chain decorations in (3+1)D FSPT phases. 
    The first region, $\mathcal R_1$, represents an FSPT state with symmetry $\Z_2^f \times G$, featuring Majorana chain decorations $n_2$ through each triangle. Complex fermion decorations, being present across the lattice, are omitted here for clarity. The second region, $\mathcal R_2$, is obtained from $\mathcal R_1$ by applying an FSLU $\mathcal U_1^f$, which enlarges the symmetry from $G$ to $G'$. The decoration data in this region is given by a 2-coboundary $n_2' = \dd m_1'$. Finally, the third region, $\mathcal R_3$, is obtained by applying another FSLU transformation, $\mathcal U_2^f$, and represents a $\Z_2^f \times G'$-FSPT state with only complex fermion decorations.}
    \label{fig:lattice3dkitaev}
\end{figure*}
\end{widetext}

As in the previous sections, the pullback construction and coboundary transformations for Majorana chain decorations can be explicitly described on the lattice. This is schematically shown in Fig.~\ref{fig:lattice3dkitaev}. The diagram consists of three regions ($\mathcal{R}_1$, $\mathcal{R}_2$, and $\mathcal{R}_3$), with two gapped interfaces ($\mathcal{I}_1$ and $\mathcal{I}_2$) separating them. For simplicity, we represent each region of the 3D spatial triangulation with a single tetrahedron, corresponding to an FSPT state. Since the focus here is on the Majorana chain decoration, we omit the construction for complex fermion decoration, which are covered in detail in the previous section (see Fig.~\ref{fig:lattice3d}).

The first region, $\mathcal{R}_1$, is constructed as a $(\mathbb{Z}_2^f \times G)$-FSPT, featuring a nontrivial Majorana chain decoration data $n_2 \in H^2(G, \mathbb{Z}_2)$. This means that we assign $|G|$ Majorana fermion modes (depicted as blue dots in Fig.~\ref{fig:lattice3dkitaev}) to the two sides of each triangle in the triangulation of the 3D spatial lattice. The Majorana fermions are in nontrivial pairing if and only if $n_2 = 1$ for a given triangle. Physically, this means there is a Majorana chain running through the triangle (represented by the red line in region $\mathcal{R}_1$ of Fig.~\ref{fig:lattice3dkitaev}) if the corresponding $n_2 = 1$.
The Majorana fermions are paired with respect to the \emph{local Kasteleyn orientations} defined in Refs.~\onlinecite{wang2017towards,wang2018construction}.
In addition to the Majorana chain, there are also, in principle, complex fermions located at the center of each tetrahedron, as specified by $n_3$. However, for clarity, we omit these details in the figure. The ground state wavefunction is a superposition of all possible vertex configurations ${g_i}$, each corresponding to its own Majorana chain and complex fermion decoration.

In the second region, $\mathcal{R}_2$, shown in the middle of Fig.~\ref{fig:lattice3dkitaev}, we have a 2-coboundary given by $n_2' = \dd m_1'$, which is obtained by extending the symmetry from $G$ to a larger group $G'$. This extension trivializes the 2-cocycle $n_2$ via the short exact sequence: 
\begin{align}
1 \rightarrow A \rightarrow G' \overset{\pi}{\rightarrow} G \rightarrow 1.
\end{align} 
Here, $A$ is a normal subgroup of $G'$, and $\pi$ is the projection map. The extension is chosen so that the 2-cocycle $n_2\in H^2(G,\Z_2)$ is pulled back to a trivial 2-cocycle, $n_2' = \pi^\ast(n_2) = \dd m_1'$, in $H^2(G', \mathbb{Z}_2)$. This allows us to define an FSLU transformation, $\mathcal{U}_1^f$. 
The trivialization equation $n_2'(g_0,g_1,g_2)=\dd m_1'(g_0,g_1,g_2)=m_1'(g_1,g_2)-m_1'(g_0,g_2)+m_1'(g_0,g_1)$ contains three terms, which suggests a physical interpretation where the Majorana chain described by $n_2' = 1$ splits into three parallel Majorana chains (possibly trivial) near the edges of the triangle $\langle 012 \rangle$. These new chains are governed by the cochain $m_1'$, and their pairing directions (i.e., the orientations of Majorana fermion couplings) remain the same with the original chain in the middle of the triangle.
This implies that for each link $\langle ij \rangle$ in the triangulation, we introduce $|G'|$ complex fermion modes per triangle where this link serves as the boundary. These complex fermions are subsequently decomposed into $2|G'|$ Majorana fermions, which are then distributed on the two sides of each triangle.
Since $m_1'(g_i, g_j)=0, 1$ determines whether the Majorana chain near the edge $\langle ij \rangle$ is in a trivial or nontrivial pairing configuration — depending only on $g_i$ and $g_j$ — a Majorana chain loop must encircle the link $\langle ij \rangle$ whenever $m_1'(g_i, g_j) = 1$. This results in the small red loops depicted in the middle of Fig.~\ref{fig:lattice3dkitaev}.
All small Majorana chain loops are Kasteleyn-oriented, ensuring a fermion parity-even state. This is ensured by construction, as the link orientations satisfy the local Kasteleyn condition defined in Ref.~\onlinecite{wang2017towards}.
In summary, we can use a FSLU transformation to annihilate the Majorana chain running through a triangle (represented by the red line in region $\mathcal{R}_1$ of Fig.~\ref{fig:lattice3dkitaev}) and creates small Majorana chain loops
(small red loops in region $\mathcal{R}_2$ of Fig.~\ref{fig:lattice3dkitaev}) surrounding each link if the corresponding $m_1' = 1$. In this way, the second region $\mathcal{R}_2$ is a $(\mathbb{Z}_2^f \times G')$-FSPT with a coboundary Majorana chain decoration data $n_2' = \dd m_1'$. There is a gapped interface, $\mathcal{I}_1$, between this FSPT and the original $(\mathbb{Z}_2^f \times G)$-FSPT with the nontrivial Majorana chain decoration data $n_2$.

The third region, $\mathcal{R}_3$, is obtained by applying another FSLU transformation, $\mathcal{U}_2^f$, to the second region. This FSLU transformation acts as a ``gauge transformation'' that removes the coboundary decoration data $n_2' = \dd m_1'$. The basic idea behind this second FSLU is to shrink the small Majorana chain loops (depicted as red loops in Fig.~\ref{fig:lattice3dkitaev}) to the vacuum. In other words, we place all Majorana fermions around each link in the triangulation into trivial vacuum pairings. This process preserves fermion parity due to the local Kasteleyn orientations on 3D lattice introduced in Ref.~\onlinecite{wang2017towards}. After applying this FSLU, the state in region $\mathcal{R}_3$ becomes an FSPT with only complex fermion decorations and no Majorana chain decorations. There is a gapped interface, $\mathcal{I}_2$, between this $(\mathbb{Z}_2^f \times G')$-FSPT with complex fermion decorations in region $\mathcal{R}_3$ and the previous $(\mathbb{Z}_2^f \times G')$-FSPT in region $\mathcal{R}_2$.

The next step in the pullback construction, which involves the complex fermion decoration data $n_3$ and the bosonic phase factor $\nu_4$, follows a similar process as described in the previous section, so we will omit further details here. 
By shrinking all intermediate SPT phases and interfaces, we ultimately obtain a symmetric gapped boundary for the original $(\mathbb{Z}_2^f \times G)$-FSPT $(n_2, n_3, \nu_4)$ in region $\mathcal{R}_1$. 

In the following, we will provide a step-by-step construction along with the corresponding formulas for the pullback trivializations.

\subsection{Majorana chain decoration $n_2$}

For a given $(\mathbb{Z}_2^f \times G)$-FSPT with decoration data $(n_2, n_3, \nu_4)$, the first step is to trivialize the Majorana chain decoration data $n_2 \in H^2(G, \mathbb{Z}_2)$. This is achieved by considering the group extension:
\begin{align}\label{3Dn2}
1 \rightarrow A \rightarrow G' \overset{\pi}{\rightarrow} G \rightarrow 1.
\end{align}
Here, we require the pullback 2-cocycle $n_2' = \pi^\ast(n_2)$ to be trivial, expressed as
\begin{align}
n'_2 \={2} \dd m'_1,
\end{align}
where $m_1' \in C^1(G', \mathbb{Z}_2)$ is a 1-cochain. Note that this equation holds only modulo 2, since both $n_2'$ and $m_1'$ take values in $\{0, 1\}$. However, in the obstruction function \eq{dnu4_}, $n_2'$ must be lifted to a $\mathbb{Z}$-valued cochain becuase of the term $-\frac{1}{4}n_2'\,\beta_2n_2'$. Therefore, we refine the trivialization equation to a $\mathbb{Z}$-valued form as follows:
\begin{align}\label{n2}
    n'_2 \={\Z} \dd m'_1 - 2 \Sq^1(m'_1) + 4m'_1 \smile_1 (m'_1)^2,
\end{align}
where $\Sq^1$ is the cochain-level Steenrod square, and the equation is detailed in Appendix~\ref{sec:triveqz}.

We would like to emphasize that there is a canonical choice of the extension \eq{3Dn2} for trivializing $n_2$. This is achieved by selecting the normal subgroup $A = \mathbb{Z}_2$ and using the 2-cocycle $\phi_2 = n_2 \in H^2(G, \mathbb{Z}_2)$ to specify the extension. With this choice, one can show that $n_2$ is pulled back to $n_2' = \pi^\ast(n_2)$, which becomes a coboundary in $H^2(G', \mathbb{Z}_2)$. This follows because the differential on the second page of the LHSSS is precisely given by $\phi_2$, and its image covers $n_2$ at the $E_2^{2,0}$ position of the spectral sequence.

%we want to emphasize that there is a canonical choice of the extnesion \eq{3Dn2} to trivialize $n_2$. this is done by choosing the normal subgroup to be $A=\Z_2$ and the 2-cocycle specifing this extension to be $\phi_2=n_2\in H^2(G,\Z_2)$. In this way, one can show that $n_2$ will be pullback to $n_2'=\pi^\ast(n_2)$ to be a coboundary in $H^2(G',\Z_2)$. this is becasue the differential on the second page of the LHSSS is given precesely by $\phi_2$ and has a image that trivialize the $n_2$ at the $E_2^{2,0}$ position of the spectral sequence.

In this way, we construct a gapped interface $\mathcal{I}_1$ between the original $(\mathbb{Z}_2^f \times G)$-FSPT phase, labeled by $(n_2, n_3, \nu_4)$ in region $\mathcal{R}_1$, and the $(\mathbb{Z}_2^f \times G')$-FSPT phase, labeled by $(\dd m_1', n_3', \nu_4')$ in region $\mathcal{R}_2$, as illustrated in Fig.~\ref{fig:lattice3dkitaev}.

Now, we want to perform a FSLU or ``gauge transformation'' to make the Majorana chain decoration data $\dd m_1'$ vanish at the cochain level. This procedure results in the interface $\mathcal{I}_2$ shown in Fig.~\ref{fig:lattice3dkitaev}, and simultaneously twists the complex fermion decoration data $n_3'$ as well as the bosonic phase factor layer $\nu_4'$.

For the extended group $G'$, the consistency condition \eq{dn3} for the 3-cochain $n_3' = \pi^\ast(n_3)$, which governs the complex fermion decoration, becomes 
\begin{align}\label{n3cobdy}
\dd n'_3 &\={2} n'_2 \smile n'_2
\={2} \dd m'_1 \smile \dd m'_1
\={2} \dd (m'_1 \smile \dd m'_1).
\end{align}
We can then define a $\mathbb{Z}_2$-valued 3-cocycle for $G'$ as
\begin{align}\label{N3}
N'_3 :\={2} n'_3 + m'_1 \smile \dd m'_1.
\end{align}
This ensures that $N'_3$ satisfies the desired consistency equation
\begin{align}
\dd N_3'\={2}0.
\end{align}
It corresponds to an FSPT phase with labels $(0, N'_3, \mathcal{V}_4')$, where the Majorana chain decoration data has been trivialized to $N'_2 = 0$ at the cochain level.

Now, we want to determine the bosonic phase factor $\mathcal{V}_4'$ after performing the FSLU or ``gauge transformation'' in trivializing the Majorana chain decoration data at the cochain level. In the extended group $G'$, the obstruction for the 4-cochain $\nu_4$ in \eq{dnu4} is pulled back as follows:
\begin{align}\label{dnu4pb}\nonumber
\dd \nu'_4
&\={1} \mathcal O_5[n'_2,n'_3]\\
&\={1} \frac12 \Sq^2(n'_3)-\frac14 n'_2\smile\beta_2 n'_2
+\frac12(n'_2)^2(02345)(n'_2)^2(01235).
\end{align}
Using the newly defined 3-cocycle $N_3'$ in \eq{N3}, we can transform the obstruction function $\mathcal{O}_5[n'_2, n'_3]$ on the right-hand side of the above equation to the obstruction function $\mathcal{O}_5[0, N_3']$, up to a coboundary term. A rather lengthy computation, which we relegate to Appendix~\ref{sec:O5pb}, shows that by defining the new 4-cochain
\begin{align}\label{N4nu4}
\mathcal V'_4 &:\={1} \nu'_4 -f_4[m'_1,n'_3],\\
\nonumber
    f_4[m_1',n_3']
    &:\={1}\frac14 (\dd m_1') \Sq^1(m_1')
    +\frac12 \dd(m_1' \dd m_1')\smile_3 (n'_3+m_1' \dd m_1')
    +\frac12 (m_1' \dd m_1')\smile_2 (n'_3+m_1' \dd m_1')\\&\quad
    +\frac12(\dd m_1')(m_1'^2\smile_2 \dd m_1')+\frac12 m_1'^2 (\dd m_1'\smile_1 m_1')
    +\frac12m_1'(01)m_1'(12)m_1'(14)(\dd m_1')(234),
\end{align}
we obtain the result that the new bosonic phase factor $\mathcal{V}_4'$ satisfies the consistency equation
\begin{align}\label{twistc}
\dd\mathcal V'_4 \={1} \frac12 \Sq^2(N'_3).
\end{align}
This provides the second consistency equation for the $(\mathbb{Z}_2^f \times G')$-FSPT with labels $(0, N_3', \mathcal{V}_4')$, where the Majorana chain decoration data has been trivialized to $N'_2 = 0$ in the cochain level and only complex fermion decorations remain.

In this way, we construct a gapped interface $\mathcal{I}_2$ between the $(\mathbb{Z}_2^f \times G)$-FSPT phase, labeled by $(\dd m_1', n_3', \nu_4')$ in region $\mathcal{R}_2$, and the $(\mathbb{Z}_2^f \times G')$-FSPT phase, labeled by $(0, N_3', \mathcal{V}_4')$ in region $\mathcal{R}_3$, as illustrated in Fig.~\ref{fig:lattice3dkitaev}. The new decoration data $N_3'$ and $\mathcal{V}_4'$ are twisted by $m_1'$ from the original pulled-back cochains $n_3'$ and $\nu_4'$.

\subsection{Complex fermion decoration $n_3$}

After trivializing the Majorana chain decoration data $n_2'$, we now have an FSPT phase labeled by $(0, N_3', \mathcal{V}_4')$, which only contains complex fermion decorations. The pullback trivialization for this FSPT phase follows a similar procedure as in the previous section. Specifically, we use a second central extension
\begin{align}
1 \rightarrow A' \rightarrow G'' \overset{\pi'}{\rightarrow} G' \rightarrow 1.
\end{align}
We can trivialize $N'_3$ by pulling it back via $\pi'$. More explicitly, we define a 2-cochain $m''_2$ of $G''$ and express the trivial 3-cocycle $N''_3:\={2}\pi'^\ast(N'_3)$ as a coboundary
\begin{align}\label{N3=dm2}
N''_3 \={2} \dd m''_2.
\end{align}
In other words, the trivialization equation for the pullback of the original 3-cochain $n_3$ becomes
\begin{align}\label{n3}
n''_3 \={2} \dd m''_2 + m''_1 \smile \dd m''_1,
\end{align}
where $m''_1$ and $n''_3$ are the pullbacks of $m'_1$ and $n_3'$ by $\pi'$, respectively.

Using the equation for $\mathcal V_4'$ and $\nu_4'$ from \eq{N4nu4}, the bosonic phase factor 4-cochain is also pulled back with $\pi'^\ast$, taking the form
\begin{align}
\mathcal V''_4 :\={1} \nu''_4 -f_4[m''_1,n''_3].
\end{align}
By pulling back the twisted cocycle condition from \eq{twistc} and using the trivialization condition for $N_3''$ in \eq{N3=dm2}, we obtain
\begin{align}
    \dd\mathcal V''_4
    \={1} \frac12 \Sq^2(N''_3)
    \={1} \frac12 \Sq^2(\dd m''_2)
    \={1} \frac12 \dd\,\Sq^2(m''_2).
\end{align}
Therefore, by redefining the phase factor as:
\begin{align}
\mathscr V''_4 :\={1} \mathcal V''_4 -\frac12 \Sq^2(m''_2),
\end{align}
we perform a FSLU or ``gauge transformation'' that trivializes the complex fermion decoration data $N_3''$ at the cochain level. As a result, we are left with a bosonic 4-cocycle $\mathscr{V}_4''$ satisfying
\begin{align}
\dd \mathscr V_4''\={1}0,
\end{align}
which corresponds to a (3+1)D $G''$-BSPT phase.

In this way, we have constructed a gapped interface between the $(\mathbb{Z}_2^f \times G')$-FSPT phase, labeled by $(0, N_3', \mathcal{V}_4')$ with only complex fermion decorations, and the $G''$-BSPT phase, labeled by the 4-cocycle $\mathscr{V}_4''\in H^4[G'',\U]$.

\subsection{Bosonic SPT $\nu_4$}

The 4-cocycle $\mathscr{V}_4''$ of the $G''$ group is not generally in the trivial class of $H^4[G'', \U]$. To trivialize it, we introduce a third extension
\begin{align}
1 \rightarrow A'' \rightarrow G''' \overset{\pi''}{\rightarrow} G'' \rightarrow 1
\end{align}
and pull back $\mathscr{V}_4''$ to $H^4[G''', \U]$ via $\pi''$, defining the new 4-cocycle as $\mathscr{V}_4''' := \pi''^\ast(\mathscr{V}_4'')$. We then introduce a 3-cochain $\mu_3''' \in C^3[G''', \U]$ such that
\begin{align}
\mathscr V'''_4 \={1} \dd \mu'''_3.
\end{align}
With this trivialized 4-coboundary $\mathscr{V}_4'''$, we obtain a trivial $G'''$-BSPT phase.

In conclusion, after pulling back to the larger group $G'''$, the original $(\mathbb{Z}_2^f \times G)$-FSPT phase with labels $(n_2, n_3, \nu_4)$ becomes $(n_2''', n_3''', \nu_4''')$. This new phase has a symmetric gapped boundary when appropriate cocycles are made to be coboundaries. The corresponding pulled-back cochains are
\begin{align}
(n'''_2,n'''_3,\nu'''_4)
=(\dd m'''_1,\dd m'''_2+m'''_1\smile \dd m'''_1, \dd \mu'''_3+ \mathcal F_4[m'''_1,m'''_2]).
\end{align}
The most nontrivial 4-cochain is
\begin{align}\label{f4maj}\nonumber
\mathcal F_4[m_1''',m_2''']
&\={1} -\frac14(\dd m_1''')\Sq^1(m_1''')
+\frac12 [\Sq^2(m_2''') + \dd m_2'''\smile_2 (m_1''' \dd m_1''')]
+\frac12(\dd m_1''')\sqb{(m_1''')^2\smile_2 \dd m_1'''}\\
&\quad
+\frac12 (m_1''')^2 (\dd m_1'''\smile_1 m_1''')
+\frac12m_1'''(01)m_1'''(12)m_1'''(14)\dd m_1'''(234),
\end{align}
where the cochains $m_1'''$ and $m_2'''$ are the pullbacks of $m_1''$ and $m_2''$, respectively.

\subsection{Example: $(\Z_2^f\times\Z_4\times\Z_4)$-FSPT, $(\Zf\times\mathtt{SmallGroup}(32,2))$-FSPT and $\mathtt{SmallGroup}(64,23)$-BSPT}

In this subsection, we will demonstrate the pullback construction of a (3+1)D FSPT state with Majorana chain decorations, using a concrete example and explicit cochains.

\subsubsection{The nontrivial $(\Zf\times \Z_4\times\Z_4)$-FSPT with Majorana chain decorations}

For the symmetry $G_f = \mathbb{Z}_2 \times G = \mathbb{Z}_2 \times (\mathbb{Z}_4 \times \mathbb{Z}_4)$, a (3+1)D FSPT state decorated with Majorana chains can be constructed. It can be shown that for Abelian groups $G$ with order smaller than 16, no FSPT states with symmetry $\mathbb{Z}_2^f \times G$ exist that have Majorana chain decorations \cite{ZWWG2019}. The decoration data for the state are given by:
\begin{align}\label{n2example}
n_2 &\={2} n_1^{(1)} n_1^{(2)},\\\label{n3example}
n_3 &\={2} n_1^{(1)}\sqb{n_1^{(1)}\smile_1n_1^{(2)}}n_1^{(2)} + x_1^{(1)}\sqb{n_1^{(2)}}^2,
\end{align}
Here, the $\mathbb{Z}_2$-valued 1-cochain $x_1^{(i)}$ is defined as
\begin{align}
    x_1^{(i)}(\mb{g}) :\={2} \frac{[\mb{g}_i]_4-[\mb{g}_i]_2}{2}, \quad \forall \mb{g}=(\mb g_1,\mb g_2)\in \Z_4\times \Z_4,
\end{align}
and the $\mathbb{Z}_2$-valued 1-cocycle $n_1^{(i)} \in H^1(\mathbb{Z}_4 \times \mathbb{Z}_4, \mathbb{Z}_2)$ is given by
\begin{align}
    n_1^{(i)}(\mb{g})=[\mb{g}_i]_2, \quad \forall \mb{g}=(\mb g_1,\mb g_2)\in \Z_4\times \Z_4.
\end{align} 
These cochains satisfy the relation
\begin{align}
\dd x_1^{(i)} \={2}\left[n_1^{(i)}\right]^2.
\end{align}
Using this relation, we can show that $n_3$ in \eq{n3example} satisfies the obstruction equation $\dd n_3 \={2} n_2\smile n_2$. However, the explicit expression for the bosonic phase factor $\nu_4$ is quite complicated, and we omit it in this paper for simplicity.

\subsubsection{Trivialization of $n_2$ to $(\Zf\times\mathtt{SmallGroup}(32,2))$-FSPT}

To trivialize $n_2$ in \eq{n2example}, let us first consider the central extension
\begin{align}\label{G'}
0 \rightarrow (A=\Z_2) \rightarrow G' \rightarrow (G=\Z_4\times\Z_4) \rightarrow 0,
\end{align}
which is specified by a 2-cocycle $\phi_2\in H^2(G,A)=H^2(\Z_4\times\Z_4,\Z_2)=(\Z_2)^3$. As discussed below \eq{n2}, we can choose a canonical $\phi_2$ to trivialize $n_2$ as
\begin{align}
\phi_2 (\mb{g,h}) = n_2 (\mb{g,h}) = \left(n_1^{(1)} n_1^{(2)}\right)(\mb{g,h}) = \mb g_1 \mb h_2\ (\mathrm{mod}\ 2),
\end{align}
for $\mb g=(\mb g_1,\mb g_2),\;\mb h=(\mb h_1,\mb h_2)\in G=\Z_4\times\Z_4$.
In the following, we denote elements in $G'=\Z_2\times_{\phi_2}(\Z_4)^2$ by $\mb g=(\mb g_0,\mb g_1,\mb g_2)$, where $\mb g_0\in A=\Z_2$ and $(\mb g_1,\mb g_2)\in G=\Z_4\times\Z_4$. Using the expression for $\phi_2$, the multiplication rule in $G'$ is given by
\begin{align}\nonumber
\mb g\times_{G'} \mb h 
&= (\mb g_0,\mb g_1,\mb g_2)\times_{G'} (\mb h_0,\mb h_1,\mb h_2)\\
&
= \left([\mb g_0+\mb h_0+\mb g_1\mb h_2]_2,[\mb g_1+\mb h_1]_4,[\mb g_2+\mb h_2]_4\right),
\end{align}
where $[x]_N$ denotes $x$ modulo $N$, as previously defined.

In this way, the larger group $G'=\Z_2\times_{\phi_2}(\Z_4)^2$ has a presentation
\begin{align}
\langle a,b,c\, |\, a^2=b^4=c^4=1, ab=ba, ac=ca, bc=acb \rangle,
\end{align}
%$G' = \langle a,b,c | a^2=b^4=c^4=1, ab=ba, ac=ca, bcb^{-1}c^{-1}=a \rangle$. 
where $a,b,c$ are generators of $A=\Z_2$ and $G=\Z_4\times\Z_4$ respectively.
The group $G'$ can be identified as
\begin{align}
G'=\Z_2\times_{\phi_2}(\Z_4)^2=\mathtt{SmallGroup}(32,2),
\end{align}
which is the second group of order 32 in the GAP small group library \cite{SmallGrp}.

When $n_2 \in H^2(G, \mathbb{Z}_2)$ is pulled back through the short exact sequence in \eq{G'}, it can be shown to be a coboundary. If we define the $\Z_2$-valued 1-cochain of $G'$ as
\begin{align}
m'_1(\mb g)=[\mb g_0]_2,\quad \forall \mb g\in G',
\end{align}
then its differential is
\begin{align}\nonumber
\dd m'_1(\mb{g},\mb{h})
&= m'_1(\mb {h}) - m'_1(\mb{g}\times_{G'}\mb{h}) + m'_1(\mb{g}) \\\nonumber
&
= [\mb h_0]_2 - [\mb g_0+\mb h_0+\mb g_1\mb h_2]_2+ [\mb g_0]_2\\
&
\={2} \mb g_1 \mb h_2.
\end{align}
This matches precisely with $n'_2(\mathbf{g}, \mathbf{h}) = \mathbf{g}_1 \mathbf{h}_2$, which is the pullback of $n_2 = n_1^{(1)} n_1^{(2)}$. Therefore, $n'_2 \in H^2(G', \mathbb{Z}_2)$ is indeed trivial, with the trivialization equation
\begin{align}
n'_2 \={2} \dd m'_1.
\end{align}

For the group $G'=\mathtt{SmallGroup}(32,2)=\Z_2\times_{\phi_2}(\Z_4)^2$, the 3-cocycle of \eq{N3} in $H^3(G',\Z_2)$ becomes
\begin{align}\label{N3'}\nonumber
N'_3(\mb g,\mb h,\mb k)
&\={2} n_3'(\mb{g},\mb{h},\mb{k}) + m'_1(\mb{g}) n_2'(\mb{h},\mb{k})\\\nonumber
&\={2} n_3[(\mb g_1,\mb g_2),(\mb h_1,\mb h_2),(\mb k_1,\mb k_2)] 
+ m_1'(\mb{g}) n_2[(\mb h_1,\mb h_2),(\mb k_1,\mb k_2)]\\
&\={2} \mb g_1 \mb h_1 \mb h_2 \mb k_2 + \frac{[\mb g_1]_4-[\mb g_1]_2}{2} \mb h_2 \mb k_2  + \mb g_0 \mb h_1 \mb k_2,
\end{align}
for $\mb g,\mb h,\mb k\in G'$. Here, the expression for $n_3 \in H^3(G, \mathbb{Z}_2)$ from Eq.~\eqref{n3example} is used to derive the last line. It can be shown that $N'_3$ is a nontrivial element in $H^3(G',\Z_2)=(\Z_2)^5$.

Thus, we have shown that for the $(\mathbb{Z}_2^f \times \mathbb{Z}_4 \times \mathbb{Z}_4)$-FSPT phase with the previously defined decoration data $(n_2, n_3, \nu_4)$, it is possible to construct a gapped interface between this phase and another FSPT phase with decoration data $(0, N_3', \mathcal{V}'_4)$, where no Majorana chain decorations are present. The symmetry group of the latter phase is $\mathbb{Z}_2^f \times G' = \mathbb{Z}_2^f \times \mathtt{SmallGroup}(32,2)$.

\begin{widetext}
\subsubsection{Trivialization of $n_3$ to $\mathtt{SmallGroup}(64,23)$-BSPT}
Now, in order to further trivialize $N'_3$, we introduce a second central extension
\begin{align}\label{sesG''}
1 \rightarrow (A'=\Z_2) \rightarrow G'' \overset{\pi'}{\rightarrow} [G'=\Z_2\times_{\phi_2}(\Z_4)^2] \rightarrow 1.
\end{align} 
We choose the extension 2-cocycle $\phi_2'\in H^2(G',A')$ to be
\begin{align}
\phi_2'(\mb g,\mb h)
=
\mb g_1\mb h_1\mb h_2+\frac{[\mb g_1]_4-[\mb g_1]_2}{2} \mb h_2+\mb g_0\mb h_1\ (\mathrm{mod}\ 2), \quad \forall \mb g,\mb h\in G'.
\end{align}
% \begin{align}
% \phi_2'(\mb g,\mb h)=\left[\mb g_1\mb h_1\mb h_2+\frac{[\mb g_1]_4-[\mb g_1]_2}{2} \mb h_2+\mb g_0\mb h_1\right]_2.
% \end{align}
One can verify that $\phi'_2$ is indeed a 2-cocycle in $H^2(G',\Z_2)$. With this choice, the multiplication rule in $G''$ is given by
\begin{align}\label{xG''}\nonumber
\ag{a}{g}\times_{G''}\ag{b}{h}
&=
\left(a+b+\phi_2'(\mb g,\mb h)\right)_{\mb g\times_{G'}\mb h}\\
&=
\left(\left[a+b+\mb g_1\mb h_1\mb h_2+\frac{[\mb g_1]_4-[\mb g_1]_2}{2} \mb h_2+\mb g_0\mb h_1\right]_2\right)_{([\mb g_0+\mb h_0+\mb g_1\mb h_2]_2,[\mb g_1+\mb h_1]_4,[\mb g_2+\mb h_2]_4)},
\end{align}
where $\ag{a}{g}$ denotes an element in $G''$ with $a\in A'=\Z_2$ and $\mb g=(\mb g_0,\mb g_1,\mb g_2)\in G'=\Z_2\times_{\phi_2}(\Z_4)^2$.

Using the multiplication rule \eq{xG''}, the group $G''$ can be presented as
\begin{align}
\langle a,b,c,d\, |\, a^2=b^2=c^4=d^4=1, ab=ba, ac=ca, ad=da,
bc=acb, bd=db,  
cd=dcb\rangle,
\end{align}
where $a$ and $b$ are the generators of $A' = \mathbb{Z}_2$ and $A = \mathbb{Z}_2$, respectively, and $c$ and $d$ are the generators of the two $\Z_4$ subgroups of $G = \mathbb{Z}_4 \times \mathbb{Z}_4$. The group $G''$ can be identified as
\begin{align}
G''=\Z_2\times_{\phi_2'}\left[\Z_2\times_{\phi_2}(\Z_4)^2\right] = \mathtt{SmallGroup}(64,23),
\end{align}
which is the 23rd group of order 64 in the GAP small group library \cite{SmallGrp}. Additional information about $G''=\mathtt{SmallGroup}(64,23)$ can be found, for example, in \refn{matydGroupNames}.

Our next goal is to show that $N''_3 = \pi'^\ast(N'_3)$ is trivialized, i.e., that it is a 3-coboundary in $H^3(G'', \mathbb{Z}_2)$. This can be demonstrated in two different ways.

The first approach to demonstrate that $N''_3$ is a coboundary is to use the LHSSS 
$E_2^{p,q}=H^p(G',H^q(A',\Z_2)) \Rightarrow H^{p+q}(G'',\Z_2)$
associated with the short exact sequence \eq{sesG''}. The cocycle $N''_3$, as the pullback of $N_3'$ for $G'$, is located at 
\begin{align}
    E_2^{3,0}=H^3(G',H^0(A',\Z_2))=H^3(G',\Z_2)
\end{align}
of the second page of the LHSSS. This cocycle becomes a coboundary if it lies in the image of the differential $d_2: E_2^{1,1}\rightarrow E_2^{3,0}$. Here, we can show that the domain of this $d_2$ is
\begin{align}
    E_2^{1,1}=H^1(G',H^1(A',\Z_2))=H^1(G',\Z_2)=(\Z_2)^2.
\end{align} 
We choose the term located at $E_2^{1,1}$ as
\begin{align}\label{F11}
F_{1,1}(a,\mb h) = a\mb h_2\ (\mathrm{mod}\ 2),\quad \forall a\in A',\;\mb h\in G'.
\end{align}
Using the results from Ref.~\onlinecite{wang2021domain}, the differential $d_2$ maps this $F_{1,1}\in E_2^{1,1}$ to a term in $E_2^{3,0}$ given by
\begin{align}\nonumber
d_2(F_{1,1})(\mb g,\mb h,\mb k) 
&= F_{1,1}(\phi_2'(\mb{g},\mb{h}),\mb{k})\\\nonumber
&=\phi_2'(\mb{g},\mb{h})\mb{k}_2\\
&=
\left(\mb g_1\mb h_1\mb h_2+\frac{[\mb g_1]_4-[\mb g_1]_2}{2} \mb h_2+\mb g_0\mb h_1\right)\mb k_2,\quad \forall \mb{g,h,k}\in G'.
\end{align}
This expression matches exactly with $N''_3 \in E_3^{3, 0} = H^3(G'', \mathbb{Z}_2)$ with the explicit form Eq.~\eqref{N3'}. Therefore, we see that $N''_3$ becomes trivial on the third page of the LHSSS, confirming that it is a coboundary in $H^3(G'',\Z_2)$.

The second approach to show that $N_3''$ is a coboundary is more explicit. Motivated by the definition of $F_{1,1}$ in \eq{F11} above, we define a 2-cochain on $G''$ as follows:
\begin{align}
m_2''(\ag{a}{g},\ag{b}{h})=a\mb h_2\ (\mathrm{mod}\ 2),\quad \forall a\in A',\;\mb h\in G'.
\end{align}
We can then calculate the differential of $m_2''$ by definition
\begin{align}\nonumber
\dd m_2''(\ag{a}{g},\ag{b}{h},\ag{c}{k}) 
&:= m_2''(\ag{b}{h},\ag{c}{k}) - m_2''(\ag{a}{g}\times_{G''}\ag{b}{h},\ag{c}{k}) + m_2''(\ag{a}{g},\ag{b}{h}\times_{G''}\ag{c}{k}) - m_2''(\ag{a}{g},\ag{b}{h}) \\\nonumber
&\={2}
b\mb k_2
-\left(a+b+\mb g_1\mb h_1\mb h_2+\frac{[\mb g_1]_4-[\mb g_1]_2}{2} \mb h_2+\mb g_0\mb h_1\right)\mb k_2
+a(\mb h_2+\mb k_2)
-a\mb h_2\\
&\={2}
\left(\mb g_1\mb h_1\mb h_2+\frac{[\mb g_1]_4-[\mb g_1]_2}{2} \mb h_2+\mb g_0\mb h_1\right)\mb k_2, \quad\forall \ag{a}{g},\ag{b}{h},\ag{c}{k}\in G'',
\end{align}
\end{widetext}
which is again $N_3'' \in H^3(G'', \mathbb{Z}_2)$ as the pullback of $N_3' \in H^3(G', \mathbb{Z}_2)$ in Eq.~\eqref{N3'}. Therefore, we conclude that 
\begin{align} N_3'' = \dd m_2'', 
\end{align} 
confirming that $N_3''$ is a coboundary for $G''$.

In this way, we have shown that for the $(\mathbb{Z}_2^f \times \mathbb{Z}_4 \times \mathbb{Z}_4)$-FSPT phase with nontrivial Majorana chain decoration and label $(n_2, n_3, \nu_4)$ as given above, there exists a gapped interface between this phase and a bosonic SPT phase with symmetry $G'' = \mathtt{SmallGroup}(64, 23)$.

\section{(3+1)D $(\mathbb Z_2^f \times_{\om_2} G)$-FSPT states with nontrivial extension symmetry group}
\label{sec:3Dbdy3}

In previous sections, we mainly focused on (3+1)D FSPT phases with a direct product symmetry group $G_f = \mathbb{Z}_2^f \times G$. This means that the central extension \eq{Gf} is trivial. However, if the extension is nontrivial, we have a nontrivial 2-cocycle $\omega_2 \in H^2(G, \mathbb{Z}_2)$. In this case, the total symmetry group becomes $G_f=\mathbb Z_2^f \times_{\om_2} G$, where $\omega_2$ specifies the nontrivial extension.

With the inclusion of the nontrivial $\omega_2$, we describe all the data characterizing a (3+1)D $G_f$-FSPT state as $(\omega_2; n_2, n_3, \nu_4)$ according to the general supercohomology theory. These data satisfy more complicated consistency equations compared to the trivial $\omega_2$ case \cite{wang2018construction}:
\begin{align}
\dd \om_2 &\={2} 0,\\\label{dn2om}
\dd n_2 &\={2} 0,\\\label{dn3om}
\dd n_3 &\={2} (\om_2+n_2)\smile n_2,\\\nonumber\label{dnu4om}
\dd \nu_4
&\={1}
\frac12\!\left[\om_2\smile n_3+\Sq^2(n_3)\right]
-\frac14(\om_2+n_2)\smile\beta_2  n_2
+\frac12 [\om_2\smile_1(\om_2+n_2)]\smile n_2\\
&\quad +\frac12 (\om_2\smile_2 n_2)\smile(n_2\smile_1 n_2)
+\frac12
(\om_2+n_2)(012)(\om_2+n_2)(023)n_2(235)n_2(345).
\end{align}
The expression on the right-hand side of \eq{dnu4om} for $\dd \nu_4$ is identical to the one given in \refn{wang2018construction}, and further details can be found in Appendix \ref{sec:O5om}. Note that when $\omega_2 = 0$, all of the above consistency equations reduce to the simpler equations, as described in \eqss{dn2}{dnu4_}.

The basic idea for handling the nontrivial $\omega_2$ case is to perform a pullback construction for this 2-cocycle $\omega_2$, which results in a trivial $\omega_2$ symmetry group, $G_f' = \Zf \times G'$. After this construction, the remaining problem reduces to the pullback trivialization construction discussed in Sec.~\ref{sec:3Dbdy2}. To summarize, the construction process is as follows:
\begin{align}\nonumber
(\om_2;n_2,n_3,\nu_4)
&\,\overset{\text{  pullback  }}{\sim} (\om'_2;n'_2,n'_3,\nu'_4) = (\dd \tau'_1;n'_2,n'_3,\nu'_4)\\\nonumber
&\overset{\text{coboundary}}{\sim} (0;N'_2,N'_3,\mathcal V'_4)\\\nonumber
&\,\overset{\text{  pullback  }}{\sim} (0;N''_2,N''_3,\mathcal V''_4) = (0;\dd m''_1,N''_3,\mathcal V''_4)\\\nonumber
&\overset{\text{coboundary}}{\sim} (0;0,\mathscr N''_3,\mathscr V''_4)\\\nonumber
&\,\overset{\text{  pullback  }}{\sim} (0;0,\mathscr N'''_3,\mathscr V'''_4) = (0;0,\dd m'''_2,\mathscr V'''_4)\\\nonumber
&\overset{\text{coboundary}}{\sim} (0;0,0,\mathbb V'''_4)\\\nonumber
&\,\overset{\text{  pullback  }}{\sim} (0;0,0,\mathbb V''''_4)=(0;0,0,\dd\mu''''_3)\\
&\overset{\text{coboundary}}{\sim} (0;0,0,0).
\end{align}
Here, ``$\sim$'' denotes the presence of a gapped interface between the two SPT phases on either side. The only new steps compared to the previous section are in the first and second lines, which will be the primary focus of this section.

Before going into the details of trivializing $\omega_2$ via the pullback construction, let us first discuss the physical interpretation of trivializing $\omega_2$. For a bosonic symmetry group $G$, a nontrivial $\omega_2$ implies symmetry fractionalization when the fermion parity symmetry is gauged, reducing the problem to that of an SET phase. This can be understood as endowing the anyons with fractional quantum numbers under $G$ (see also \refns{Aasen2021, Bulmash2021frac, Bulmash2021anomaly} for relevant discussions). A group element $g_f$ of $G_f$ takes the form $g_f = (n(g), g)$, where $(P_f)^n \in \mathbb{Z}_2^f$ and $g \in G$, with $n(g) = 0, 1$. The multiplication rule in $G_f$ is given by
\begin{align}\label{fmult}
    g_f\cdot h_f=(n(g),g)\cdot (n(h),h):= (n(g)+n(h)+\om_2(g,h),gh).
\end{align}
By modifying $\omega_2 \rightarrow \omega_2 + \dd \tau_1$, where $\tau_1 \in C^1(G, \mathbb{Z}_2 = \{0, 1\})$, we effectively implement a relabeling of $g_f$ in $G_f$ as
\begin{align}
    (0,g)\rightarrow (0,g)'=(\tau_1(g),g).
\end{align}
After relabeling, we obtain a new symmetry group $G_f'$ generated by $ (0, g)'$, which is isomorphic to $G_f' = \mathbb{Z}_2^f \times G$ without extension. Physically, this relabeling modifies the $G$-gauge flux excitation if the $G$ symmetry is gauged, attaching a fermion parity flux determined by $\tau_1(g)$. Although this modification redefines the $G$ flux by incorporating fermion parity, the classification of $G_f$-FSPT and $G_f'$-FSPT phases remains identical because the two symmetry groups are isomorphic. However, the corresponding FSPT states exhibit a nontrivial correspondence due to the 1-cochain $\tau_1$ twist.

\subsection{Pullback trivialization of $\om_2$}

In the following, we provide a detailed construction for trivializing $\omega_2$ via pullback construction. To begin with, consider the group extension
\begin{align}\label{extomega}
1 \rightarrow A \rightarrow G' \overset{\pi}{\rightarrow} G \rightarrow 1.
\end{align}
Suppose this extension trivializes the cocycle $\omega_2$ by pulling it back, i.e., $\omega_2' = \pi^\ast (\omega_2) \in H^2(G', \mathbb{Z}_2)$, as
\begin{align}
\om'_2 \={2} \dd \tau'_1.
\end{align}
However, there is a $\frac{1}{4}$-term in Eq.~(\ref{dnu4om}) that depends on $\omega_2'$, and it should take values in $\{0, 1\}$. To ensure that $\dd \tau_1'$ takes values in $\mathbb{Z}_2 = \{0, 1\}$, we adopt the (mod 4) trivialization equation for $\omega_2'$ as follows [see also \eq{om2triv2d}]:
\begin{align}\label{om2triv}
\om'_2 \={4} \dd \tau'_1-2\Sq^1(\tau_1'),
\end{align}
where the detailed derivation of this result can be found in Appendix~\ref{sec:triveqz}.

Since we have the equivalence class $[\omega_2'] = [0]$ in $H^2(G', \mathbb{Z}_2)$, we can define an FSPT phase with a trivial $\omega_2$ symmetry group, $G_f' = \mathbb{Z}_2^f \times G'$. To realize this phase, we apply FSLU or ``gauge transformations'' to the obstruction functions associated with the decoration data $n_2'$, $n_3'$, and $\nu_4'$, as follows:
\begin{align}
    (\mathcal O_3[\om'_2],\mathcal O_4[\om'_2;n_2'],\mathcal O_5[\om'_2;n_2',n_3'])
    \rightarrow
    (\Tilde{\mathcal O}_3[0],\Tilde{\mathcal O}_4[0;N_2'],\Tilde{\mathcal O}_5[0;N_2',N_3'])
\end{align}
where the transformations are applied to the obstruction functions on the right-hand side of Eqs.~(\ref{dn2om})-(\ref{dnu4om}). This adjustment ensures that the consistency equations for the respective cocycles have no dependence on $\omega_2'$, effectively trivializing its contribution to the FSPT phase.

\begin{widetext}
From \eqs{dn2om}{dn3om}, it is useful to define the transformed 2- and 3-cocycles as follows:
\begin{align}\label{gaugen2}
    N'_2 &\={2} n'_2,\\\label{gaugen3}
    N'_3 &\={2} n'_3+\tau'_1n'_2.
\end{align}
These transformed cocycles satisfy the desired consistency conditions:
\begin{align}\label{gaugen2woom2}
    \dd N'_2 &\={2} \Tilde{\mathcal O}_3[0] \={2}0,\\\label{gaugen3woom2}
    \dd N'_3 &\={2} \Tilde{\mathcal O}_4[0,N_2'] \={2} n'_2\smile n'_2.
\end{align}
This ensures that the new cocycles have trivial dependence on $\omega_2'$, aligning with the intended construction.

The FSLU or ``gauge transformation'' of the pulled-back obstruction function $\mathcal{O}_5$ is more complicated. Starting from equation \eq{dnu4om}, we have
\begin{align}\nonumber
    \mathcal O_5[\om_2';n'_2,n'_3] 
    &\={1} 
    \frac12\left[\om_2'\smile n_3'+\Sq^2(n_3')\right]
    -\frac14(\om_2'+n_2')\smile\beta_2  n_2'
    +\frac12 [\om_2'\smile_1(\om_2'+n_2')]\smile n_2'\\
    &\quad +\frac12 (\om_2'\smile_2 n_2')\smile(n_2'\smile_1 n_2')
    +\frac12(\om_2'+n_2')(012)(\om_2'+n_2')(023)n_2'(235)n_2'(345).
\end{align}
Next, we express $\om_2'$ in terms of $\tau_1'$ as in \eq{om2triv}, which transforms $\mathcal{O}_5$ into the following form:
\begin{align}\label{O5wom2}\nonumber
    \mathcal O_5[\tau_1';n'_2,n'_3] 
    &\={1} \frac12\dd\tau_1'\smile n_3'+\frac12\Sq^2(n_3')
    -\frac14(\dd\tau_1'+n_2')\smile\beta_2  n_2'
    +\frac12 \Sq^1(\tau_1')\smile (n_2'\smile_1 n_2')\\\nonumber
    &\quad +\frac12 (\dd\tau_1'\smile_1\dd\tau_1')\smile n_2'
    +\frac12 (\dd\tau_1'\smile_1n_2')\smile n_2'
    +\frac12 (\dd\tau_1'\smile_2 n_2')\smile(n_2'\smile_1 n_2')\\
    &\quad+\frac12(\dd\tau_1'+n_2')(012)(\dd\tau_1'+n_2')(023)n_2'(235)n_2'(345).
\end{align}
To obtain an FSPT with direct product symmetry group, we aim for an expression that depends only on the transformed 2- and 3-cocycles $N'_2$ and $N'_3$, as defined in \eq{gaugen2} and \eq{gaugen3}. That means, we should have
\begin{align}
    \Tilde{\mathcal O}_5[0;N'_2,N'_3] &\={1} \frac12\Sq^2(n'_3+\tau'_1n'_2)
    -\frac14 n_2'\smile\beta_2  n_2'
    +\frac12 n_2'(012) n_2'(023)n_2'(235)n_2'(345)\\\label{gaugeO5}\nonumber
    &\={1} \frac12\Sq^2(n'_3)+\frac12\Sq^2(\tau'_1n'_2)+\frac12 \dd n_3' \smile_3\dd(\tau'_1n'_2)
    +\frac12 \dd[n_3' \smile_2 (\tau'_1n'_2)]+\frac12 \dd[\dd n_3'\smile_3(\tau'_1n'_2)]\\
    &\quad
    -\frac14 n_2'\smile\beta_2  n_2'
    +\frac12 n_2'(012) n_2'(023)n_2'(235)n_2'(345).
\end{align}
In the first line of \eq{gaugeO5}, we have applied the action of the Steenrod square $\Sq^2$ to the sum $n'_3 + \tau'_1 n'_2$. Subtracting \eq{gaugeO5} from \eq{O5wom2} yields the following expression for $B_5:=\tilde{\mathcal O}_5-\mathcal O_5$:
\begin{align}\nonumber
    B_5 &\={1} 
    \dd\!\left[ \frac12\tau_1'n_3'+\frac12 n_3'\smile_2 (\tau_1'n_2')
    +\frac12 (\tau_1'n_2')\smile_3(n_2')^2+\frac12 \Sq^1(\tau_1')n_2'+\frac14 \tau_1'\smile \beta_2 n_2'\right]\\\nonumber
    &\quad
    +\frac12 \left[\Sq^1(\tau_1')+\dd \tau_1'\smile_2n_2'\right]\smile (n_2'\smile_1 n_2')
    +\frac12\left[\tau_1'(\dd\tau_1'+n_2')
    +(\dd\tau_1'\smile_1 n_2')\right]n_2'\\\nonumber
    &\quad
    +\frac12(\dd\tau_1' n_2')\smile_3(n_2')^2
    +\frac12 (\tau_1'n_2')\smile_1 (\tau_1'n_2')
    +\frac12 (\dd\tau_1'n_2')\smile_2(\tau_1'n_2')\\
    &\quad+\frac12\left[\dd\tau_1'(012)\dd\tau_1'(023)+\dd\tau_1'(012)n_2'(023)+n_2'(012)\dd\tau_1'(023)\right]n_2'(235)n_2'(345).
\end{align}
Upon numerical verification, it is found that the last three lines of this expression also form a coboundary. This gives us:
\begin{align}\nonumber
    B_5 &\={1} 
    \dd\!\left[ \frac12\tau_1'n_3'+\frac12 n_3'\smile_2 (\tau_1'n_2')
    +\frac12 (\tau_1'n_2')\smile_3(n_2')^2+\frac12 \Sq^1(\tau_1')n_2'+\frac14 \tau_1'\smile \beta_2 n_2'\right]\\
    &\quad+ \dd\!\left[\frac12 n'_2(134)n'_2(123)\tau'_1(01)+\frac12[1+n'_2(124)]n'_2(234)\tau'_1(01)\tau'_1(02)\right].
\end{align}
By extracting the 4-cochain $C_4$ from the coboundary $B_5=\dd C_4$, we define a new 4-cochain
\begin{align}\nonumber
    \mathcal V'_4 &\={1} \nu'_4+\frac12\tau'_1n'_3
    +\frac12n'_3\smile_2(\tau'_1n'_2)
    +\frac12 (\tau'_1n'_2)\smile_3 (n_2')^2
    +\frac12\Sq^1(\tau'_1)n'_2\\
    &\quad
    +\frac14 \tau'_1\smile\beta_2 n'_2
    + \frac12 n'_2(134)n'_2(123)\tau'_1(01)+\frac12[1+n'_2(124)]n'_2(234)\tau'_1(01)\tau'_1(02),
\end{align}
which obeys the twisted cocycle equation
\begin{align}\label{gaugenu4woom2}
    \dd \mathcal V'_4 &\={1} \frac12 \Sq^2(n'_3)-\frac14 n'_2\smile \beta_2  n'_2
    +\frac12(n'_2)^2(02345)(n'_2)^2(01235).
\end{align}
with trivial extension $\omega_2=0$.
\end{widetext}

The set of gauge-transformed consistency equations in Eqs.~(\ref{gaugen2woom2}), (\ref{gaugen3woom2}), and (\ref{gaugenu4woom2}) takes the same form as those in \eqss{dn2}{dnu4_} with the trivial extension $\omega_2=0$. This allows us to apply the pullback trivialization procedure, as discussed in the previous section for the case of a trivial extension 2-cocycle $\omega_2$. Thus, we can complete the construction of the gapped boundary for the $(\mathbb Z_2^f \times_{\om_2} G)$-FSPT phase.

In summary, by performing the symmetry extension \eq{extomega} and the subsequent FSLU (``gauge transformations''), we can construct a gapped interface between two phases: the (3+1)D $(\Zf\times_{\omega_2} G)$-FSPT with the extended symmetry group $G_f=\Zf \times_{\omega_2} G$, and the (3+1)D $(\Zf\times G')$-FSPT with the direct-product symmetric group $G_f' = \Zf \times G'$. Furthermore, the latter direct product symmetry FSPT can be further trivialized to a product state using the pullback construction, as detailed in the previous section.

\section{Conclusions and Discussions}
\label{sec:conclusion}

In this paper, we employ the pullback construction to systematically construct gapped interfaces between FSPT phases with different decoration layers by extending the symmetry groups to larger ones. The decoration cocycles are trivialized layer by layer through the pullback. After each trivialization step, we apply an FSLU transformation or ``gauge transformation'' to ensure that the relevant coboundaries are zero at the cochain level. These transformations, in turn, twist other decoration layers with nontrivial cochains. By collapsing all intermediate layers, we ultimately achieve a gapped boundary for the original FSPT states. While this paper primarily focuses on (2+1)D and (3+1)D cases, the general procedure is applicable to arbitrary dimensions within the framework of general group supercohomology theory.

There are several directions for future exploration. The first is the explicit construction of lattice models. While we have briefly explained the lattice construction for the gapped interfaces, it would be valuable to explicitly write down the lattice Hamiltonians and wavefunctions for systems with an interface between two FSPTs with different symmetry groups.

Another immediate question concerns the relationship between our pullback construction method and the general framework of SET phases on the (2+1)D boundary \cite{etingof_fusion_2009, BBCW2019,Bulmash2021anomaly}. It remains unclear how to connect our construction with the understanding of how the original $G$ symmetry acts on the boundary topological orders, particularly in terms of anyon permutation and symmetry fractionalization on a modular tensor category. Moreover, higher-dimensional generalizations remain an open problem, as the classification of SETs in those dimensions is not yet complete.

It would also be interesting to investigate FSPT phases with point/space group symmetries \cite{HuangPRB2017,PhysRevB.99.115116,pointgroup,PhysRevResearch.4.033081,PhysRevB.101.165129} and continuous Lie group symmetries \cite{U1SET,PhysRevB.107.075112,IEI}. To date, systematic constructions for gapped boundaries via pullback trivialization in these settings are not well understood.

\section*{Acknowledgments}
This project is supported by the National Natural Science Foundation of China (Grant No. 12274250). K.L. thanks the Chinese University of Hong Kong for hospitality during the ``Tensor Network States: Algorithms and Applications (TNSAA) 2024'' workshop, where part of this work was completed and presented.

%\bibliography{boundary.bib}

\begin{thebibliography}{191}%
\makeatletter
\providecommand \@ifxundefined [1]{%
 \@ifx{#1\undefined}
}%
\providecommand \@ifnum [1]{%
 \ifnum #1\expandafter \@firstoftwo
 \else \expandafter \@secondoftwo
 \fi
}%
\providecommand \@ifx [1]{%
 \ifx #1\expandafter \@firstoftwo
 \else \expandafter \@secondoftwo
 \fi
}%
\providecommand \natexlab [1]{#1}%
\providecommand \enquote  [1]{``#1''}%
\providecommand \bibnamefont  [1]{#1}%
\providecommand \bibfnamefont [1]{#1}%
\providecommand \citenamefont [1]{#1}%
\providecommand \href@noop [0]{\@secondoftwo}%
\providecommand \href [0]{\begingroup \@sanitize@url \@href}%
\providecommand \@href[1]{\@@startlink{#1}\@@href}%
\providecommand \@@href[1]{\endgroup#1\@@endlink}%
\providecommand \@sanitize@url [0]{\catcode `\\12\catcode `\$12\catcode `\&12\catcode `\#12\catcode `\^12\catcode `\_12\catcode `\%12\relax}%
\providecommand \@@startlink[1]{}%
\providecommand \@@endlink[0]{}%
\providecommand \url  [0]{\begingroup\@sanitize@url \@url }%
\providecommand \@url [1]{\endgroup\@href {#1}{\urlprefix }}%
\providecommand \urlprefix  [0]{URL }%
\providecommand \Eprint [0]{\href }%
\providecommand \doibase [0]{http://dx.doi.org/}%
\providecommand \selectlanguage [0]{\@gobble}%
\providecommand \bibinfo  [0]{\@secondoftwo}%
\providecommand \bibfield  [0]{\@secondoftwo}%
\providecommand \translation [1]{[#1]}%
\providecommand \BibitemOpen [0]{}%
\providecommand \bibitemStop [0]{}%
\providecommand \bibitemNoStop [0]{.\EOS\space}%
\providecommand \EOS [0]{\spacefactor3000\relax}%
\providecommand \BibitemShut  [1]{\csname bibitem#1\endcsname}%
\let\auto@bib@innerbib\@empty
%</preamble>
\bibitem [{\citenamefont {Wen}(1989)}]{wen1989}%
  \BibitemOpen
  \bibfield  {author} {\bibinfo {author} {\bibfnamefont {X.~G.}\ \bibnamefont {Wen}},\ }\href {\doibase 10.1103/PhysRevB.40.7387} {\bibfield  {journal} {\bibinfo  {journal} {Phys. Rev. B}\ }\textbf {\bibinfo {volume} {40}},\ \bibinfo {pages} {7387} (\bibinfo {year} {1989})}\BibitemShut {NoStop}%
\bibitem [{\citenamefont {Wen}(1990)}]{Wen1990to}%
  \BibitemOpen
  \bibfield  {author} {\bibinfo {author} {\bibfnamefont {X.-G.}\ \bibnamefont {Wen}},\ }\href {https://api.semanticscholar.org/CorpusID:120441771} {\bibfield  {journal} {\bibinfo  {journal} {International Journal of Modern Physics B}\ }\textbf {\bibinfo {volume} {4}},\ \bibinfo {pages} {239} (\bibinfo {year} {1990})}\BibitemShut {NoStop}%
\bibitem [{\citenamefont {Levin}\ and\ \citenamefont {Wen}(2006)}]{Levin2006to}%
  \BibitemOpen
  \bibfield  {author} {\bibinfo {author} {\bibfnamefont {M.}~\bibnamefont {Levin}}\ and\ \bibinfo {author} {\bibfnamefont {X.-G.}\ \bibnamefont {Wen}},\ }\href {\doibase 10.1103/PhysRevLett.96.110405} {\bibfield  {journal} {\bibinfo  {journal} {Phys. Rev. Lett.}\ }\textbf {\bibinfo {volume} {96}},\ \bibinfo {pages} {110405} (\bibinfo {year} {2006})}\BibitemShut {NoStop}%
\bibitem [{\citenamefont {Chen}\ \emph {et~al.}(2010)\citenamefont {Chen}, \citenamefont {Gu},\ and\ \citenamefont {Wen}}]{Chen2010lu}%
  \BibitemOpen
  \bibfield  {author} {\bibinfo {author} {\bibfnamefont {X.}~\bibnamefont {Chen}}, \bibinfo {author} {\bibfnamefont {Z.~C.}\ \bibnamefont {Gu}}, \ and\ \bibinfo {author} {\bibfnamefont {X.~G.}\ \bibnamefont {Wen}},\ }\href {\doibase 10.1103/PhysRevB.82.155138} {\bibfield  {journal} {\bibinfo  {journal} {Phys. Rev. B}\ }\textbf {\bibinfo {volume} {82}},\ \bibinfo {pages} {155138} (\bibinfo {year} {2010})}\BibitemShut {NoStop}%
\bibitem [{\citenamefont {Gu}\ and\ \citenamefont {Wen}(2009)}]{Gu2009spt}%
  \BibitemOpen
  \bibfield  {author} {\bibinfo {author} {\bibfnamefont {Z.-C.}\ \bibnamefont {Gu}}\ and\ \bibinfo {author} {\bibfnamefont {X.-G.}\ \bibnamefont {Wen}},\ }\href {\doibase 10.1103/PhysRevB.80.155131} {\bibfield  {journal} {\bibinfo  {journal} {Phys. Rev. B}\ }\textbf {\bibinfo {volume} {80}},\ \bibinfo {pages} {155131} (\bibinfo {year} {2009})}\BibitemShut {NoStop}%
\bibitem [{\citenamefont {Chiu}\ \emph {et~al.}(2016)\citenamefont {Chiu}, \citenamefont {Teo}, \citenamefont {Schnyder},\ and\ \citenamefont {Ryu}}]{CTSR}%
  \BibitemOpen
  \bibfield  {author} {\bibinfo {author} {\bibfnamefont {C.-K.}\ \bibnamefont {Chiu}}, \bibinfo {author} {\bibfnamefont {J.~C.~Y.}\ \bibnamefont {Teo}}, \bibinfo {author} {\bibfnamefont {A.~P.}\ \bibnamefont {Schnyder}}, \ and\ \bibinfo {author} {\bibfnamefont {S.}~\bibnamefont {Ryu}},\ }\href {\doibase 10.1103/RevModPhys.88.035005} {\bibfield  {journal} {\bibinfo  {journal} {Rev. Mod. Phys.}\ }\textbf {\bibinfo {volume} {88}},\ \bibinfo {pages} {035005} (\bibinfo {year} {2016})}\BibitemShut {NoStop}%
\bibitem [{\citenamefont {Ludwig}(2015)}]{Ludwig}%
  \BibitemOpen
  \bibfield  {author} {\bibinfo {author} {\bibfnamefont {A.~W.~W.}\ \bibnamefont {Ludwig}},\ }\href {\doibase 10.1088/0031-8949/2015/t168/014001} {\bibfield  {journal} {\bibinfo  {journal} {Physica Scripta}\ }\textbf {\bibinfo {volume} {T168}},\ \bibinfo {pages} {014001} (\bibinfo {year} {2015})}\BibitemShut {NoStop}%
\bibitem [{\citenamefont {Senthil}(2015)}]{Senthil2015}%
  \BibitemOpen
  \bibfield  {author} {\bibinfo {author} {\bibfnamefont {T.}~\bibnamefont {Senthil}},\ }\href {\doibase 10.1146/annurev-conmatphys-031214-014740} {\bibfield  {journal} {\bibinfo  {journal} {Annual Review of Condensed Matter Physics}\ }\textbf {\bibinfo {volume} {6}},\ \bibinfo {pages} {299–324} (\bibinfo {year} {2015})}\BibitemShut {NoStop}%
\bibitem [{\citenamefont {Wen}(2017{\natexlab{a}})}]{wen2017review}%
  \BibitemOpen
  \bibfield  {author} {\bibinfo {author} {\bibfnamefont {X.-G.}\ \bibnamefont {Wen}},\ }\href {\doibase 10.1103/revmodphys.89.041004} {\bibfield  {journal} {\bibinfo  {journal} {Reviews of Modern Physics}\ }\textbf {\bibinfo {volume} {89}},\ \bibinfo {pages} {041004} (\bibinfo {year} {2017}{\natexlab{a}})}\BibitemShut {NoStop}%
\bibitem [{\citenamefont {Hasan}\ and\ \citenamefont {Kane}(2010)}]{Hasan_2010}%
  \BibitemOpen
  \bibfield  {author} {\bibinfo {author} {\bibfnamefont {M.~Z.}\ \bibnamefont {Hasan}}\ and\ \bibinfo {author} {\bibfnamefont {C.~L.}\ \bibnamefont {Kane}},\ }\href {\doibase 10.1103/revmodphys.82.3045} {\bibfield  {journal} {\bibinfo  {journal} {Reviews of Modern Physics}\ }\textbf {\bibinfo {volume} {82}},\ \bibinfo {pages} {3045–3067} (\bibinfo {year} {2010})}\BibitemShut {NoStop}%
\bibitem [{\citenamefont {Qi}\ and\ \citenamefont {Zhang}(2011)}]{Qi_2011}%
  \BibitemOpen
  \bibfield  {author} {\bibinfo {author} {\bibfnamefont {X.-L.}\ \bibnamefont {Qi}}\ and\ \bibinfo {author} {\bibfnamefont {S.-C.}\ \bibnamefont {Zhang}},\ }\href {\doibase 10.1103/revmodphys.83.1057} {\bibfield  {journal} {\bibinfo  {journal} {Reviews of Modern Physics}\ }\textbf {\bibinfo {volume} {83}},\ \bibinfo {pages} {1057–1110} (\bibinfo {year} {2011})}\BibitemShut {NoStop}%
\bibitem [{\citenamefont {Hasan}\ and\ \citenamefont {Moore}(2011)}]{Hasan_2011}%
  \BibitemOpen
  \bibfield  {author} {\bibinfo {author} {\bibfnamefont {M.~Z.}\ \bibnamefont {Hasan}}\ and\ \bibinfo {author} {\bibfnamefont {J.~E.}\ \bibnamefont {Moore}},\ }\href {\doibase 10.1146/annurev-conmatphys-062910-140432} {\bibfield  {journal} {\bibinfo  {journal} {Annual Review of Condensed Matter Physics}\ }\textbf {\bibinfo {volume} {2}},\ \bibinfo {pages} {55–78} (\bibinfo {year} {2011})}\BibitemShut {NoStop}%
\bibitem [{\citenamefont {Kitaev}(2009)}]{Kitaev:2009mg}%
  \BibitemOpen
  \bibfield  {author} {\bibinfo {author} {\bibfnamefont {A.}~\bibnamefont {Kitaev}},\ }\href {\doibase 10.1063/1.3149495} {\bibfield  {journal} {\bibinfo  {journal} {AIP Conf. Proc.}\ }\textbf {\bibinfo {volume} {1134}},\ \bibinfo {pages} {22} (\bibinfo {year} {2009})}\BibitemShut {NoStop}%
\bibitem [{\citenamefont {Ryu}\ \emph {et~al.}(2010)\citenamefont {Ryu}, \citenamefont {Schnyder}, \citenamefont {Furusaki},\ and\ \citenamefont {Ludwig}}]{Ryu_2010}%
  \BibitemOpen
  \bibfield  {author} {\bibinfo {author} {\bibfnamefont {S.}~\bibnamefont {Ryu}}, \bibinfo {author} {\bibfnamefont {A.~P.}\ \bibnamefont {Schnyder}}, \bibinfo {author} {\bibfnamefont {A.}~\bibnamefont {Furusaki}}, \ and\ \bibinfo {author} {\bibfnamefont {A.~W.~W.}\ \bibnamefont {Ludwig}},\ }\href {\doibase 10.1088/1367-2630/12/6/065010} {\bibfield  {journal} {\bibinfo  {journal} {New Journal of Physics}\ }\textbf {\bibinfo {volume} {12}},\ \bibinfo {pages} {065010} (\bibinfo {year} {2010})}\BibitemShut {NoStop}%
\bibitem [{\citenamefont {Wen}(2012)}]{Wen2011free}%
  \BibitemOpen
  \bibfield  {author} {\bibinfo {author} {\bibfnamefont {X.-G.}\ \bibnamefont {Wen}},\ }\href {\doibase 10.1103/PhysRevB.85.085103} {\bibfield  {journal} {\bibinfo  {journal} {Phys. Rev. B}\ }\textbf {\bibinfo {volume} {85}},\ \bibinfo {pages} {085103} (\bibinfo {year} {2012})}\BibitemShut {NoStop}%
\bibitem [{\citenamefont {Kruthoff}\ \emph {et~al.}(2017)\citenamefont {Kruthoff}, \citenamefont {de~Boer}, \citenamefont {van Wezel}, \citenamefont {Kane},\ and\ \citenamefont {Slager}}]{KdvKS2017}%
  \BibitemOpen
  \bibfield  {author} {\bibinfo {author} {\bibfnamefont {J.}~\bibnamefont {Kruthoff}}, \bibinfo {author} {\bibfnamefont {J.}~\bibnamefont {de~Boer}}, \bibinfo {author} {\bibfnamefont {J.}~\bibnamefont {van Wezel}}, \bibinfo {author} {\bibfnamefont {C.~L.}\ \bibnamefont {Kane}}, \ and\ \bibinfo {author} {\bibfnamefont {R.-J.}\ \bibnamefont {Slager}},\ }\href {\doibase 10.1103/PhysRevX.7.041069} {\bibfield  {journal} {\bibinfo  {journal} {Phys. Rev. X}\ }\textbf {\bibinfo {volume} {7}},\ \bibinfo {pages} {041069} (\bibinfo {year} {2017})}\BibitemShut {NoStop}%
\bibitem [{\citenamefont {Chen}\ \emph {et~al.}(2012)\citenamefont {Chen}, \citenamefont {Gu}, \citenamefont {Liu},\ and\ \citenamefont {Wen}}]{chen2013bspt}%
  \BibitemOpen
  \bibfield  {author} {\bibinfo {author} {\bibfnamefont {X.}~\bibnamefont {Chen}}, \bibinfo {author} {\bibfnamefont {Z.-C.}\ \bibnamefont {Gu}}, \bibinfo {author} {\bibfnamefont {Z.-X.}\ \bibnamefont {Liu}}, \ and\ \bibinfo {author} {\bibfnamefont {X.-G.}\ \bibnamefont {Wen}},\ }\href {\doibase 10.1126/science.1227224} {\bibfield  {journal} {\bibinfo  {journal} {Science}\ }\textbf {\bibinfo {volume} {338}},\ \bibinfo {pages} {1604} (\bibinfo {year} {2012})}\BibitemShut {NoStop}%
\bibitem [{\citenamefont {Chen}\ \emph {et~al.}(2013)\citenamefont {Chen}, \citenamefont {Gu}, \citenamefont {Liu},\ and\ \citenamefont {Wen}}]{Chen:2011pg}%
  \BibitemOpen
  \bibfield  {author} {\bibinfo {author} {\bibfnamefont {X.}~\bibnamefont {Chen}}, \bibinfo {author} {\bibfnamefont {Z.-C.}\ \bibnamefont {Gu}}, \bibinfo {author} {\bibfnamefont {Z.-X.}\ \bibnamefont {Liu}}, \ and\ \bibinfo {author} {\bibfnamefont {X.-G.}\ \bibnamefont {Wen}},\ }\href {\doibase 10.1103/PhysRevB.87.155114} {\bibfield  {journal} {\bibinfo  {journal} {Phys. Rev. B}\ }\textbf {\bibinfo {volume} {87}},\ \bibinfo {pages} {155114} (\bibinfo {year} {2013})}\BibitemShut {NoStop}%
\bibitem [{\citenamefont {Wen}(2015)}]{Wen2014nlsm}%
  \BibitemOpen
  \bibfield  {author} {\bibinfo {author} {\bibfnamefont {X.-G.}\ \bibnamefont {Wen}},\ }\href {\doibase 10.1103/PhysRevB.91.205101} {\bibfield  {journal} {\bibinfo  {journal} {Phys. Rev. B}\ }\textbf {\bibinfo {volume} {91}},\ \bibinfo {pages} {205101} (\bibinfo {year} {2015})}\BibitemShut {NoStop}%
\bibitem [{\citenamefont {Kapustin}(2014)}]{Kapustin:2014tfa}%
  \BibitemOpen
  \bibfield  {author} {\bibinfo {author} {\bibfnamefont {A.}~\bibnamefont {Kapustin}},\ }\href@noop {} {\  (\bibinfo {year} {2014})},\ \Eprint {http://arxiv.org/abs/1403.1467} {arXiv:1403.1467 [cond-mat.str-el]} \BibitemShut {NoStop}%
\bibitem [{\citenamefont {Kapustin}\ \emph {et~al.}(2015)\citenamefont {Kapustin}, \citenamefont {Thorngren}, \citenamefont {Turzillo},\ and\ \citenamefont {Wang}}]{KTTW2015}%
  \BibitemOpen
  \bibfield  {author} {\bibinfo {author} {\bibfnamefont {A.}~\bibnamefont {Kapustin}}, \bibinfo {author} {\bibfnamefont {R.}~\bibnamefont {Thorngren}}, \bibinfo {author} {\bibfnamefont {A.}~\bibnamefont {Turzillo}}, \ and\ \bibinfo {author} {\bibfnamefont {Z.}~\bibnamefont {Wang}},\ }\href {\doibase 10.1007/JHEP12(2015)052} {\bibfield  {journal} {\bibinfo  {journal} {JHEP}\ }\textbf {\bibinfo {volume} {12}},\ \bibinfo {pages} {052} (\bibinfo {year} {2015})}\BibitemShut {NoStop}%
\bibitem [{\citenamefont {Yonekura}(2019)}]{Yonekura2018}%
  \BibitemOpen
  \bibfield  {author} {\bibinfo {author} {\bibfnamefont {K.}~\bibnamefont {Yonekura}},\ }\href {\doibase 10.1007/s00220-019-03439-y} {\bibfield  {journal} {\bibinfo  {journal} {Commun. Math. Phys.}\ }\textbf {\bibinfo {volume} {368}},\ \bibinfo {pages} {1121} (\bibinfo {year} {2019})}\BibitemShut {NoStop}%
\bibitem [{\citenamefont {Wang}\ \emph {et~al.}(2015{\natexlab{a}})\citenamefont {Wang}, \citenamefont {Gu},\ and\ \citenamefont {Wen}}]{WGW2014}%
  \BibitemOpen
  \bibfield  {author} {\bibinfo {author} {\bibfnamefont {J.~C.}\ \bibnamefont {Wang}}, \bibinfo {author} {\bibfnamefont {Z.-C.}\ \bibnamefont {Gu}}, \ and\ \bibinfo {author} {\bibfnamefont {X.-G.}\ \bibnamefont {Wen}},\ }\href {\doibase 10.1103/PhysRevLett.114.031601} {\bibfield  {journal} {\bibinfo  {journal} {Phys. Rev. Lett.}\ }\textbf {\bibinfo {volume} {114}},\ \bibinfo {pages} {031601} (\bibinfo {year} {2015}{\natexlab{a}})}\BibitemShut {NoStop}%
\bibitem [{\citenamefont {Putrov}\ \emph {et~al.}(2017)\citenamefont {Putrov}, \citenamefont {Wang},\ and\ \citenamefont {Yau}}]{PWY2016}%
  \BibitemOpen
  \bibfield  {author} {\bibinfo {author} {\bibfnamefont {P.}~\bibnamefont {Putrov}}, \bibinfo {author} {\bibfnamefont {J.}~\bibnamefont {Wang}}, \ and\ \bibinfo {author} {\bibfnamefont {S.-T.}\ \bibnamefont {Yau}},\ }\href {\doibase 10.1016/j.aop.2017.06.019} {\bibfield  {journal} {\bibinfo  {journal} {Annals Phys.}\ }\textbf {\bibinfo {volume} {384}},\ \bibinfo {pages} {254} (\bibinfo {year} {2017})}\BibitemShut {NoStop}%
\bibitem [{\citenamefont {Gaiotto}\ and\ \citenamefont {Kapustin}(2016)}]{GK2016}%
  \BibitemOpen
  \bibfield  {author} {\bibinfo {author} {\bibfnamefont {D.}~\bibnamefont {Gaiotto}}\ and\ \bibinfo {author} {\bibfnamefont {A.}~\bibnamefont {Kapustin}},\ }\href {\doibase 10.1142/s0217751x16450445} {\bibfield  {journal} {\bibinfo  {journal} {International Journal of Modern Physics A}\ }\textbf {\bibinfo {volume} {31}},\ \bibinfo {pages} {1645044} (\bibinfo {year} {2016})}\BibitemShut {NoStop}%
\bibitem [{\citenamefont {Freed}\ and\ \citenamefont {Hopkins}(2021)}]{FH2016}%
  \BibitemOpen
  \bibfield  {author} {\bibinfo {author} {\bibfnamefont {D.~S.}\ \bibnamefont {Freed}}\ and\ \bibinfo {author} {\bibfnamefont {M.~J.}\ \bibnamefont {Hopkins}},\ }\href {\doibase 10.2140/gt.2021.25.1165} {\bibfield  {journal} {\bibinfo  {journal} {Geom. Topol.}\ }\textbf {\bibinfo {volume} {25}},\ \bibinfo {pages} {1165} (\bibinfo {year} {2021})}\BibitemShut {NoStop}%
\bibitem [{\citenamefont {Bhardwaj}\ \emph {et~al.}(2017)\citenamefont {Bhardwaj}, \citenamefont {Gaiotto},\ and\ \citenamefont {Kapustin}}]{BGK2016}%
  \BibitemOpen
  \bibfield  {author} {\bibinfo {author} {\bibfnamefont {L.}~\bibnamefont {Bhardwaj}}, \bibinfo {author} {\bibfnamefont {D.}~\bibnamefont {Gaiotto}}, \ and\ \bibinfo {author} {\bibfnamefont {A.}~\bibnamefont {Kapustin}},\ }\href {\doibase 10.1007/JHEP04(2017)096} {\bibfield  {journal} {\bibinfo  {journal} {JHEP}\ }\textbf {\bibinfo {volume} {04}},\ \bibinfo {pages} {096} (\bibinfo {year} {2017})}\BibitemShut {NoStop}%
\bibitem [{\citenamefont {Tiwari}\ \emph {et~al.}(2018)\citenamefont {Tiwari}, \citenamefont {Chen}, \citenamefont {Shiozaki},\ and\ \citenamefont {Ryu}}]{TCSR2017}%
  \BibitemOpen
  \bibfield  {author} {\bibinfo {author} {\bibfnamefont {A.}~\bibnamefont {Tiwari}}, \bibinfo {author} {\bibfnamefont {X.}~\bibnamefont {Chen}}, \bibinfo {author} {\bibfnamefont {K.}~\bibnamefont {Shiozaki}}, \ and\ \bibinfo {author} {\bibfnamefont {S.}~\bibnamefont {Ryu}},\ }\href {\doibase 10.1103/PhysRevB.97.245133} {\bibfield  {journal} {\bibinfo  {journal} {Phys. Rev. B}\ }\textbf {\bibinfo {volume} {97}},\ \bibinfo {pages} {245133} (\bibinfo {year} {2018})}\BibitemShut {NoStop}%
\bibitem [{\citenamefont {Wang}\ \emph {et~al.}(2019{\natexlab{a}})\citenamefont {Wang}, \citenamefont {Cheng}, \citenamefont {Wang},\ and\ \citenamefont {Gu}}]{WCWG2018}%
  \BibitemOpen
  \bibfield  {author} {\bibinfo {author} {\bibfnamefont {Q.-R.}\ \bibnamefont {Wang}}, \bibinfo {author} {\bibfnamefont {M.}~\bibnamefont {Cheng}}, \bibinfo {author} {\bibfnamefont {C.}~\bibnamefont {Wang}}, \ and\ \bibinfo {author} {\bibfnamefont {Z.-C.}\ \bibnamefont {Gu}},\ }\href {\doibase 10.1103/PhysRevB.99.235137} {\bibfield  {journal} {\bibinfo  {journal} {Phys. Rev. B}\ }\textbf {\bibinfo {volume} {99}},\ \bibinfo {pages} {235137} (\bibinfo {year} {2019}{\natexlab{a}})}\BibitemShut {NoStop}%
\bibitem [{\citenamefont {Wang}\ \emph {et~al.}(2018{\natexlab{a}})\citenamefont {Wang}, \citenamefont {Ohmori}, \citenamefont {Putrov}, \citenamefont {Zheng}, \citenamefont {Wan}, \citenamefont {Guo}, \citenamefont {Lin}, \citenamefont {Gao},\ and\ \citenamefont {Yau}}]{WOPZ2018}%
  \BibitemOpen
  \bibfield  {author} {\bibinfo {author} {\bibfnamefont {J.}~\bibnamefont {Wang}}, \bibinfo {author} {\bibfnamefont {K.}~\bibnamefont {Ohmori}}, \bibinfo {author} {\bibfnamefont {P.}~\bibnamefont {Putrov}}, \bibinfo {author} {\bibfnamefont {Y.}~\bibnamefont {Zheng}}, \bibinfo {author} {\bibfnamefont {Z.}~\bibnamefont {Wan}}, \bibinfo {author} {\bibfnamefont {M.}~\bibnamefont {Guo}}, \bibinfo {author} {\bibfnamefont {H.}~\bibnamefont {Lin}}, \bibinfo {author} {\bibfnamefont {P.}~\bibnamefont {Gao}}, \ and\ \bibinfo {author} {\bibfnamefont {S.-T.}\ \bibnamefont {Yau}},\ }\href {\doibase 10.1093/ptep/pty051} {\bibfield  {journal} {\bibinfo  {journal} {PTEP}\ }\textbf {\bibinfo {volume} {2018}},\ \bibinfo {pages} {053A01} (\bibinfo {year} {2018}{\natexlab{a}})}\BibitemShut {NoStop}%
\bibitem [{\citenamefont {Ning}\ \emph {et~al.}(2021{\natexlab{a}})\citenamefont {Ning}, \citenamefont {Wang}, \citenamefont {Wang},\ and\ \citenamefont {Gu}}]{NWWG2021}%
  \BibitemOpen
  \bibfield  {author} {\bibinfo {author} {\bibfnamefont {S.-Q.}\ \bibnamefont {Ning}}, \bibinfo {author} {\bibfnamefont {C.}~\bibnamefont {Wang}}, \bibinfo {author} {\bibfnamefont {Q.-R.}\ \bibnamefont {Wang}}, \ and\ \bibinfo {author} {\bibfnamefont {Z.-C.}\ \bibnamefont {Gu}},\ }\href {\doibase 10.1103/physrevb.104.075151} {\bibfield  {journal} {\bibinfo  {journal} {Physical Review B}\ }\textbf {\bibinfo {volume} {104}},\ \bibinfo {pages} {075151} (\bibinfo {year} {2021}{\natexlab{a}})}\BibitemShut {NoStop}%
\bibitem [{\citenamefont {Gaiotto}\ and\ \citenamefont {Johnson-Freyd}(2019)}]{GJ2017}%
  \BibitemOpen
  \bibfield  {author} {\bibinfo {author} {\bibfnamefont {D.}~\bibnamefont {Gaiotto}}\ and\ \bibinfo {author} {\bibfnamefont {T.}~\bibnamefont {Johnson-Freyd}},\ }\href {\doibase 10.1007/JHEP05(2019)007} {\bibfield  {journal} {\bibinfo  {journal} {JHEP}\ }\textbf {\bibinfo {volume} {05}},\ \bibinfo {pages} {007} (\bibinfo {year} {2019})}\BibitemShut {NoStop}%
\bibitem [{\citenamefont {Chen}\ \emph {et~al.}(2014{\natexlab{a}})\citenamefont {Chen}, \citenamefont {Lu},\ and\ \citenamefont {Vishwanath}}]{Chen2014dw}%
  \BibitemOpen
  \bibfield  {author} {\bibinfo {author} {\bibfnamefont {X.}~\bibnamefont {Chen}}, \bibinfo {author} {\bibfnamefont {Y.-M.}\ \bibnamefont {Lu}}, \ and\ \bibinfo {author} {\bibfnamefont {A.}~\bibnamefont {Vishwanath}},\ }\href {\doibase 10.1038/ncomms4507} {\bibfield  {journal} {\bibinfo  {journal} {Nature Communications}\ }\textbf {\bibinfo {volume} {5}},\ \bibinfo {pages} {3507} (\bibinfo {year} {2014}{\natexlab{a}})}\BibitemShut {NoStop}%
\bibitem [{\citenamefont {Wang}\ \emph {et~al.}(2018{\natexlab{b}})\citenamefont {Wang}, \citenamefont {Ning},\ and\ \citenamefont {Chen}}]{WNC2018}%
  \BibitemOpen
  \bibfield  {author} {\bibinfo {author} {\bibfnamefont {Z.}~\bibnamefont {Wang}}, \bibinfo {author} {\bibfnamefont {S.-Q.}\ \bibnamefont {Ning}}, \ and\ \bibinfo {author} {\bibfnamefont {X.}~\bibnamefont {Chen}},\ }\href {\doibase 10.1103/PhysRevB.98.094502} {\bibfield  {journal} {\bibinfo  {journal} {Phys. Rev. B}\ }\textbf {\bibinfo {volume} {98}},\ \bibinfo {pages} {094502} (\bibinfo {year} {2018}{\natexlab{b}})}\BibitemShut {NoStop}%
\bibitem [{\citenamefont {Yang}\ \emph {et~al.}(2018)\citenamefont {Yang}, \citenamefont {Jiang}, \citenamefont {Vishwanath},\ and\ \citenamefont {Ran}}]{Yang_LSMSPT}%
  \BibitemOpen
  \bibfield  {author} {\bibinfo {author} {\bibfnamefont {X.}~\bibnamefont {Yang}}, \bibinfo {author} {\bibfnamefont {S.}~\bibnamefont {Jiang}}, \bibinfo {author} {\bibfnamefont {A.}~\bibnamefont {Vishwanath}}, \ and\ \bibinfo {author} {\bibfnamefont {Y.}~\bibnamefont {Ran}},\ }\href {\doibase 10.1103/PhysRevB.98.125120} {\bibfield  {journal} {\bibinfo  {journal} {Phys. Rev. B}\ }\textbf {\bibinfo {volume} {98}},\ \bibinfo {pages} {125120} (\bibinfo {year} {2018})}\BibitemShut {NoStop}%
\bibitem [{\citenamefont {Huang}\ \emph {et~al.}(2017)\citenamefont {Huang}, \citenamefont {Song}, \citenamefont {Huang},\ and\ \citenamefont {Hermele}}]{HuangPRB2017}%
  \BibitemOpen
  \bibfield  {author} {\bibinfo {author} {\bibfnamefont {S.-J.}\ \bibnamefont {Huang}}, \bibinfo {author} {\bibfnamefont {H.}~\bibnamefont {Song}}, \bibinfo {author} {\bibfnamefont {Y.-P.}\ \bibnamefont {Huang}}, \ and\ \bibinfo {author} {\bibfnamefont {M.}~\bibnamefont {Hermele}},\ }\href {\doibase 10.1103/PhysRevB.96.205106} {\bibfield  {journal} {\bibinfo  {journal} {Phys. Rev. B}\ }\textbf {\bibinfo {volume} {96}},\ \bibinfo {pages} {205106} (\bibinfo {year} {2017})}\BibitemShut {NoStop}%
\bibitem [{\citenamefont {Kou}\ and\ \citenamefont {Wen}(2009)}]{Kou:2009vea}%
  \BibitemOpen
  \bibfield  {author} {\bibinfo {author} {\bibfnamefont {S.~P.}\ \bibnamefont {Kou}}\ and\ \bibinfo {author} {\bibfnamefont {X.-G.}\ \bibnamefont {Wen}},\ }\href {\doibase 10.1103/PhysRevB.80.224406} {\bibfield  {journal} {\bibinfo  {journal} {Phys. Rev. B}\ }\textbf {\bibinfo {volume} {80}},\ \bibinfo {pages} {224406} (\bibinfo {year} {2009})}\BibitemShut {NoStop}%
\bibitem [{\citenamefont {Chen}\ \emph {et~al.}(2011{\natexlab{a}})\citenamefont {Chen}, \citenamefont {Gu},\ and\ \citenamefont {Wen}}]{Chen:2010zpc}%
  \BibitemOpen
  \bibfield  {author} {\bibinfo {author} {\bibfnamefont {X.}~\bibnamefont {Chen}}, \bibinfo {author} {\bibfnamefont {Z.-C.}\ \bibnamefont {Gu}}, \ and\ \bibinfo {author} {\bibfnamefont {X.-G.}\ \bibnamefont {Wen}},\ }\href {\doibase 10.1103/PhysRevB.83.035107} {\bibfield  {journal} {\bibinfo  {journal} {Phys. Rev. B}\ }\textbf {\bibinfo {volume} {83}},\ \bibinfo {pages} {035107} (\bibinfo {year} {2011}{\natexlab{a}})}\BibitemShut {NoStop}%
\bibitem [{\citenamefont {Chen}\ \emph {et~al.}(2011{\natexlab{b}})\citenamefont {Chen}, \citenamefont {Gu},\ and\ \citenamefont {Wen}}]{CGW2011}%
  \BibitemOpen
  \bibfield  {author} {\bibinfo {author} {\bibfnamefont {X.}~\bibnamefont {Chen}}, \bibinfo {author} {\bibfnamefont {Z.-C.}\ \bibnamefont {Gu}}, \ and\ \bibinfo {author} {\bibfnamefont {X.-G.}\ \bibnamefont {Wen}},\ }\href {\doibase 10.1103/physrevb.84.235128} {\bibfield  {journal} {\bibinfo  {journal} {Physical Review B}\ }\textbf {\bibinfo {volume} {84}},\ \bibinfo {pages} {235128} (\bibinfo {year} {2011}{\natexlab{b}})}\BibitemShut {NoStop}%
\bibitem [{\citenamefont {Schuch}\ \emph {et~al.}(2011)\citenamefont {Schuch}, \citenamefont {Pérez-García},\ and\ \citenamefont {Cirac}}]{SPC2011}%
  \BibitemOpen
  \bibfield  {author} {\bibinfo {author} {\bibfnamefont {N.}~\bibnamefont {Schuch}}, \bibinfo {author} {\bibfnamefont {D.}~\bibnamefont {Pérez-García}}, \ and\ \bibinfo {author} {\bibfnamefont {I.}~\bibnamefont {Cirac}},\ }\href {\doibase 10.1103/physrevb.84.165139} {\bibfield  {journal} {\bibinfo  {journal} {Physical Review B}\ }\textbf {\bibinfo {volume} {84}},\ \bibinfo {pages} {165139} (\bibinfo {year} {2011})}\BibitemShut {NoStop}%
\bibitem [{\citenamefont {Fidkowski}\ and\ \citenamefont {Kitaev}(2011)}]{kitaev2011}%
  \BibitemOpen
  \bibfield  {author} {\bibinfo {author} {\bibfnamefont {L.}~\bibnamefont {Fidkowski}}\ and\ \bibinfo {author} {\bibfnamefont {A.}~\bibnamefont {Kitaev}},\ }\href {\doibase 10.1103/PhysRevB.83.075103} {\bibfield  {journal} {\bibinfo  {journal} {Phys. Rev. B}\ }\textbf {\bibinfo {volume} {83}},\ \bibinfo {pages} {075103} (\bibinfo {year} {2011})}\BibitemShut {NoStop}%
\bibitem [{\citenamefont {Lam}(2024)}]{Lam2023}%
  \BibitemOpen
  \bibfield  {author} {\bibinfo {author} {\bibfnamefont {H.~T.}\ \bibnamefont {Lam}},\ }\href {\doibase 10.1103/PhysRevB.109.115142} {\bibfield  {journal} {\bibinfo  {journal} {Phys. Rev. B}\ }\textbf {\bibinfo {volume} {109}},\ \bibinfo {pages} {115142} (\bibinfo {year} {2024})}\BibitemShut {NoStop}%
\bibitem [{\citenamefont {Lan}\ \emph {et~al.}(2017{\natexlab{a}})\citenamefont {Lan}, \citenamefont {Kong},\ and\ \citenamefont {Wen}}]{Lan:2016rcq}%
  \BibitemOpen
  \bibfield  {author} {\bibinfo {author} {\bibfnamefont {T.}~\bibnamefont {Lan}}, \bibinfo {author} {\bibfnamefont {L.}~\bibnamefont {Kong}}, \ and\ \bibinfo {author} {\bibfnamefont {X.-G.}\ \bibnamefont {Wen}},\ }\href {\doibase 10.1007/s00220-016-2748-y} {\bibfield  {journal} {\bibinfo  {journal} {Commun. Math. Phys.}\ }\textbf {\bibinfo {volume} {351}},\ \bibinfo {pages} {709} (\bibinfo {year} {2017}{\natexlab{a}})}\BibitemShut {NoStop}%
\bibitem [{\citenamefont {Lan}\ \emph {et~al.}(2017{\natexlab{b}})\citenamefont {Lan}, \citenamefont {Kong},\ and\ \citenamefont {Wen}}]{Lan:2016seg}%
  \BibitemOpen
  \bibfield  {author} {\bibinfo {author} {\bibfnamefont {T.}~\bibnamefont {Lan}}, \bibinfo {author} {\bibfnamefont {L.}~\bibnamefont {Kong}}, \ and\ \bibinfo {author} {\bibfnamefont {X.-G.}\ \bibnamefont {Wen}},\ }\href {\doibase 10.1103/PhysRevB.95.235140} {\bibfield  {journal} {\bibinfo  {journal} {Phys. Rev. B}\ }\textbf {\bibinfo {volume} {95}},\ \bibinfo {pages} {235140} (\bibinfo {year} {2017}{\natexlab{b}})}\BibitemShut {NoStop}%
\bibitem [{\citenamefont {Kong}\ \emph {et~al.}(2020{\natexlab{a}})\citenamefont {Kong}, \citenamefont {Lan}, \citenamefont {Wen}, \citenamefont {Zhang},\ and\ \citenamefont {Zheng}}]{Kong:2020cie}%
  \BibitemOpen
  \bibfield  {author} {\bibinfo {author} {\bibfnamefont {L.}~\bibnamefont {Kong}}, \bibinfo {author} {\bibfnamefont {T.}~\bibnamefont {Lan}}, \bibinfo {author} {\bibfnamefont {X.-G.}\ \bibnamefont {Wen}}, \bibinfo {author} {\bibfnamefont {Z.-H.}\ \bibnamefont {Zhang}}, \ and\ \bibinfo {author} {\bibfnamefont {H.}~\bibnamefont {Zheng}},\ }\href {\doibase 10.1103/PhysRevResearch.2.043086} {\bibfield  {journal} {\bibinfo  {journal} {Phys. Rev. Res.}\ }\textbf {\bibinfo {volume} {2}},\ \bibinfo {pages} {043086} (\bibinfo {year} {2020}{\natexlab{a}})}\BibitemShut {NoStop}%
\bibitem [{\citenamefont {Kong}\ \emph {et~al.}(2022)\citenamefont {Kong}, \citenamefont {Wen},\ and\ \citenamefont {Zheng}}]{Kong:2021equ}%
  \BibitemOpen
  \bibfield  {author} {\bibinfo {author} {\bibfnamefont {L.}~\bibnamefont {Kong}}, \bibinfo {author} {\bibfnamefont {X.-G.}\ \bibnamefont {Wen}}, \ and\ \bibinfo {author} {\bibfnamefont {H.}~\bibnamefont {Zheng}},\ }\href {\doibase 10.1007/JHEP03(2022)022} {\bibfield  {journal} {\bibinfo  {journal} {JHEP}\ }\textbf {\bibinfo {volume} {03}},\ \bibinfo {pages} {022} (\bibinfo {year} {2022})}\BibitemShut {NoStop}%
\bibitem [{\citenamefont {Levin}\ and\ \citenamefont {Gu}(2012)}]{Levin2012spt}%
  \BibitemOpen
  \bibfield  {author} {\bibinfo {author} {\bibfnamefont {M.}~\bibnamefont {Levin}}\ and\ \bibinfo {author} {\bibfnamefont {Z.-C.}\ \bibnamefont {Gu}},\ }\href {\doibase 10.1103/PhysRevB.86.115109} {\bibfield  {journal} {\bibinfo  {journal} {Phys. Rev. B}\ }\textbf {\bibinfo {volume} {86}},\ \bibinfo {pages} {115109} (\bibinfo {year} {2012})}\BibitemShut {NoStop}%
\bibitem [{\citenamefont {Wang}\ and\ \citenamefont {Levin}(2014)}]{WL2014}%
  \BibitemOpen
  \bibfield  {author} {\bibinfo {author} {\bibfnamefont {C.}~\bibnamefont {Wang}}\ and\ \bibinfo {author} {\bibfnamefont {M.}~\bibnamefont {Levin}},\ }\href {\doibase 10.1103/PhysRevLett.113.080403} {\bibfield  {journal} {\bibinfo  {journal} {Phys. Rev. Lett.}\ }\textbf {\bibinfo {volume} {113}},\ \bibinfo {pages} {080403} (\bibinfo {year} {2014})}\BibitemShut {NoStop}%
\bibitem [{\citenamefont {Jiang}\ \emph {et~al.}(2014)\citenamefont {Jiang}, \citenamefont {Mesaros},\ and\ \citenamefont {Ran}}]{JMR2014}%
  \BibitemOpen
  \bibfield  {author} {\bibinfo {author} {\bibfnamefont {S.}~\bibnamefont {Jiang}}, \bibinfo {author} {\bibfnamefont {A.}~\bibnamefont {Mesaros}}, \ and\ \bibinfo {author} {\bibfnamefont {Y.}~\bibnamefont {Ran}},\ }\href {\doibase 10.1103/PhysRevX.4.031048} {\bibfield  {journal} {\bibinfo  {journal} {Phys. Rev. X}\ }\textbf {\bibinfo {volume} {4}},\ \bibinfo {pages} {031048} (\bibinfo {year} {2014})}\BibitemShut {NoStop}%
\bibitem [{\citenamefont {Lin}\ and\ \citenamefont {Levin}(2015)}]{LL2015}%
  \BibitemOpen
  \bibfield  {author} {\bibinfo {author} {\bibfnamefont {C.-H.}\ \bibnamefont {Lin}}\ and\ \bibinfo {author} {\bibfnamefont {M.}~\bibnamefont {Levin}},\ }\href {\doibase 10.1103/PhysRevB.92.035115} {\bibfield  {journal} {\bibinfo  {journal} {Phys. Rev. B}\ }\textbf {\bibinfo {volume} {92}},\ \bibinfo {pages} {035115} (\bibinfo {year} {2015})}\BibitemShut {NoStop}%
\bibitem [{\citenamefont {Wang}(2016)}]{wang2016bs}%
  \BibitemOpen
  \bibfield  {author} {\bibinfo {author} {\bibfnamefont {C.}~\bibnamefont {Wang}},\ }\href {\doibase 10.1103/PhysRevB.94.085130} {\bibfield  {journal} {\bibinfo  {journal} {Phys. Rev. B}\ }\textbf {\bibinfo {volume} {94}},\ \bibinfo {pages} {085130} (\bibinfo {year} {2016})}\BibitemShut {NoStop}%
\bibitem [{\citenamefont {Cheng}\ \emph {et~al.}(2018{\natexlab{a}})\citenamefont {Cheng}, \citenamefont {Tantivasadakarn},\ and\ \citenamefont {Wang}}]{CTW2018}%
  \BibitemOpen
  \bibfield  {author} {\bibinfo {author} {\bibfnamefont {M.}~\bibnamefont {Cheng}}, \bibinfo {author} {\bibfnamefont {N.}~\bibnamefont {Tantivasadakarn}}, \ and\ \bibinfo {author} {\bibfnamefont {C.}~\bibnamefont {Wang}},\ }\href {\doibase 10.1103/PhysRevX.8.011054} {\bibfield  {journal} {\bibinfo  {journal} {Phys. Rev. X}\ }\textbf {\bibinfo {volume} {8}},\ \bibinfo {pages} {011054} (\bibinfo {year} {2018}{\natexlab{a}})}\BibitemShut {NoStop}%
\bibitem [{\citenamefont {Zhou}\ \emph {et~al.}(2021)\citenamefont {Zhou}, \citenamefont {Wang}, \citenamefont {Wang},\ and\ \citenamefont {Gu}}]{ZWWG2019}%
  \BibitemOpen
  \bibfield  {author} {\bibinfo {author} {\bibfnamefont {J.-R.}\ \bibnamefont {Zhou}}, \bibinfo {author} {\bibfnamefont {Q.-R.}\ \bibnamefont {Wang}}, \bibinfo {author} {\bibfnamefont {C.}~\bibnamefont {Wang}}, \ and\ \bibinfo {author} {\bibfnamefont {Z.-C.}\ \bibnamefont {Gu}},\ }\href {\doibase 10.1038/s41467-021-23309-3} {\bibfield  {journal} {\bibinfo  {journal} {Nature Commun.}\ }\textbf {\bibinfo {volume} {12}},\ \bibinfo {pages} {3191} (\bibinfo {year} {2021})}\BibitemShut {NoStop}%
\bibitem [{\citenamefont {Zhang}\ \emph {et~al.}(2023{\natexlab{a}})\citenamefont {Zhang}, \citenamefont {Wang},\ and\ \citenamefont {Ye}}]{Non-Abelian}%
  \BibitemOpen
  \bibfield  {author} {\bibinfo {author} {\bibfnamefont {Z.-F.}\ \bibnamefont {Zhang}}, \bibinfo {author} {\bibfnamefont {Q.-R.}\ \bibnamefont {Wang}}, \ and\ \bibinfo {author} {\bibfnamefont {P.}~\bibnamefont {Ye}},\ }\href {\doibase 10.1103/PhysRevB.107.165117} {\bibfield  {journal} {\bibinfo  {journal} {Phys. Rev. B}\ }\textbf {\bibinfo {volume} {107}},\ \bibinfo {pages} {165117} (\bibinfo {year} {2023}{\natexlab{a}})}\BibitemShut {NoStop}%
\bibitem [{\citenamefont {Zhang}\ \emph {et~al.}(2023{\natexlab{b}})\citenamefont {Zhang}, \citenamefont {Wang},\ and\ \citenamefont {Ye}}]{fbraiding}%
  \BibitemOpen
  \bibfield  {author} {\bibinfo {author} {\bibfnamefont {Z.-F.}\ \bibnamefont {Zhang}}, \bibinfo {author} {\bibfnamefont {Q.-R.}\ \bibnamefont {Wang}}, \ and\ \bibinfo {author} {\bibfnamefont {P.}~\bibnamefont {Ye}},\ }\href {\doibase 10.1103/PhysRevResearch.5.043111} {\bibfield  {journal} {\bibinfo  {journal} {Phys. Rev. Res.}\ }\textbf {\bibinfo {volume} {5}},\ \bibinfo {pages} {043111} (\bibinfo {year} {2023}{\natexlab{b}})}\BibitemShut {NoStop}%
\bibitem [{\citenamefont {Hai}\ \emph {et~al.}(2023)\citenamefont {Hai}, \citenamefont {Zhang}, \citenamefont {Zheng}, \citenamefont {Kong}, \citenamefont {Wu},\ and\ \citenamefont {Yu}}]{Hai:2023osv}%
  \BibitemOpen
  \bibfield  {author} {\bibinfo {author} {\bibfnamefont {Y.-J.}\ \bibnamefont {Hai}}, \bibinfo {author} {\bibfnamefont {Z.}~\bibnamefont {Zhang}}, \bibinfo {author} {\bibfnamefont {H.}~\bibnamefont {Zheng}}, \bibinfo {author} {\bibfnamefont {L.}~\bibnamefont {Kong}}, \bibinfo {author} {\bibfnamefont {J.}~\bibnamefont {Wu}}, \ and\ \bibinfo {author} {\bibfnamefont {D.}~\bibnamefont {Yu}},\ }\href {\doibase 10.1093/nsr/nwac264} {\bibfield  {journal} {\bibinfo  {journal} {Natl. Sci. Rev.}\ }\textbf {\bibinfo {volume} {10}},\ \bibinfo {pages} {nwac264} (\bibinfo {year} {2023})}\BibitemShut {NoStop}%
\bibitem [{\citenamefont {Ye}\ and\ \citenamefont {Wang}(2013)}]{YW2013}%
  \BibitemOpen
  \bibfield  {author} {\bibinfo {author} {\bibfnamefont {P.}~\bibnamefont {Ye}}\ and\ \bibinfo {author} {\bibfnamefont {J.}~\bibnamefont {Wang}},\ }\href {\doibase 10.1103/PhysRevB.88.235109} {\bibfield  {journal} {\bibinfo  {journal} {Phys. Rev. B}\ }\textbf {\bibinfo {volume} {88}},\ \bibinfo {pages} {235109} (\bibinfo {year} {2013})}\BibitemShut {NoStop}%
\bibitem [{\citenamefont {Cheng}\ and\ \citenamefont {Gu}(2014)}]{cg2014}%
  \BibitemOpen
  \bibfield  {author} {\bibinfo {author} {\bibfnamefont {M.}~\bibnamefont {Cheng}}\ and\ \bibinfo {author} {\bibfnamefont {Z.-C.}\ \bibnamefont {Gu}},\ }\href {\doibase 10.1103/PhysRevLett.112.141602} {\bibfield  {journal} {\bibinfo  {journal} {Phys. Rev. Lett.}\ }\textbf {\bibinfo {volume} {112}},\ \bibinfo {pages} {141602} (\bibinfo {year} {2014})}\BibitemShut {NoStop}%
\bibitem [{\citenamefont {Fidkowski}\ and\ \citenamefont {Kitaev}(2010)}]{kitaev2010free}%
  \BibitemOpen
  \bibfield  {author} {\bibinfo {author} {\bibfnamefont {L.}~\bibnamefont {Fidkowski}}\ and\ \bibinfo {author} {\bibfnamefont {A.}~\bibnamefont {Kitaev}},\ }\href {\doibase 10.1103/PhysRevB.81.134509} {\bibfield  {journal} {\bibinfo  {journal} {Phys. Rev. B}\ }\textbf {\bibinfo {volume} {81}},\ \bibinfo {pages} {134509} (\bibinfo {year} {2010})}\BibitemShut {NoStop}%
\bibitem [{\citenamefont {Hung}\ and\ \citenamefont {Wen}(2012)}]{Hung:2012dx}%
  \BibitemOpen
  \bibfield  {author} {\bibinfo {author} {\bibfnamefont {L.-Y.}\ \bibnamefont {Hung}}\ and\ \bibinfo {author} {\bibfnamefont {X.-G.}\ \bibnamefont {Wen}},\ }\href@noop {} {\  (\bibinfo {year} {2012})},\ \Eprint {http://arxiv.org/abs/1211.2767} {arXiv:1211.2767 [cond-mat.str-el]} \BibitemShut {NoStop}%
\bibitem [{\citenamefont {Wen}(2014)}]{Wen:2013ue}%
  \BibitemOpen
  \bibfield  {author} {\bibinfo {author} {\bibfnamefont {X.-G.}\ \bibnamefont {Wen}},\ }\href {\doibase 10.1103/PhysRevB.89.035147} {\bibfield  {journal} {\bibinfo  {journal} {Phys. Rev. B}\ }\textbf {\bibinfo {volume} {89}},\ \bibinfo {pages} {035147} (\bibinfo {year} {2014})}\BibitemShut {NoStop}%
\bibitem [{\citenamefont {Hung}\ and\ \citenamefont {Wen}(2014)}]{Hung:2013cda}%
  \BibitemOpen
  \bibfield  {author} {\bibinfo {author} {\bibfnamefont {L.-Y.}\ \bibnamefont {Hung}}\ and\ \bibinfo {author} {\bibfnamefont {X.-G.}\ \bibnamefont {Wen}},\ }\href {\doibase 10.1103/PhysRevB.89.075121} {\bibfield  {journal} {\bibinfo  {journal} {Phys. Rev. B}\ }\textbf {\bibinfo {volume} {89}},\ \bibinfo {pages} {075121} (\bibinfo {year} {2014})}\BibitemShut {NoStop}%
\bibitem [{\citenamefont {Qi}(2013)}]{Qi2013}%
  \BibitemOpen
  \bibfield  {author} {\bibinfo {author} {\bibfnamefont {X.-L.}\ \bibnamefont {Qi}},\ }\href {\doibase 10.1088/1367-2630/15/6/065002} {\bibfield  {journal} {\bibinfo  {journal} {New Journal of Physics}\ }\textbf {\bibinfo {volume} {15}},\ \bibinfo {pages} {065002} (\bibinfo {year} {2013})}\BibitemShut {NoStop}%
\bibitem [{\citenamefont {Santos}\ and\ \citenamefont {Wang}(2014)}]{Santos:2013uda}%
  \BibitemOpen
  \bibfield  {author} {\bibinfo {author} {\bibfnamefont {L.~H.}\ \bibnamefont {Santos}}\ and\ \bibinfo {author} {\bibfnamefont {J.}~\bibnamefont {Wang}},\ }\href {\doibase 10.1103/PhysRevB.89.195122} {\bibfield  {journal} {\bibinfo  {journal} {Phys. Rev. B}\ }\textbf {\bibinfo {volume} {89}},\ \bibinfo {pages} {195122} (\bibinfo {year} {2014})}\BibitemShut {NoStop}%
\bibitem [{\citenamefont {Freed}(2014)}]{Freed2014}%
  \BibitemOpen
  \bibfield  {author} {\bibinfo {author} {\bibfnamefont {D.~S.}\ \bibnamefont {Freed}},\ }\href@noop {} {\  (\bibinfo {year} {2014})},\ \Eprint {http://arxiv.org/abs/1406.7278} {arXiv:1406.7278 [cond-mat.str-el]} \BibitemShut {NoStop}%
\bibitem [{\citenamefont {Gu}\ \emph {et~al.}(2016)\citenamefont {Gu}, \citenamefont {Wang},\ and\ \citenamefont {Wen}}]{Gu:2015lfa}%
  \BibitemOpen
  \bibfield  {author} {\bibinfo {author} {\bibfnamefont {Z.-C.}\ \bibnamefont {Gu}}, \bibinfo {author} {\bibfnamefont {J.~C.}\ \bibnamefont {Wang}}, \ and\ \bibinfo {author} {\bibfnamefont {X.-G.}\ \bibnamefont {Wen}},\ }\href {\doibase 10.1103/PhysRevB.93.115136} {\bibfield  {journal} {\bibinfo  {journal} {Phys. Rev. B}\ }\textbf {\bibinfo {volume} {93}},\ \bibinfo {pages} {115136} (\bibinfo {year} {2016})}\BibitemShut {NoStop}%
\bibitem [{\citenamefont {Wang}\ and\ \citenamefont {Levin}(2015)}]{WangLevin2015}%
  \BibitemOpen
  \bibfield  {author} {\bibinfo {author} {\bibfnamefont {C.}~\bibnamefont {Wang}}\ and\ \bibinfo {author} {\bibfnamefont {M.}~\bibnamefont {Levin}},\ }\href {\doibase 10.1103/physrevb.91.165119} {\bibfield  {journal} {\bibinfo  {journal} {Physical Review B}\ }\textbf {\bibinfo {volume} {91}},\ \bibinfo {pages} {165119} (\bibinfo {year} {2015})}\BibitemShut {NoStop}%
\bibitem [{\citenamefont {Tantivasadakarn}(2017)}]{Tantivasadakarn_2017}%
  \BibitemOpen
  \bibfield  {author} {\bibinfo {author} {\bibfnamefont {N.}~\bibnamefont {Tantivasadakarn}},\ }\href {\doibase 10.1103/PhysRevB.96.195101} {\bibfield  {journal} {\bibinfo  {journal} {Phys. Rev. B}\ }\textbf {\bibinfo {volume} {96}},\ \bibinfo {pages} {195101} (\bibinfo {year} {2017})}\BibitemShut {NoStop}%
\bibitem [{\citenamefont {Shiozaki}\ \emph {et~al.}(2018)\citenamefont {Shiozaki}, \citenamefont {Shapourian}, \citenamefont {Gomi},\ and\ \citenamefont {Ryu}}]{SSGR2017}%
  \BibitemOpen
  \bibfield  {author} {\bibinfo {author} {\bibfnamefont {K.}~\bibnamefont {Shiozaki}}, \bibinfo {author} {\bibfnamefont {H.}~\bibnamefont {Shapourian}}, \bibinfo {author} {\bibfnamefont {K.}~\bibnamefont {Gomi}}, \ and\ \bibinfo {author} {\bibfnamefont {S.}~\bibnamefont {Ryu}},\ }\href {\doibase 10.1103/PhysRevB.98.035151} {\bibfield  {journal} {\bibinfo  {journal} {Phys. Rev. B}\ }\textbf {\bibinfo {volume} {98}},\ \bibinfo {pages} {035151} (\bibinfo {year} {2018})}\BibitemShut {NoStop}%
\bibitem [{\citenamefont {Chen}\ \emph {et~al.}(2019)\citenamefont {Chen}, \citenamefont {Kapustin}, \citenamefont {Turzillo},\ and\ \citenamefont {You}}]{CKTY2018}%
  \BibitemOpen
  \bibfield  {author} {\bibinfo {author} {\bibfnamefont {Y.-A.}\ \bibnamefont {Chen}}, \bibinfo {author} {\bibfnamefont {A.}~\bibnamefont {Kapustin}}, \bibinfo {author} {\bibfnamefont {A.}~\bibnamefont {Turzillo}}, \ and\ \bibinfo {author} {\bibfnamefont {M.}~\bibnamefont {You}},\ }\href {\doibase 10.1103/PhysRevB.100.195128} {\bibfield  {journal} {\bibinfo  {journal} {Phys. Rev. B}\ }\textbf {\bibinfo {volume} {100}},\ \bibinfo {pages} {195128} (\bibinfo {year} {2019})}\BibitemShut {NoStop}%
\bibitem [{\citenamefont {Inamura}\ and\ \citenamefont {Wen}(2023)}]{Inamura:2023ldn}%
  \BibitemOpen
  \bibfield  {author} {\bibinfo {author} {\bibfnamefont {K.}~\bibnamefont {Inamura}}\ and\ \bibinfo {author} {\bibfnamefont {X.-G.}\ \bibnamefont {Wen}},\ }\href@noop {} {\  (\bibinfo {year} {2023})},\ \Eprint {http://arxiv.org/abs/2310.05790} {arXiv:2310.05790 [cond-mat.str-el]} \BibitemShut {NoStop}%
\bibitem [{\citenamefont {Kobayashi}\ \emph {et~al.}(2024)\citenamefont {Kobayashi}, \citenamefont {Zhang}, \citenamefont {Wang},\ and\ \citenamefont {Barkeshli}}]{Kobayashi:2024bts}%
  \BibitemOpen
  \bibfield  {author} {\bibinfo {author} {\bibfnamefont {R.}~\bibnamefont {Kobayashi}}, \bibinfo {author} {\bibfnamefont {Y.}~\bibnamefont {Zhang}}, \bibinfo {author} {\bibfnamefont {Y.-Q.}\ \bibnamefont {Wang}}, \ and\ \bibinfo {author} {\bibfnamefont {M.}~\bibnamefont {Barkeshli}},\ }\href@noop {} {\  (\bibinfo {year} {2024})},\ \Eprint {http://arxiv.org/abs/2403.18887} {arXiv:2403.18887 [cond-mat.str-el]} \BibitemShut {NoStop}%
\bibitem [{\citenamefont {Wen}(2013)}]{Wen2013a}%
  \BibitemOpen
  \bibfield  {author} {\bibinfo {author} {\bibfnamefont {X.-G.}\ \bibnamefont {Wen}},\ }\href {\doibase 10.1103/physrevd.88.045013} {\bibfield  {journal} {\bibinfo  {journal} {Physical Review D}\ }\textbf {\bibinfo {volume} {88}},\ \bibinfo {pages} {045103} (\bibinfo {year} {2013})}\BibitemShut {NoStop}%
\bibitem [{\citenamefont {Wang}\ \emph {et~al.}(2015{\natexlab{b}})\citenamefont {Wang}, \citenamefont {Santos},\ and\ \citenamefont {Wen}}]{WSW2014}%
  \BibitemOpen
  \bibfield  {author} {\bibinfo {author} {\bibfnamefont {J.}~\bibnamefont {Wang}}, \bibinfo {author} {\bibfnamefont {L.~H.}\ \bibnamefont {Santos}}, \ and\ \bibinfo {author} {\bibfnamefont {X.-G.}\ \bibnamefont {Wen}},\ }\href {\doibase 10.1103/PhysRevB.91.195134} {\bibfield  {journal} {\bibinfo  {journal} {Phys. Rev. B}\ }\textbf {\bibinfo {volume} {91}},\ \bibinfo {pages} {195134} (\bibinfo {year} {2015}{\natexlab{b}})}\BibitemShut {NoStop}%
\bibitem [{\citenamefont {Kapustin}\ and\ \citenamefont {Thorngren}(2014)}]{KT2014anomaly}%
  \BibitemOpen
  \bibfield  {author} {\bibinfo {author} {\bibfnamefont {A.}~\bibnamefont {Kapustin}}\ and\ \bibinfo {author} {\bibfnamefont {R.}~\bibnamefont {Thorngren}},\ }\href {\doibase 10.1103/physrevlett.112.231602} {\bibfield  {journal} {\bibinfo  {journal} {Physical Review Letters}\ }\textbf {\bibinfo {volume} {112}},\ \bibinfo {pages} {231602} (\bibinfo {year} {2014})}\BibitemShut {NoStop}%
\bibitem [{\citenamefont {Witten}(2016{\natexlab{a}})}]{Witten2016parity}%
  \BibitemOpen
  \bibfield  {author} {\bibinfo {author} {\bibfnamefont {E.}~\bibnamefont {Witten}},\ }\href {\doibase 10.1103/PhysRevB.94.195150} {\bibfield  {journal} {\bibinfo  {journal} {Phys. Rev. B}\ }\textbf {\bibinfo {volume} {94}},\ \bibinfo {pages} {195150} (\bibinfo {year} {2016}{\natexlab{a}})}\BibitemShut {NoStop}%
\bibitem [{\citenamefont {Witten}(2016{\natexlab{b}})}]{Witten2016f}%
  \BibitemOpen
  \bibfield  {author} {\bibinfo {author} {\bibfnamefont {E.}~\bibnamefont {Witten}},\ }\href {\doibase 10.1103/RevModPhys.88.035001} {\bibfield  {journal} {\bibinfo  {journal} {Rev. Mod. Phys.}\ }\textbf {\bibinfo {volume} {88}},\ \bibinfo {pages} {035001} (\bibinfo {year} {2016}{\natexlab{b}})}\BibitemShut {NoStop}%
\bibitem [{\citenamefont {Bulmash}\ and\ \citenamefont {Barkeshli}(2022{\natexlab{a}})}]{Bulmash2021anomaly}%
  \BibitemOpen
  \bibfield  {author} {\bibinfo {author} {\bibfnamefont {D.}~\bibnamefont {Bulmash}}\ and\ \bibinfo {author} {\bibfnamefont {M.}~\bibnamefont {Barkeshli}},\ }\href {\doibase 10.1103/PhysRevB.105.155126} {\bibfield  {journal} {\bibinfo  {journal} {Phys. Rev. B}\ }\textbf {\bibinfo {volume} {105}},\ \bibinfo {pages} {155126} (\bibinfo {year} {2022}{\natexlab{a}})}\BibitemShut {NoStop}%
\bibitem [{\citenamefont {Lu}(2024)}]{Lu_LSMSPT}%
  \BibitemOpen
  \bibfield  {author} {\bibinfo {author} {\bibfnamefont {Y.-M.}\ \bibnamefont {Lu}},\ }\href {\doibase 10.1016/j.aop.2024.169806} {\bibfield  {journal} {\bibinfo  {journal} {Annals of Physics}\ }\textbf {\bibinfo {volume} {470}},\ \bibinfo {pages} {169806} (\bibinfo {year} {2024})}\BibitemShut {NoStop}%
\bibitem [{\citenamefont {Cho}\ \emph {et~al.}(2017)\citenamefont {Cho}, \citenamefont {Hsieh},\ and\ \citenamefont {Ryu}}]{ChoPRB2017}%
  \BibitemOpen
  \bibfield  {author} {\bibinfo {author} {\bibfnamefont {G.~Y.}\ \bibnamefont {Cho}}, \bibinfo {author} {\bibfnamefont {C.-T.}\ \bibnamefont {Hsieh}}, \ and\ \bibinfo {author} {\bibfnamefont {S.}~\bibnamefont {Ryu}},\ }\href {\doibase 10.1103/PhysRevB.96.195105} {\bibfield  {journal} {\bibinfo  {journal} {Phys. Rev. B}\ }\textbf {\bibinfo {volume} {96}},\ \bibinfo {pages} {195105} (\bibinfo {year} {2017})}\BibitemShut {NoStop}%
\bibitem [{\citenamefont {Cheng}(2019)}]{Cheng2018f}%
  \BibitemOpen
  \bibfield  {author} {\bibinfo {author} {\bibfnamefont {M.}~\bibnamefont {Cheng}},\ }\href {\doibase 10.1103/PhysRevB.99.075143} {\bibfield  {journal} {\bibinfo  {journal} {Phys. Rev. B}\ }\textbf {\bibinfo {volume} {99}},\ \bibinfo {pages} {075143} (\bibinfo {year} {2019})}\BibitemShut {NoStop}%
\bibitem [{\citenamefont {Jian}\ \emph {et~al.}(2018)\citenamefont {Jian}, \citenamefont {Bi},\ and\ \citenamefont {Xu}}]{JianPRB2018}%
  \BibitemOpen
  \bibfield  {author} {\bibinfo {author} {\bibfnamefont {C.-M.}\ \bibnamefont {Jian}}, \bibinfo {author} {\bibfnamefont {Z.}~\bibnamefont {Bi}}, \ and\ \bibinfo {author} {\bibfnamefont {C.}~\bibnamefont {Xu}},\ }\href {\doibase 10.1103/PhysRevB.97.054412} {\bibfield  {journal} {\bibinfo  {journal} {Phys. Rev. B}\ }\textbf {\bibinfo {volume} {97}},\ \bibinfo {pages} {054412} (\bibinfo {year} {2018})}\BibitemShut {NoStop}%
\bibitem [{\citenamefont {Jiang}\ \emph {et~al.}(2021)\citenamefont {Jiang}, \citenamefont {Cheng}, \citenamefont {Qi},\ and\ \citenamefont {Lu}}]{JCQL2019}%
  \BibitemOpen
  \bibfield  {author} {\bibinfo {author} {\bibfnamefont {S.}~\bibnamefont {Jiang}}, \bibinfo {author} {\bibfnamefont {M.}~\bibnamefont {Cheng}}, \bibinfo {author} {\bibfnamefont {Y.}~\bibnamefont {Qi}}, \ and\ \bibinfo {author} {\bibfnamefont {Y.-M.}\ \bibnamefont {Lu}},\ }\href {\doibase 10.21468/SciPostPhys.11.2.024} {\bibfield  {journal} {\bibinfo  {journal} {SciPost Phys.}\ }\textbf {\bibinfo {volume} {11}},\ \bibinfo {pages} {024} (\bibinfo {year} {2021})}\BibitemShut {NoStop}%
\bibitem [{\citenamefont {Kobayashi}\ and\ \citenamefont {Barkeshli}(2024)}]{Kobayashi:2021jsc}%
  \BibitemOpen
  \bibfield  {author} {\bibinfo {author} {\bibfnamefont {R.}~\bibnamefont {Kobayashi}}\ and\ \bibinfo {author} {\bibfnamefont {M.}~\bibnamefont {Barkeshli}},\ }\href {\doibase 10.1103/PhysRevB.110.155140} {\bibfield  {journal} {\bibinfo  {journal} {Phys. Rev. B}\ }\textbf {\bibinfo {volume} {110}},\ \bibinfo {pages} {155140} (\bibinfo {year} {2024})}\BibitemShut {NoStop}%
\bibitem [{\citenamefont {You}\ \emph {et~al.}(2018)\citenamefont {You}, \citenamefont {Devakul}, \citenamefont {Burnell},\ and\ \citenamefont {Sondhi}}]{You:2018oai}%
  \BibitemOpen
  \bibfield  {author} {\bibinfo {author} {\bibfnamefont {Y.}~\bibnamefont {You}}, \bibinfo {author} {\bibfnamefont {T.}~\bibnamefont {Devakul}}, \bibinfo {author} {\bibfnamefont {F.~J.}\ \bibnamefont {Burnell}}, \ and\ \bibinfo {author} {\bibfnamefont {S.~L.}\ \bibnamefont {Sondhi}},\ }\href {\doibase 10.1103/PhysRevB.98.035112} {\bibfield  {journal} {\bibinfo  {journal} {Phys. Rev. B}\ }\textbf {\bibinfo {volume} {98}},\ \bibinfo {pages} {035112} (\bibinfo {year} {2018})}\BibitemShut {NoStop}%
\bibitem [{\citenamefont {Devakul}\ \emph {et~al.}(2018)\citenamefont {Devakul}, \citenamefont {Williamson},\ and\ \citenamefont {You}}]{Devakul:2018fhz}%
  \BibitemOpen
  \bibfield  {author} {\bibinfo {author} {\bibfnamefont {T.}~\bibnamefont {Devakul}}, \bibinfo {author} {\bibfnamefont {D.~J.}\ \bibnamefont {Williamson}}, \ and\ \bibinfo {author} {\bibfnamefont {Y.}~\bibnamefont {You}},\ }\href {\doibase 10.1103/PhysRevB.98.235121} {\bibfield  {journal} {\bibinfo  {journal} {Phys. Rev. B}\ }\textbf {\bibinfo {volume} {98}},\ \bibinfo {pages} {235121} (\bibinfo {year} {2018})}\BibitemShut {NoStop}%
\bibitem [{\citenamefont {Devakul}\ \emph {et~al.}(2020)\citenamefont {Devakul}, \citenamefont {Shirley},\ and\ \citenamefont {Wang}}]{Devakul:2019duj}%
  \BibitemOpen
  \bibfield  {author} {\bibinfo {author} {\bibfnamefont {T.}~\bibnamefont {Devakul}}, \bibinfo {author} {\bibfnamefont {W.}~\bibnamefont {Shirley}}, \ and\ \bibinfo {author} {\bibfnamefont {J.}~\bibnamefont {Wang}},\ }\href {\doibase 10.1103/PhysRevResearch.2.012059} {\bibfield  {journal} {\bibinfo  {journal} {Phys. Rev. Res.}\ }\textbf {\bibinfo {volume} {2}},\ \bibinfo {pages} {012059} (\bibinfo {year} {2020})}\BibitemShut {NoStop}%
\bibitem [{\citenamefont {Burnell}\ \emph {et~al.}(2022)\citenamefont {Burnell}, \citenamefont {Devakul}, \citenamefont {Gorantla}, \citenamefont {Lam},\ and\ \citenamefont {Shao}}]{Burnell:2021reh}%
  \BibitemOpen
  \bibfield  {author} {\bibinfo {author} {\bibfnamefont {F.~J.}\ \bibnamefont {Burnell}}, \bibinfo {author} {\bibfnamefont {T.}~\bibnamefont {Devakul}}, \bibinfo {author} {\bibfnamefont {P.}~\bibnamefont {Gorantla}}, \bibinfo {author} {\bibfnamefont {H.~T.}\ \bibnamefont {Lam}}, \ and\ \bibinfo {author} {\bibfnamefont {S.-H.}\ \bibnamefont {Shao}},\ }\href {\doibase 10.1103/PhysRevB.106.085113} {\bibfield  {journal} {\bibinfo  {journal} {Phys. Rev. B}\ }\textbf {\bibinfo {volume} {106}},\ \bibinfo {pages} {085113} (\bibinfo {year} {2022})}\BibitemShut {NoStop}%
\bibitem [{\citenamefont {Zhou}\ \emph {et~al.}(2022)\citenamefont {Zhou}, \citenamefont {Li}, \citenamefont {Yan}, \citenamefont {Ye},\ and\ \citenamefont {Meng}}]{Zhou:2022eig}%
  \BibitemOpen
  \bibfield  {author} {\bibinfo {author} {\bibfnamefont {C.}~\bibnamefont {Zhou}}, \bibinfo {author} {\bibfnamefont {M.-Y.}\ \bibnamefont {Li}}, \bibinfo {author} {\bibfnamefont {Z.}~\bibnamefont {Yan}}, \bibinfo {author} {\bibfnamefont {P.}~\bibnamefont {Ye}}, \ and\ \bibinfo {author} {\bibfnamefont {Z.~Y.}\ \bibnamefont {Meng}},\ }\href {\doibase 10.1103/PhysRevB.106.214428} {\bibfield  {journal} {\bibinfo  {journal} {Phys. Rev. B}\ }\textbf {\bibinfo {volume} {106}},\ \bibinfo {pages} {214428} (\bibinfo {year} {2022})}\BibitemShut {NoStop}%
\bibitem [{\citenamefont {You}(2024)}]{You:2024syf}%
  \BibitemOpen
  \bibfield  {author} {\bibinfo {author} {\bibfnamefont {Y.}~\bibnamefont {You}},\ }\href {\doibase 10.1088/1367-2630/ad78f9} {\bibfield  {journal} {\bibinfo  {journal} {New J. Phys.}\ }\textbf {\bibinfo {volume} {26}},\ \bibinfo {pages} {093028} (\bibinfo {year} {2024})}\BibitemShut {NoStop}%
\bibitem [{\citenamefont {{Thorngren}}\ and\ \citenamefont {{von Keyserlingk}}(2015)}]{Thorngren2015}%
  \BibitemOpen
  \bibfield  {author} {\bibinfo {author} {\bibfnamefont {R.}~\bibnamefont {{Thorngren}}}\ and\ \bibinfo {author} {\bibfnamefont {C.}~\bibnamefont {{von Keyserlingk}}},\ }\href@noop {} {\  (\bibinfo {year} {2015})},\ \Eprint {http://arxiv.org/abs/1511.02929} {arXiv:1511.02929} \BibitemShut {NoStop}%
\bibitem [{\citenamefont {Tachikawa}(2020)}]{Tachikawa2017}%
  \BibitemOpen
  \bibfield  {author} {\bibinfo {author} {\bibfnamefont {Y.}~\bibnamefont {Tachikawa}},\ }\href {\doibase 10.21468/SciPostPhys.8.1.015} {\bibfield  {journal} {\bibinfo  {journal} {SciPost Phys.}\ }\textbf {\bibinfo {volume} {8}},\ \bibinfo {pages} {15} (\bibinfo {year} {2020})}\BibitemShut {NoStop}%
\bibitem [{\citenamefont {Wan}\ and\ \citenamefont {Wang}(2019)}]{WW2018}%
  \BibitemOpen
  \bibfield  {author} {\bibinfo {author} {\bibfnamefont {Z.}~\bibnamefont {Wan}}\ and\ \bibinfo {author} {\bibfnamefont {J.}~\bibnamefont {Wang}},\ }\href {\doibase 10.4310/AMSA.2019.v4.n2.a2} {\bibfield  {journal} {\bibinfo  {journal} {Ann. Math. Sci. Appl.}\ }\textbf {\bibinfo {volume} {4}},\ \bibinfo {pages} {107} (\bibinfo {year} {2019})}\BibitemShut {NoStop}%
\bibitem [{\citenamefont {Tsui}\ and\ \citenamefont {Wen}(2020)}]{Tsui:2019ykk}%
  \BibitemOpen
  \bibfield  {author} {\bibinfo {author} {\bibfnamefont {L.}~\bibnamefont {Tsui}}\ and\ \bibinfo {author} {\bibfnamefont {X.-G.}\ \bibnamefont {Wen}},\ }\href {\doibase 10.1103/PhysRevB.101.035101} {\bibfield  {journal} {\bibinfo  {journal} {Phys. Rev. B}\ }\textbf {\bibinfo {volume} {101}},\ \bibinfo {pages} {035101} (\bibinfo {year} {2020})}\BibitemShut {NoStop}%
\bibitem [{\citenamefont {Kong}\ \emph {et~al.}(2020{\natexlab{b}})\citenamefont {Kong}, \citenamefont {Lan}, \citenamefont {Wen}, \citenamefont {Zhang},\ and\ \citenamefont {Zheng}}]{KLWZZ2020class}%
  \BibitemOpen
  \bibfield  {author} {\bibinfo {author} {\bibfnamefont {L.}~\bibnamefont {Kong}}, \bibinfo {author} {\bibfnamefont {T.}~\bibnamefont {Lan}}, \bibinfo {author} {\bibfnamefont {X.-G.}\ \bibnamefont {Wen}}, \bibinfo {author} {\bibfnamefont {Z.-H.}\ \bibnamefont {Zhang}}, \ and\ \bibinfo {author} {\bibfnamefont {H.}~\bibnamefont {Zheng}},\ }\href {\doibase 10.1007/JHEP09(2020)093} {\bibfield  {journal} {\bibinfo  {journal} {JHEP}\ }\textbf {\bibinfo {volume} {09}},\ \bibinfo {pages} {093} (\bibinfo {year} {2020}{\natexlab{b}})}\BibitemShut {NoStop}%
\bibitem [{\citenamefont {Inamura}(2021)}]{Inamura2021}%
  \BibitemOpen
  \bibfield  {author} {\bibinfo {author} {\bibfnamefont {K.}~\bibnamefont {Inamura}},\ }\href {\doibase 10.1007/JHEP05(2021)204} {\bibfield  {journal} {\bibinfo  {journal} {JHEP}\ }\textbf {\bibinfo {volume} {05}},\ \bibinfo {pages} {204} (\bibinfo {year} {2021})}\BibitemShut {NoStop}%
\bibitem [{\citenamefont {Inamura}\ and\ \citenamefont {Ohyama}(2024)}]{IO2024}%
  \BibitemOpen
  \bibfield  {author} {\bibinfo {author} {\bibfnamefont {K.}~\bibnamefont {Inamura}}\ and\ \bibinfo {author} {\bibfnamefont {S.}~\bibnamefont {Ohyama}},\ }\href@noop {} {\  (\bibinfo {year} {2024})},\ \Eprint {http://arxiv.org/abs/2408.15960} {arXiv:2408.15960 [cond-mat.str-el]} \BibitemShut {NoStop}%
\bibitem [{\citenamefont {Seifnashri}\ and\ \citenamefont {Shao}(2024)}]{SS2024}%
  \BibitemOpen
  \bibfield  {author} {\bibinfo {author} {\bibfnamefont {S.}~\bibnamefont {Seifnashri}}\ and\ \bibinfo {author} {\bibfnamefont {S.-H.}\ \bibnamefont {Shao}},\ }\href {\doibase 10.1103/PhysRevLett.133.116601} {\bibfield  {journal} {\bibinfo  {journal} {Phys. Rev. Lett.}\ }\textbf {\bibinfo {volume} {133}},\ \bibinfo {pages} {116601} (\bibinfo {year} {2024})}\BibitemShut {NoStop}%
\bibitem [{\citenamefont {Moy}\ and\ \citenamefont {Fradkin}(2024)}]{MF2024}%
  \BibitemOpen
  \bibfield  {author} {\bibinfo {author} {\bibfnamefont {B.}~\bibnamefont {Moy}}\ and\ \bibinfo {author} {\bibfnamefont {E.}~\bibnamefont {Fradkin}},\ }\href@noop {} {\  (\bibinfo {year} {2024})},\ \Eprint {http://arxiv.org/abs/2412.02748} {arXiv:2412.02748 [cond-mat.str-el]} \BibitemShut {NoStop}%
\bibitem [{\citenamefont {Li}\ \emph {et~al.}(2023)\citenamefont {Li}, \citenamefont {Oshikawa},\ and\ \citenamefont {Zheng}}]{LOZ2023}%
  \BibitemOpen
  \bibfield  {author} {\bibinfo {author} {\bibfnamefont {L.}~\bibnamefont {Li}}, \bibinfo {author} {\bibfnamefont {M.}~\bibnamefont {Oshikawa}}, \ and\ \bibinfo {author} {\bibfnamefont {Y.}~\bibnamefont {Zheng}},\ }\href {\doibase 10.1103/PhysRevB.108.214429} {\bibfield  {journal} {\bibinfo  {journal} {Phys. Rev. B}\ }\textbf {\bibinfo {volume} {108}},\ \bibinfo {pages} {214429} (\bibinfo {year} {2023})}\BibitemShut {NoStop}%
\bibitem [{\citenamefont {Meng}\ \emph {et~al.}(2024)\citenamefont {Meng}, \citenamefont {Yang}, \citenamefont {Lan},\ and\ \citenamefont {Gu}}]{Meng:2024nxx}%
  \BibitemOpen
  \bibfield  {author} {\bibinfo {author} {\bibfnamefont {C.}~\bibnamefont {Meng}}, \bibinfo {author} {\bibfnamefont {X.}~\bibnamefont {Yang}}, \bibinfo {author} {\bibfnamefont {T.}~\bibnamefont {Lan}}, \ and\ \bibinfo {author} {\bibfnamefont {Z.}~\bibnamefont {Gu}},\ }\href@noop {} {\  (\bibinfo {year} {2024})},\ \Eprint {http://arxiv.org/abs/2412.20546} {arXiv:2412.20546 [cond-mat.str-el]} \BibitemShut {NoStop}%
\bibitem [{\citenamefont {Aksoy}\ and\ \citenamefont {Wen}(2025)}]{Aksoy:2025rmg}%
  \BibitemOpen
  \bibfield  {author} {\bibinfo {author} {\bibfnamefont {O.~M.}\ \bibnamefont {Aksoy}}\ and\ \bibinfo {author} {\bibfnamefont {X.-G.}\ \bibnamefont {Wen}},\ }\href@noop {} {\  (\bibinfo {year} {2025})},\ \Eprint {http://arxiv.org/abs/2503.21764} {arXiv:2503.21764 [cond-mat.str-el]} \BibitemShut {NoStop}%
\bibitem [{\citenamefont {Gu}\ and\ \citenamefont {Wen}(2014)}]{GuWen2012}%
  \BibitemOpen
  \bibfield  {author} {\bibinfo {author} {\bibfnamefont {Z.-C.}\ \bibnamefont {Gu}}\ and\ \bibinfo {author} {\bibfnamefont {X.-G.}\ \bibnamefont {Wen}},\ }\href {\doibase 10.1103/PhysRevB.90.115141} {\bibfield  {journal} {\bibinfo  {journal} {Phys. Rev. B}\ }\textbf {\bibinfo {volume} {90}},\ \bibinfo {pages} {115141} (\bibinfo {year} {2014})}\BibitemShut {NoStop}%
\bibitem [{\citenamefont {Wang}\ and\ \citenamefont {Gu}(2018)}]{wang2017towards}%
  \BibitemOpen
  \bibfield  {author} {\bibinfo {author} {\bibfnamefont {Q.-R.}\ \bibnamefont {Wang}}\ and\ \bibinfo {author} {\bibfnamefont {Z.-C.}\ \bibnamefont {Gu}},\ }\href {\doibase 10.1103/PhysRevX.8.011055} {\bibfield  {journal} {\bibinfo  {journal} {Phys. Rev. X}\ }\textbf {\bibinfo {volume} {8}},\ \bibinfo {pages} {011055} (\bibinfo {year} {2018})}\BibitemShut {NoStop}%
\bibitem [{\citenamefont {Wang}\ and\ \citenamefont {Gu}(2020)}]{wang2018construction}%
  \BibitemOpen
  \bibfield  {author} {\bibinfo {author} {\bibfnamefont {Q.-R.}\ \bibnamefont {Wang}}\ and\ \bibinfo {author} {\bibfnamefont {Z.-C.}\ \bibnamefont {Gu}},\ }\href {\doibase 10.1103/PhysRevX.10.031055} {\bibfield  {journal} {\bibinfo  {journal} {Phys. Rev. X}\ }\textbf {\bibinfo {volume} {10}},\ \bibinfo {pages} {031055} (\bibinfo {year} {2020})}\BibitemShut {NoStop}%
\bibitem [{\citenamefont {Wang}\ \emph {et~al.}(2021)\citenamefont {Wang}, \citenamefont {Ning},\ and\ \citenamefont {Cheng}}]{wang2021domain}%
  \BibitemOpen
  \bibfield  {author} {\bibinfo {author} {\bibfnamefont {Q.-R.}\ \bibnamefont {Wang}}, \bibinfo {author} {\bibfnamefont {S.-Q.}\ \bibnamefont {Ning}}, \ and\ \bibinfo {author} {\bibfnamefont {M.}~\bibnamefont {Cheng}},\ }\href@noop {} {\  (\bibinfo {year} {2021})},\ \Eprint {http://arxiv.org/abs/2104.13233} {arXiv:2104.13233 [cond-mat.str-el]} \BibitemShut {NoStop}%
\bibitem [{\citenamefont {Ouyang}\ \emph {et~al.}(2021)\citenamefont {Ouyang}, \citenamefont {Wang}, \citenamefont {Gu},\ and\ \citenamefont {Qi}}]{Ouyang2020}%
  \BibitemOpen
  \bibfield  {author} {\bibinfo {author} {\bibfnamefont {Y.}~\bibnamefont {Ouyang}}, \bibinfo {author} {\bibfnamefont {Q.-R.}\ \bibnamefont {Wang}}, \bibinfo {author} {\bibfnamefont {Z.-C.}\ \bibnamefont {Gu}}, \ and\ \bibinfo {author} {\bibfnamefont {Y.}~\bibnamefont {Qi}},\ }\href {\doibase 10.1088/0256-307X/38/12/127101} {\bibfield  {journal} {\bibinfo  {journal} {Chin. Phys. Lett.}\ }\textbf {\bibinfo {volume} {38}},\ \bibinfo {pages} {127101} (\bibinfo {year} {2021})}\BibitemShut {NoStop}%
\bibitem [{\citenamefont {Ryu}\ and\ \citenamefont {Zhang}(2012)}]{ryuzhang2012}%
  \BibitemOpen
  \bibfield  {author} {\bibinfo {author} {\bibfnamefont {S.}~\bibnamefont {Ryu}}\ and\ \bibinfo {author} {\bibfnamefont {S.-C.}\ \bibnamefont {Zhang}},\ }\href {\doibase 10.1103/PhysRevB.85.245132} {\bibfield  {journal} {\bibinfo  {journal} {Phys. Rev. B}\ }\textbf {\bibinfo {volume} {85}},\ \bibinfo {pages} {245132} (\bibinfo {year} {2012})}\BibitemShut {NoStop}%
\bibitem [{\citenamefont {Yao}\ and\ \citenamefont {Ryu}(2013)}]{yaoryu2013}%
  \BibitemOpen
  \bibfield  {author} {\bibinfo {author} {\bibfnamefont {H.}~\bibnamefont {Yao}}\ and\ \bibinfo {author} {\bibfnamefont {S.}~\bibnamefont {Ryu}},\ }\href {\doibase 10.1103/PhysRevB.88.064507} {\bibfield  {journal} {\bibinfo  {journal} {Phys. Rev. B}\ }\textbf {\bibinfo {volume} {88}},\ \bibinfo {pages} {064507} (\bibinfo {year} {2013})}\BibitemShut {NoStop}%
\bibitem [{\citenamefont {Wang}\ \emph {et~al.}(2014)\citenamefont {Wang}, \citenamefont {Potter},\ and\ \citenamefont {Senthil}}]{wps2014}%
  \BibitemOpen
  \bibfield  {author} {\bibinfo {author} {\bibfnamefont {C.}~\bibnamefont {Wang}}, \bibinfo {author} {\bibfnamefont {A.~C.}\ \bibnamefont {Potter}}, \ and\ \bibinfo {author} {\bibfnamefont {T.}~\bibnamefont {Senthil}},\ }\href {\doibase 10.1126/science.1243326} {\bibfield  {journal} {\bibinfo  {journal} {Science}\ }\textbf {\bibinfo {volume} {343}},\ \bibinfo {pages} {629} (\bibinfo {year} {2014})}\BibitemShut {NoStop}%
\bibitem [{\citenamefont {Gu}\ and\ \citenamefont {Levin}(2014)}]{gulevin2014}%
  \BibitemOpen
  \bibfield  {author} {\bibinfo {author} {\bibfnamefont {Z.-C.}\ \bibnamefont {Gu}}\ and\ \bibinfo {author} {\bibfnamefont {M.}~\bibnamefont {Levin}},\ }\href {\doibase 10.1103/PhysRevB.89.201113} {\bibfield  {journal} {\bibinfo  {journal} {Phys. Rev. B}\ }\textbf {\bibinfo {volume} {89}},\ \bibinfo {pages} {201113} (\bibinfo {year} {2014})}\BibitemShut {NoStop}%
\bibitem [{\citenamefont {You}\ and\ \citenamefont {Xu}(2014)}]{youxu2014}%
  \BibitemOpen
  \bibfield  {author} {\bibinfo {author} {\bibfnamefont {Y.-Z.}\ \bibnamefont {You}}\ and\ \bibinfo {author} {\bibfnamefont {C.}~\bibnamefont {Xu}},\ }\href {\doibase 10.1103/PhysRevB.90.245120} {\bibfield  {journal} {\bibinfo  {journal} {Phys. Rev. B}\ }\textbf {\bibinfo {volume} {90}},\ \bibinfo {pages} {245120} (\bibinfo {year} {2014})}\BibitemShut {NoStop}%
\bibitem [{\citenamefont {Morimoto}\ \emph {et~al.}(2015)\citenamefont {Morimoto}, \citenamefont {Furusaki},\ and\ \citenamefont {Mudry}}]{mfm2015}%
  \BibitemOpen
  \bibfield  {author} {\bibinfo {author} {\bibfnamefont {T.}~\bibnamefont {Morimoto}}, \bibinfo {author} {\bibfnamefont {A.}~\bibnamefont {Furusaki}}, \ and\ \bibinfo {author} {\bibfnamefont {C.}~\bibnamefont {Mudry}},\ }\href {\doibase 10.1103/PhysRevB.92.125104} {\bibfield  {journal} {\bibinfo  {journal} {Phys. Rev. B}\ }\textbf {\bibinfo {volume} {92}},\ \bibinfo {pages} {125104} (\bibinfo {year} {2015})}\BibitemShut {NoStop}%
\bibitem [{\citenamefont {Tarantino}\ and\ \citenamefont {Fidkowski}(2016)}]{TF2016}%
  \BibitemOpen
  \bibfield  {author} {\bibinfo {author} {\bibfnamefont {N.}~\bibnamefont {Tarantino}}\ and\ \bibinfo {author} {\bibfnamefont {L.}~\bibnamefont {Fidkowski}},\ }\href {\doibase 10.1103/PhysRevB.94.115115} {\bibfield  {journal} {\bibinfo  {journal} {Phys. Rev. B}\ }\textbf {\bibinfo {volume} {94}},\ \bibinfo {pages} {115115} (\bibinfo {year} {2016})}\BibitemShut {NoStop}%
\bibitem [{\citenamefont {Wang}\ \emph {et~al.}(2017)\citenamefont {Wang}, \citenamefont {Lin},\ and\ \citenamefont {Gu}}]{wlg2017}%
  \BibitemOpen
  \bibfield  {author} {\bibinfo {author} {\bibfnamefont {C.}~\bibnamefont {Wang}}, \bibinfo {author} {\bibfnamefont {C.-H.}\ \bibnamefont {Lin}}, \ and\ \bibinfo {author} {\bibfnamefont {Z.-C.}\ \bibnamefont {Gu}},\ }\href {\doibase 10.1103/PhysRevB.95.195147} {\bibfield  {journal} {\bibinfo  {journal} {Phys. Rev. B}\ }\textbf {\bibinfo {volume} {95}},\ \bibinfo {pages} {195147} (\bibinfo {year} {2017})}\BibitemShut {NoStop}%
\bibitem [{\citenamefont {Kapustin}\ and\ \citenamefont {Thorngren}(2017)}]{KT2017}%
  \BibitemOpen
  \bibfield  {author} {\bibinfo {author} {\bibfnamefont {A.}~\bibnamefont {Kapustin}}\ and\ \bibinfo {author} {\bibfnamefont {R.}~\bibnamefont {Thorngren}},\ }\href {\doibase 10.1007/JHEP10(2017)080} {\bibfield  {journal} {\bibinfo  {journal} {JHEP}\ }\textbf {\bibinfo {volume} {10}},\ \bibinfo {pages} {080} (\bibinfo {year} {2017})}\BibitemShut {NoStop}%
\bibitem [{\citenamefont {Guo}\ \emph {et~al.}(2018)\citenamefont {Guo}, \citenamefont {Putrov},\ and\ \citenamefont {Wang}}]{GPW2017}%
  \BibitemOpen
  \bibfield  {author} {\bibinfo {author} {\bibfnamefont {M.}~\bibnamefont {Guo}}, \bibinfo {author} {\bibfnamefont {P.}~\bibnamefont {Putrov}}, \ and\ \bibinfo {author} {\bibfnamefont {J.}~\bibnamefont {Wang}},\ }\href {\doibase 10.1016/j.aop.2018.04.025} {\bibfield  {journal} {\bibinfo  {journal} {Annals Phys.}\ }\textbf {\bibinfo {volume} {394}},\ \bibinfo {pages} {244} (\bibinfo {year} {2018})}\BibitemShut {NoStop}%
\bibitem [{\citenamefont {Cheng}\ \emph {et~al.}(2018{\natexlab{b}})\citenamefont {Cheng}, \citenamefont {Bi}, \citenamefont {You},\ and\ \citenamefont {Gu}}]{cbyg2018}%
  \BibitemOpen
  \bibfield  {author} {\bibinfo {author} {\bibfnamefont {M.}~\bibnamefont {Cheng}}, \bibinfo {author} {\bibfnamefont {Z.}~\bibnamefont {Bi}}, \bibinfo {author} {\bibfnamefont {Y.-Z.}\ \bibnamefont {You}}, \ and\ \bibinfo {author} {\bibfnamefont {Z.-C.}\ \bibnamefont {Gu}},\ }\href {\doibase 10.1103/physrevb.97.205109} {\bibfield  {journal} {\bibinfo  {journal} {Physical Review B}\ }\textbf {\bibinfo {volume} {97}},\ \bibinfo {pages} {205109} (\bibinfo {year} {2018}{\natexlab{b}})}\BibitemShut {NoStop}%
\bibitem [{\citenamefont {Cheng}\ and\ \citenamefont {Wang}(2022)}]{CW2018}%
  \BibitemOpen
  \bibfield  {author} {\bibinfo {author} {\bibfnamefont {M.}~\bibnamefont {Cheng}}\ and\ \bibinfo {author} {\bibfnamefont {C.}~\bibnamefont {Wang}},\ }\href {\doibase 10.1103/PhysRevB.105.195154} {\bibfield  {journal} {\bibinfo  {journal} {Phys. Rev. B}\ }\textbf {\bibinfo {volume} {105}},\ \bibinfo {pages} {195154} (\bibinfo {year} {2022})}\BibitemShut {NoStop}%
\bibitem [{\citenamefont {Thorngren}\ and\ \citenamefont {Else}(2018)}]{ThorngrenPRX2018}%
  \BibitemOpen
  \bibfield  {author} {\bibinfo {author} {\bibfnamefont {R.}~\bibnamefont {Thorngren}}\ and\ \bibinfo {author} {\bibfnamefont {D.~V.}\ \bibnamefont {Else}},\ }\href {\doibase 10.1103/PhysRevX.8.011040} {\bibfield  {journal} {\bibinfo  {journal} {Phys. Rev. X}\ }\textbf {\bibinfo {volume} {8}},\ \bibinfo {pages} {011040} (\bibinfo {year} {2018})}\BibitemShut {NoStop}%
\bibitem [{\citenamefont {Wang}\ \emph {et~al.}(2019{\natexlab{b}})\citenamefont {Wang}, \citenamefont {Qi},\ and\ \citenamefont {Gu}}]{WangASPT}%
  \BibitemOpen
  \bibfield  {author} {\bibinfo {author} {\bibfnamefont {Q.-R.}\ \bibnamefont {Wang}}, \bibinfo {author} {\bibfnamefont {Y.}~\bibnamefont {Qi}}, \ and\ \bibinfo {author} {\bibfnamefont {Z.-C.}\ \bibnamefont {Gu}},\ }\href {\doibase 10.1103/PhysRevLett.123.207003} {\bibfield  {journal} {\bibinfo  {journal} {Phys. Rev. Lett.}\ }\textbf {\bibinfo {volume} {123}},\ \bibinfo {pages} {207003} (\bibinfo {year} {2019}{\natexlab{b}})}\BibitemShut {NoStop}%
\bibitem [{\citenamefont {Prakash}\ and\ \citenamefont {Wang}(2021{\natexlab{a}})}]{PW2020letter}%
  \BibitemOpen
  \bibfield  {author} {\bibinfo {author} {\bibfnamefont {A.}~\bibnamefont {Prakash}}\ and\ \bibinfo {author} {\bibfnamefont {J.}~\bibnamefont {Wang}},\ }\href {\doibase 10.1103/PhysRevLett.126.236802} {\bibfield  {journal} {\bibinfo  {journal} {Phys. Rev. Lett.}\ }\textbf {\bibinfo {volume} {126}},\ \bibinfo {pages} {236802} (\bibinfo {year} {2021}{\natexlab{a}})}\BibitemShut {NoStop}%
\bibitem [{\citenamefont {Prakash}\ and\ \citenamefont {Wang}(2021{\natexlab{b}})}]{PW2020}%
  \BibitemOpen
  \bibfield  {author} {\bibinfo {author} {\bibfnamefont {A.}~\bibnamefont {Prakash}}\ and\ \bibinfo {author} {\bibfnamefont {J.}~\bibnamefont {Wang}},\ }\href {\doibase 10.1103/PhysRevB.103.085130} {\bibfield  {journal} {\bibinfo  {journal} {Phys. Rev. B}\ }\textbf {\bibinfo {volume} {103}},\ \bibinfo {pages} {085130} (\bibinfo {year} {2021}{\natexlab{b}})}\BibitemShut {NoStop}%
\bibitem [{\citenamefont {Barkeshli}\ \emph {et~al.}(2022)\citenamefont {Barkeshli}, \citenamefont {Chen}, \citenamefont {Hsin},\ and\ \citenamefont {Manjunath}}]{Barkeshli:2021ypb}%
  \BibitemOpen
  \bibfield  {author} {\bibinfo {author} {\bibfnamefont {M.}~\bibnamefont {Barkeshli}}, \bibinfo {author} {\bibfnamefont {Y.-A.}\ \bibnamefont {Chen}}, \bibinfo {author} {\bibfnamefont {P.-S.}\ \bibnamefont {Hsin}}, \ and\ \bibinfo {author} {\bibfnamefont {N.}~\bibnamefont {Manjunath}},\ }\href {\doibase 10.1103/PhysRevB.105.235143} {\bibfield  {journal} {\bibinfo  {journal} {Phys. Rev. B}\ }\textbf {\bibinfo {volume} {105}},\ \bibinfo {pages} {235143} (\bibinfo {year} {2022})}\BibitemShut {NoStop}%
\bibitem [{\citenamefont {Zhang}\ \emph {et~al.}(2022{\natexlab{a}})\citenamefont {Zhang}, \citenamefont {Ning}, \citenamefont {Qi},\ and\ \citenamefont {Gu}}]{ZNQG2022}%
  \BibitemOpen
  \bibfield  {author} {\bibinfo {author} {\bibfnamefont {J.-H.}\ \bibnamefont {Zhang}}, \bibinfo {author} {\bibfnamefont {S.-Q.}\ \bibnamefont {Ning}}, \bibinfo {author} {\bibfnamefont {Y.}~\bibnamefont {Qi}}, \ and\ \bibinfo {author} {\bibfnamefont {Z.-C.}\ \bibnamefont {Gu}},\ }\href@noop {} {\  (\bibinfo {year} {2022}{\natexlab{a}})},\ \Eprint {http://arxiv.org/abs/2204.13558} {arXiv:2204.13558 [cond-mat.str-el]} \BibitemShut {NoStop}%
\bibitem [{\citenamefont {Haldane}(1983{\natexlab{a}})}]{haldane1983}%
  \BibitemOpen
  \bibfield  {author} {\bibinfo {author} {\bibfnamefont {F.}~\bibnamefont {Haldane}},\ }\href {\doibase https://doi.org/10.1016/0375-9601(83)90631-X} {\bibfield  {journal} {\bibinfo  {journal} {Physics Letters A}\ }\textbf {\bibinfo {volume} {93}},\ \bibinfo {pages} {464} (\bibinfo {year} {1983}{\natexlab{a}})}\BibitemShut {NoStop}%
\bibitem [{\citenamefont {Affleck}\ \emph {et~al.}(1988)\citenamefont {Affleck}, \citenamefont {Kennedy}, \citenamefont {Lieb},\ and\ \citenamefont {Tasaki}}]{AKLT1988}%
  \BibitemOpen
  \bibfield  {author} {\bibinfo {author} {\bibfnamefont {I.}~\bibnamefont {Affleck}}, \bibinfo {author} {\bibfnamefont {T.}~\bibnamefont {Kennedy}}, \bibinfo {author} {\bibfnamefont {E.~H.}\ \bibnamefont {Lieb}}, \ and\ \bibinfo {author} {\bibfnamefont {H.}~\bibnamefont {Tasaki}},\ }\href {https://api.semanticscholar.org/CorpusID:121787482} {\bibfield  {journal} {\bibinfo  {journal} {Communications in Mathematical Physics}\ }\textbf {\bibinfo {volume} {115}},\ \bibinfo {pages} {477} (\bibinfo {year} {1988})}\BibitemShut {NoStop}%
\bibitem [{\citenamefont {Hagiwara}\ \emph {et~al.}(1990)\citenamefont {Hagiwara}, \citenamefont {Katsumata}, \citenamefont {Affleck}, \citenamefont {Halperin},\ and\ \citenamefont {Renard}}]{Hagiwara1990}%
  \BibitemOpen
  \bibfield  {author} {\bibinfo {author} {\bibfnamefont {M.}~\bibnamefont {Hagiwara}}, \bibinfo {author} {\bibfnamefont {K.}~\bibnamefont {Katsumata}}, \bibinfo {author} {\bibfnamefont {I.}~\bibnamefont {Affleck}}, \bibinfo {author} {\bibfnamefont {B.~I.}\ \bibnamefont {Halperin}}, \ and\ \bibinfo {author} {\bibfnamefont {J.~P.}\ \bibnamefont {Renard}},\ }\href {\doibase 10.1103/PhysRevLett.65.3181} {\bibfield  {journal} {\bibinfo  {journal} {Phys. Rev. Lett.}\ }\textbf {\bibinfo {volume} {65}},\ \bibinfo {pages} {3181} (\bibinfo {year} {1990})}\BibitemShut {NoStop}%
\bibitem [{\citenamefont {Glarum}\ \emph {et~al.}(1991)\citenamefont {Glarum}, \citenamefont {Geschwind}, \citenamefont {Lee}, \citenamefont {Kaplan},\ and\ \citenamefont {Michel}}]{glarum1991}%
  \BibitemOpen
  \bibfield  {author} {\bibinfo {author} {\bibfnamefont {S.~H.}\ \bibnamefont {Glarum}}, \bibinfo {author} {\bibfnamefont {S.}~\bibnamefont {Geschwind}}, \bibinfo {author} {\bibfnamefont {K.~M.}\ \bibnamefont {Lee}}, \bibinfo {author} {\bibfnamefont {M.~L.}\ \bibnamefont {Kaplan}}, \ and\ \bibinfo {author} {\bibfnamefont {J.}~\bibnamefont {Michel}},\ }\href {\doibase 10.1103/PhysRevLett.67.1614} {\bibfield  {journal} {\bibinfo  {journal} {Phys. Rev. Lett.}\ }\textbf {\bibinfo {volume} {67}},\ \bibinfo {pages} {1614} (\bibinfo {year} {1991})}\BibitemShut {NoStop}%
\bibitem [{\citenamefont {Ng}(1994)}]{NG1994}%
  \BibitemOpen
  \bibfield  {author} {\bibinfo {author} {\bibfnamefont {T.-K.}\ \bibnamefont {Ng}},\ }\href {\doibase 10.1103/PhysRevB.50.555} {\bibfield  {journal} {\bibinfo  {journal} {Phys. Rev. B}\ }\textbf {\bibinfo {volume} {50}},\ \bibinfo {pages} {555} (\bibinfo {year} {1994})}\BibitemShut {NoStop}%
\bibitem [{\citenamefont {Kane}\ and\ \citenamefont {Mele}(2005{\natexlab{a}})}]{qhe2005z2}%
  \BibitemOpen
  \bibfield  {author} {\bibinfo {author} {\bibfnamefont {C.~L.}\ \bibnamefont {Kane}}\ and\ \bibinfo {author} {\bibfnamefont {E.~J.}\ \bibnamefont {Mele}},\ }\href {\doibase 10.1103/PhysRevLett.95.146802} {\bibfield  {journal} {\bibinfo  {journal} {Phys. Rev. Lett.}\ }\textbf {\bibinfo {volume} {95}},\ \bibinfo {pages} {146802} (\bibinfo {year} {2005}{\natexlab{a}})}\BibitemShut {NoStop}%
\bibitem [{\citenamefont {Kane}\ and\ \citenamefont {Mele}(2005{\natexlab{b}})}]{qhe2005}%
  \BibitemOpen
  \bibfield  {author} {\bibinfo {author} {\bibfnamefont {C.~L.}\ \bibnamefont {Kane}}\ and\ \bibinfo {author} {\bibfnamefont {E.~J.}\ \bibnamefont {Mele}},\ }\href {\doibase 10.1103/PhysRevLett.95.226801} {\bibfield  {journal} {\bibinfo  {journal} {Phys. Rev. Lett.}\ }\textbf {\bibinfo {volume} {95}},\ \bibinfo {pages} {226801} (\bibinfo {year} {2005}{\natexlab{b}})}\BibitemShut {NoStop}%
\bibitem [{\citenamefont {Bernevig}\ and\ \citenamefont {Zhang}(2006)}]{qhe2006}%
  \BibitemOpen
  \bibfield  {author} {\bibinfo {author} {\bibfnamefont {B.~A.}\ \bibnamefont {Bernevig}}\ and\ \bibinfo {author} {\bibfnamefont {S.-C.}\ \bibnamefont {Zhang}},\ }\href {\doibase 10.1103/PhysRevLett.96.106802} {\bibfield  {journal} {\bibinfo  {journal} {Phys. Rev. Lett.}\ }\textbf {\bibinfo {volume} {96}},\ \bibinfo {pages} {106802} (\bibinfo {year} {2006})}\BibitemShut {NoStop}%
\bibitem [{\citenamefont {Moore}\ and\ \citenamefont {Balents}(2007)}]{topinv2007}%
  \BibitemOpen
  \bibfield  {author} {\bibinfo {author} {\bibfnamefont {J.~E.}\ \bibnamefont {Moore}}\ and\ \bibinfo {author} {\bibfnamefont {L.}~\bibnamefont {Balents}},\ }\href {\doibase 10.1103/PhysRevB.75.121306} {\bibfield  {journal} {\bibinfo  {journal} {Phys. Rev. B}\ }\textbf {\bibinfo {volume} {75}},\ \bibinfo {pages} {121306} (\bibinfo {year} {2007})}\BibitemShut {NoStop}%
\bibitem [{\citenamefont {Fu}\ \emph {et~al.}(2007)\citenamefont {Fu}, \citenamefont {Kane},\ and\ \citenamefont {Mele}}]{ti2007}%
  \BibitemOpen
  \bibfield  {author} {\bibinfo {author} {\bibfnamefont {L.}~\bibnamefont {Fu}}, \bibinfo {author} {\bibfnamefont {C.~L.}\ \bibnamefont {Kane}}, \ and\ \bibinfo {author} {\bibfnamefont {E.~J.}\ \bibnamefont {Mele}},\ }\href {\doibase 10.1103/PhysRevLett.98.106803} {\bibfield  {journal} {\bibinfo  {journal} {Phys. Rev. Lett.}\ }\textbf {\bibinfo {volume} {98}},\ \bibinfo {pages} {106803} (\bibinfo {year} {2007})}\BibitemShut {NoStop}%
\bibitem [{\citenamefont {Qi}\ \emph {et~al.}(2008)\citenamefont {Qi}, \citenamefont {Hughes},\ and\ \citenamefont {Zhang}}]{ti2008}%
  \BibitemOpen
  \bibfield  {author} {\bibinfo {author} {\bibfnamefont {X.-L.}\ \bibnamefont {Qi}}, \bibinfo {author} {\bibfnamefont {T.~L.}\ \bibnamefont {Hughes}}, \ and\ \bibinfo {author} {\bibfnamefont {S.-C.}\ \bibnamefont {Zhang}},\ }\href {\doibase 10.1103/PhysRevB.78.195424} {\bibfield  {journal} {\bibinfo  {journal} {Phys. Rev. B}\ }\textbf {\bibinfo {volume} {78}},\ \bibinfo {pages} {195424} (\bibinfo {year} {2008})}\BibitemShut {NoStop}%
\bibitem [{\citenamefont {Chen}\ and\ \citenamefont {Wen}(2012)}]{Chen:2012hc}%
  \BibitemOpen
  \bibfield  {author} {\bibinfo {author} {\bibfnamefont {X.}~\bibnamefont {Chen}}\ and\ \bibinfo {author} {\bibfnamefont {X.-G.}\ \bibnamefont {Wen}},\ }\href {\doibase 10.1103/PhysRevB.86.235135} {\bibfield  {journal} {\bibinfo  {journal} {Phys. Rev. B}\ }\textbf {\bibinfo {volume} {86}},\ \bibinfo {pages} {235135} (\bibinfo {year} {2012})}\BibitemShut {NoStop}%
\bibitem [{\citenamefont {Liu}\ and\ \citenamefont {Wen}(2013)}]{Liu:2012taf}%
  \BibitemOpen
  \bibfield  {author} {\bibinfo {author} {\bibfnamefont {Z.-X.}\ \bibnamefont {Liu}}\ and\ \bibinfo {author} {\bibfnamefont {X.-G.}\ \bibnamefont {Wen}},\ }\href {\doibase 10.1103/PhysRevLett.110.067205} {\bibfield  {journal} {\bibinfo  {journal} {Phys. Rev. Lett.}\ }\textbf {\bibinfo {volume} {110}},\ \bibinfo {pages} {067205} (\bibinfo {year} {2013})}\BibitemShut {NoStop}%
\bibitem [{\citenamefont {'t~Hooft}(1980)}]{tHooft1979}%
  \BibitemOpen
  \bibfield  {author} {\bibinfo {author} {\bibfnamefont {G.}~\bibnamefont {'t~Hooft}},\ }\href {\doibase 10.1007/978-1-4684-7571-5_9} {\bibfield  {journal} {\bibinfo  {journal} {NATO Sci. Ser. B}\ }\textbf {\bibinfo {volume} {59}},\ \bibinfo {pages} {135} (\bibinfo {year} {1980})}\BibitemShut {NoStop}%
\bibitem [{\citenamefont {Chen}\ \emph {et~al.}(2011{\natexlab{c}})\citenamefont {Chen}, \citenamefont {Liu},\ and\ \citenamefont {Wen}}]{chen2011spt}%
  \BibitemOpen
  \bibfield  {author} {\bibinfo {author} {\bibfnamefont {X.}~\bibnamefont {Chen}}, \bibinfo {author} {\bibfnamefont {Z.-X.}\ \bibnamefont {Liu}}, \ and\ \bibinfo {author} {\bibfnamefont {X.-G.}\ \bibnamefont {Wen}},\ }\href {\doibase 10.1103/physrevb.84.235141} {\bibfield  {journal} {\bibinfo  {journal} {Physical Review B}\ }\textbf {\bibinfo {volume} {84}},\ \bibinfo {pages} {235141} (\bibinfo {year} {2011}{\natexlab{c}})}\BibitemShut {NoStop}%
\bibitem [{\citenamefont {Fidkowski}\ \emph {et~al.}(2013)\citenamefont {Fidkowski}, \citenamefont {Chen},\ and\ \citenamefont {Vishwanath}}]{fcv2013}%
  \BibitemOpen
  \bibfield  {author} {\bibinfo {author} {\bibfnamefont {L.}~\bibnamefont {Fidkowski}}, \bibinfo {author} {\bibfnamefont {X.}~\bibnamefont {Chen}}, \ and\ \bibinfo {author} {\bibfnamefont {A.}~\bibnamefont {Vishwanath}},\ }\href {\doibase 10.1103/PhysRevX.3.041016} {\bibfield  {journal} {\bibinfo  {journal} {Phys. Rev. X}\ }\textbf {\bibinfo {volume} {3}},\ \bibinfo {pages} {041016} (\bibinfo {year} {2013})}\BibitemShut {NoStop}%
\bibitem [{\citenamefont {Bonderson}\ \emph {et~al.}(2013)\citenamefont {Bonderson}, \citenamefont {Nayak},\ and\ \citenamefont {Qi}}]{BNQ2013}%
  \BibitemOpen
  \bibfield  {author} {\bibinfo {author} {\bibfnamefont {P.}~\bibnamefont {Bonderson}}, \bibinfo {author} {\bibfnamefont {C.}~\bibnamefont {Nayak}}, \ and\ \bibinfo {author} {\bibfnamefont {X.-L.}\ \bibnamefont {Qi}},\ }\href {\doibase 10.1088/1742-5468/2013/09/P09016} {\bibfield  {journal} {\bibinfo  {journal} {Journal of Statistical Mechanics: Theory and Experiment}\ }\textbf {\bibinfo {volume} {2013}},\ \bibinfo {pages} {P09016} (\bibinfo {year} {2013})}\BibitemShut {NoStop}%
\bibitem [{\citenamefont {Wang}\ \emph {et~al.}(2013)\citenamefont {Wang}, \citenamefont {Potter},\ and\ \citenamefont {Senthil}}]{wps2013}%
  \BibitemOpen
  \bibfield  {author} {\bibinfo {author} {\bibfnamefont {C.}~\bibnamefont {Wang}}, \bibinfo {author} {\bibfnamefont {A.~C.}\ \bibnamefont {Potter}}, \ and\ \bibinfo {author} {\bibfnamefont {T.}~\bibnamefont {Senthil}},\ }\href {\doibase 10.1103/physrevb.88.115137} {\bibfield  {journal} {\bibinfo  {journal} {Physical Review B}\ }\textbf {\bibinfo {volume} {88}},\ \bibinfo {pages} {115137} (\bibinfo {year} {2013})}\BibitemShut {NoStop}%
\bibitem [{\citenamefont {Vishwanath}\ and\ \citenamefont {Senthil}(2013)}]{VS2013}%
  \BibitemOpen
  \bibfield  {author} {\bibinfo {author} {\bibfnamefont {A.}~\bibnamefont {Vishwanath}}\ and\ \bibinfo {author} {\bibfnamefont {T.}~\bibnamefont {Senthil}},\ }\href {\doibase 10.1103/physrevx.3.011016} {\bibfield  {journal} {\bibinfo  {journal} {Physical Review X}\ }\textbf {\bibinfo {volume} {3}},\ \bibinfo {pages} {011016} (\bibinfo {year} {2013})}\BibitemShut {NoStop}%
\bibitem [{\citenamefont {Metlitski}\ \emph {et~al.}(2013)\citenamefont {Metlitski}, \citenamefont {Kane},\ and\ \citenamefont {Fisher}}]{MKF2013}%
  \BibitemOpen
  \bibfield  {author} {\bibinfo {author} {\bibfnamefont {M.~A.}\ \bibnamefont {Metlitski}}, \bibinfo {author} {\bibfnamefont {C.~L.}\ \bibnamefont {Kane}}, \ and\ \bibinfo {author} {\bibfnamefont {M.~P.~A.}\ \bibnamefont {Fisher}},\ }\href {\doibase 10.1103/physrevb.88.035131} {\bibfield  {journal} {\bibinfo  {journal} {Physical Review B}\ }\textbf {\bibinfo {volume} {88}},\ \bibinfo {pages} {035131} (\bibinfo {year} {2013})}\BibitemShut {NoStop}%
\bibitem [{\citenamefont {Burnell}\ \emph {et~al.}(2014)\citenamefont {Burnell}, \citenamefont {Chen}, \citenamefont {Fidkowski},\ and\ \citenamefont {Vishwanath}}]{BCFV2014}%
  \BibitemOpen
  \bibfield  {author} {\bibinfo {author} {\bibfnamefont {F.~J.}\ \bibnamefont {Burnell}}, \bibinfo {author} {\bibfnamefont {X.}~\bibnamefont {Chen}}, \bibinfo {author} {\bibfnamefont {L.}~\bibnamefont {Fidkowski}}, \ and\ \bibinfo {author} {\bibfnamefont {A.}~\bibnamefont {Vishwanath}},\ }\href {\doibase 10.1103/physrevb.90.245122} {\bibfield  {journal} {\bibinfo  {journal} {Physical Review B}\ }\textbf {\bibinfo {volume} {90}},\ \bibinfo {eid} {245122} (\bibinfo {year} {2014})}\BibitemShut {NoStop}%
\bibitem [{\citenamefont {Chen}\ \emph {et~al.}(2014{\natexlab{b}})\citenamefont {Chen}, \citenamefont {Fidkowski},\ and\ \citenamefont {Vishwanath}}]{cfv2014}%
  \BibitemOpen
  \bibfield  {author} {\bibinfo {author} {\bibfnamefont {X.}~\bibnamefont {Chen}}, \bibinfo {author} {\bibfnamefont {L.}~\bibnamefont {Fidkowski}}, \ and\ \bibinfo {author} {\bibfnamefont {A.}~\bibnamefont {Vishwanath}},\ }\href {\doibase 10.1103/PhysRevB.89.165132} {\bibfield  {journal} {\bibinfo  {journal} {Phys. Rev. B}\ }\textbf {\bibinfo {volume} {89}},\ \bibinfo {pages} {165132} (\bibinfo {year} {2014}{\natexlab{b}})}\BibitemShut {NoStop}%
\bibitem [{\citenamefont {Metlitski}\ \emph {et~al.}(2014)\citenamefont {Metlitski}, \citenamefont {Fidkowski}, \citenamefont {Chen},\ and\ \citenamefont {Vishwanath}}]{mfcv2014}%
  \BibitemOpen
  \bibfield  {author} {\bibinfo {author} {\bibfnamefont {M.~A.}\ \bibnamefont {Metlitski}}, \bibinfo {author} {\bibfnamefont {L.}~\bibnamefont {Fidkowski}}, \bibinfo {author} {\bibfnamefont {X.}~\bibnamefont {Chen}}, \ and\ \bibinfo {author} {\bibfnamefont {A.}~\bibnamefont {Vishwanath}},\ }\href@noop {} {\  (\bibinfo {year} {2014})},\ \Eprint {http://arxiv.org/abs/1406.3032} {arXiv:1406.3032 [cond-mat.str-el]} \BibitemShut {NoStop}%
\bibitem [{\citenamefont {Wang}\ and\ \citenamefont {Senthil}(2014)}]{ws2014}%
  \BibitemOpen
  \bibfield  {author} {\bibinfo {author} {\bibfnamefont {C.}~\bibnamefont {Wang}}\ and\ \bibinfo {author} {\bibfnamefont {T.}~\bibnamefont {Senthil}},\ }\href {\doibase 10.1103/PhysRevB.89.195124} {\bibfield  {journal} {\bibinfo  {journal} {Phys. Rev. B}\ }\textbf {\bibinfo {volume} {89}},\ \bibinfo {pages} {195124} (\bibinfo {year} {2014})}\BibitemShut {NoStop}%
\bibitem [{\citenamefont {Metlitski}\ \emph {et~al.}(2015)\citenamefont {Metlitski}, \citenamefont {Kane},\ and\ \citenamefont {Fisher}}]{mkf2015}%
  \BibitemOpen
  \bibfield  {author} {\bibinfo {author} {\bibfnamefont {M.~A.}\ \bibnamefont {Metlitski}}, \bibinfo {author} {\bibfnamefont {C.~L.}\ \bibnamefont {Kane}}, \ and\ \bibinfo {author} {\bibfnamefont {M.~P.~A.}\ \bibnamefont {Fisher}},\ }\href {\doibase 10.1103/PhysRevB.92.125111} {\bibfield  {journal} {\bibinfo  {journal} {Phys. Rev. B}\ }\textbf {\bibinfo {volume} {92}},\ \bibinfo {pages} {125111} (\bibinfo {year} {2015})}\BibitemShut {NoStop}%
\bibitem [{\citenamefont {Mross}\ \emph {et~al.}(2015)\citenamefont {Mross}, \citenamefont {Essin},\ and\ \citenamefont {Alicea}}]{MEA2015}%
  \BibitemOpen
  \bibfield  {author} {\bibinfo {author} {\bibfnamefont {D.~F.}\ \bibnamefont {Mross}}, \bibinfo {author} {\bibfnamefont {A.}~\bibnamefont {Essin}}, \ and\ \bibinfo {author} {\bibfnamefont {J.}~\bibnamefont {Alicea}},\ }\href {\doibase 10.1103/physrevx.5.011011} {\bibfield  {journal} {\bibinfo  {journal} {Physical Review X}\ }\textbf {\bibinfo {volume} {5}},\ \bibinfo {eid} {011011} (\bibinfo {year} {2015})}\BibitemShut {NoStop}%
\bibitem [{\citenamefont {Wang}\ \emph {et~al.}(2016)\citenamefont {Wang}, \citenamefont {Lin},\ and\ \citenamefont {Levin}}]{WLL2016}%
  \BibitemOpen
  \bibfield  {author} {\bibinfo {author} {\bibfnamefont {C.}~\bibnamefont {Wang}}, \bibinfo {author} {\bibfnamefont {C.-H.}\ \bibnamefont {Lin}}, \ and\ \bibinfo {author} {\bibfnamefont {M.}~\bibnamefont {Levin}},\ }\href {\doibase 10.1103/PhysRevX.6.021015} {\bibfield  {journal} {\bibinfo  {journal} {Physical Review X}\ }\textbf {\bibinfo {volume} {6}},\ \bibinfo {eid} {021015} (\bibinfo {year} {2016})}\BibitemShut {NoStop}%
\bibitem [{\citenamefont {Fidkowski}\ \emph {et~al.}(2018)\citenamefont {Fidkowski}, \citenamefont {Vishwanath},\ and\ \citenamefont {Metlitski}}]{FVM2018}%
  \BibitemOpen
  \bibfield  {author} {\bibinfo {author} {\bibfnamefont {L.}~\bibnamefont {Fidkowski}}, \bibinfo {author} {\bibfnamefont {A.}~\bibnamefont {Vishwanath}}, \ and\ \bibinfo {author} {\bibfnamefont {M.~A.}\ \bibnamefont {Metlitski}},\ }\href@noop {} {\  (\bibinfo {year} {2018})},\ \Eprint {http://arxiv.org/abs/1804.08628} {arXiv:1804.08628 [cond-mat.str-el]} \BibitemShut {NoStop}%
\bibitem [{\citenamefont {Ning}\ \emph {et~al.}(2021{\natexlab{b}})\citenamefont {Ning}, \citenamefont {Mao}, \citenamefont {Li},\ and\ \citenamefont {Wang}}]{NMLW2021}%
  \BibitemOpen
  \bibfield  {author} {\bibinfo {author} {\bibfnamefont {S.-Q.}\ \bibnamefont {Ning}}, \bibinfo {author} {\bibfnamefont {B.-B.}\ \bibnamefont {Mao}}, \bibinfo {author} {\bibfnamefont {Z.}~\bibnamefont {Li}}, \ and\ \bibinfo {author} {\bibfnamefont {C.}~\bibnamefont {Wang}},\ }\href {\doibase 10.1103/PhysRevB.104.075111} {\bibfield  {journal} {\bibinfo  {journal} {Phys. Rev. B}\ }\textbf {\bibinfo {volume} {104}},\ \bibinfo {pages} {075111} (\bibinfo {year} {2021}{\natexlab{b}})}\BibitemShut {NoStop}%
\bibitem [{\citenamefont {Tata}\ \emph {et~al.}(2023)\citenamefont {Tata}, \citenamefont {Kobayashi}, \citenamefont {Bulmash},\ and\ \citenamefont {Barkeshli}}]{TKBB2021}%
  \BibitemOpen
  \bibfield  {author} {\bibinfo {author} {\bibfnamefont {S.}~\bibnamefont {Tata}}, \bibinfo {author} {\bibfnamefont {R.}~\bibnamefont {Kobayashi}}, \bibinfo {author} {\bibfnamefont {D.}~\bibnamefont {Bulmash}}, \ and\ \bibinfo {author} {\bibfnamefont {M.}~\bibnamefont {Barkeshli}},\ }\href {\doibase 10.1007/s00220-022-04484-w} {\bibfield  {journal} {\bibinfo  {journal} {Commun. Math. Phys.}\ }\textbf {\bibinfo {volume} {397}},\ \bibinfo {pages} {199} (\bibinfo {year} {2023})}\BibitemShut {NoStop}%
\bibitem [{\citenamefont {Yang}\ and\ \citenamefont {Cheng}(2024)}]{YC2023}%
  \BibitemOpen
  \bibfield  {author} {\bibinfo {author} {\bibfnamefont {X.}~\bibnamefont {Yang}}\ and\ \bibinfo {author} {\bibfnamefont {M.}~\bibnamefont {Cheng}},\ }\href {\doibase 10.1103/PhysRevB.110.045137} {\bibfield  {journal} {\bibinfo  {journal} {Phys. Rev. B}\ }\textbf {\bibinfo {volume} {110}},\ \bibinfo {pages} {045137} (\bibinfo {year} {2024})}\BibitemShut {NoStop}%
\bibitem [{\citenamefont {Cheng}\ \emph {et~al.}(2024)\citenamefont {Cheng}, \citenamefont {Wang},\ and\ \citenamefont {Yang}}]{CWY2024}%
  \BibitemOpen
  \bibfield  {author} {\bibinfo {author} {\bibfnamefont {M.}~\bibnamefont {Cheng}}, \bibinfo {author} {\bibfnamefont {J.}~\bibnamefont {Wang}}, \ and\ \bibinfo {author} {\bibfnamefont {X.}~\bibnamefont {Yang}},\ }\href@noop {} {\  (\bibinfo {year} {2024})},\ \Eprint {http://arxiv.org/abs/2411.05786} {arXiv:2411.05786 [cond-mat.str-el]} \BibitemShut {NoStop}%
\bibitem [{\citenamefont {Etingof}\ \emph {et~al.}(2009)\citenamefont {Etingof}, \citenamefont {Nikshych}, \citenamefont {Ostrik},\ and\ \citenamefont {with an appendix~by Ehud~Meir}}]{etingof_fusion_2009}%
  \BibitemOpen
  \bibfield  {author} {\bibinfo {author} {\bibfnamefont {P.}~\bibnamefont {Etingof}}, \bibinfo {author} {\bibfnamefont {D.}~\bibnamefont {Nikshych}}, \bibinfo {author} {\bibfnamefont {V.}~\bibnamefont {Ostrik}}, \ and\ \bibinfo {author} {\bibnamefont {with an appendix~by Ehud~Meir}},\ }\href@noop {} {\  (\bibinfo {year} {2009})},\ \Eprint {http://arxiv.org/abs/0909.3140} {arXiv:0909.3140 [math.QA]} \BibitemShut {NoStop}%
\bibitem [{\citenamefont {Heinrich}\ \emph {et~al.}(2016)\citenamefont {Heinrich}, \citenamefont {Burnell}, \citenamefont {Fidkowski},\ and\ \citenamefont {Levin}}]{Heinrich_2016}%
  \BibitemOpen
  \bibfield  {author} {\bibinfo {author} {\bibfnamefont {C.}~\bibnamefont {Heinrich}}, \bibinfo {author} {\bibfnamefont {F.}~\bibnamefont {Burnell}}, \bibinfo {author} {\bibfnamefont {L.}~\bibnamefont {Fidkowski}}, \ and\ \bibinfo {author} {\bibfnamefont {M.}~\bibnamefont {Levin}},\ }\href {\doibase 10.1103/physrevb.94.235136} {\bibfield  {journal} {\bibinfo  {journal} {Physical Review B}\ }\textbf {\bibinfo {volume} {94}},\ \bibinfo {pages} {235136} (\bibinfo {year} {2016})}\BibitemShut {NoStop}%
\bibitem [{\citenamefont {Barkeshli}\ \emph {et~al.}(2019{\natexlab{a}})\citenamefont {Barkeshli}, \citenamefont {Bonderson}, \citenamefont {Jian}, \citenamefont {Cheng},\ and\ \citenamefont {Walker}}]{Barkeshli:2016mew}%
  \BibitemOpen
  \bibfield  {author} {\bibinfo {author} {\bibfnamefont {M.}~\bibnamefont {Barkeshli}}, \bibinfo {author} {\bibfnamefont {P.}~\bibnamefont {Bonderson}}, \bibinfo {author} {\bibfnamefont {C.-M.}\ \bibnamefont {Jian}}, \bibinfo {author} {\bibfnamefont {M.}~\bibnamefont {Cheng}}, \ and\ \bibinfo {author} {\bibfnamefont {K.}~\bibnamefont {Walker}},\ }\href {\doibase 10.1007/s00220-019-03475-8} {\bibfield  {journal} {\bibinfo  {journal} {Commun. Math. Phys.}\ }\textbf {\bibinfo {volume} {374}},\ \bibinfo {pages} {1021} (\bibinfo {year} {2019}{\natexlab{a}})}\BibitemShut {NoStop}%
\bibitem [{\citenamefont {Wen}(2017{\natexlab{b}})}]{Wen:2016cij}%
  \BibitemOpen
  \bibfield  {author} {\bibinfo {author} {\bibfnamefont {X.-G.}\ \bibnamefont {Wen}},\ }\href {\doibase 10.1103/PhysRevB.95.205142} {\bibfield  {journal} {\bibinfo  {journal} {Phys. Rev. B}\ }\textbf {\bibinfo {volume} {95}},\ \bibinfo {pages} {205142} (\bibinfo {year} {2017}{\natexlab{b}})}\BibitemShut {NoStop}%
\bibitem [{\citenamefont {Cheng}\ \emph {et~al.}(2017)\citenamefont {Cheng}, \citenamefont {Gu}, \citenamefont {Jiang},\ and\ \citenamefont {Qi}}]{Cheng_2017}%
  \BibitemOpen
  \bibfield  {author} {\bibinfo {author} {\bibfnamefont {M.}~\bibnamefont {Cheng}}, \bibinfo {author} {\bibfnamefont {Z.-C.}\ \bibnamefont {Gu}}, \bibinfo {author} {\bibfnamefont {S.}~\bibnamefont {Jiang}}, \ and\ \bibinfo {author} {\bibfnamefont {Y.}~\bibnamefont {Qi}},\ }\href {\doibase 10.1103/physrevb.96.115107} {\bibfield  {journal} {\bibinfo  {journal} {Physical Review B}\ }\textbf {\bibinfo {volume} {96}},\ \bibinfo {pages} {115107} (\bibinfo {year} {2017})}\BibitemShut {NoStop}%
\bibitem [{\citenamefont {Barkeshli}\ \emph {et~al.}(2019{\natexlab{b}})\citenamefont {Barkeshli}, \citenamefont {Bonderson}, \citenamefont {Cheng},\ and\ \citenamefont {Wang}}]{BBCW2019}%
  \BibitemOpen
  \bibfield  {author} {\bibinfo {author} {\bibfnamefont {M.}~\bibnamefont {Barkeshli}}, \bibinfo {author} {\bibfnamefont {P.}~\bibnamefont {Bonderson}}, \bibinfo {author} {\bibfnamefont {M.}~\bibnamefont {Cheng}}, \ and\ \bibinfo {author} {\bibfnamefont {Z.}~\bibnamefont {Wang}},\ }\href {\doibase 10.1103/PhysRevB.100.115147} {\bibfield  {journal} {\bibinfo  {journal} {Phys. Rev. B}\ }\textbf {\bibinfo {volume} {100}},\ \bibinfo {pages} {115147} (\bibinfo {year} {2019}{\natexlab{b}})}\BibitemShut {NoStop}%
\bibitem [{\citenamefont {Barkeshli}\ and\ \citenamefont {Cheng}(2020)}]{Barkeshli:2019vtb}%
  \BibitemOpen
  \bibfield  {author} {\bibinfo {author} {\bibfnamefont {M.}~\bibnamefont {Barkeshli}}\ and\ \bibinfo {author} {\bibfnamefont {M.}~\bibnamefont {Cheng}},\ }\href {\doibase 10.21468/SciPostPhys.8.2.028} {\bibfield  {journal} {\bibinfo  {journal} {SciPost Phys.}\ }\textbf {\bibinfo {volume} {8}},\ \bibinfo {pages} {028} (\bibinfo {year} {2020})}\BibitemShut {NoStop}%
\bibitem [{\citenamefont {Aasen}\ \emph {et~al.}(2021)\citenamefont {Aasen}, \citenamefont {Bonderson},\ and\ \citenamefont {Knapp}}]{Aasen2021}%
  \BibitemOpen
  \bibfield  {author} {\bibinfo {author} {\bibfnamefont {D.}~\bibnamefont {Aasen}}, \bibinfo {author} {\bibfnamefont {P.}~\bibnamefont {Bonderson}}, \ and\ \bibinfo {author} {\bibfnamefont {C.}~\bibnamefont {Knapp}},\ }\href@noop {} {\  (\bibinfo {year} {2021})},\ \Eprint {http://arxiv.org/abs/2109.10911} {arXiv:2109.10911 [cond-mat.str-el]} \BibitemShut {NoStop}%
\bibitem [{\citenamefont {Bulmash}\ and\ \citenamefont {Barkeshli}(2020)}]{Bulmash:2020flp}%
  \BibitemOpen
  \bibfield  {author} {\bibinfo {author} {\bibfnamefont {D.}~\bibnamefont {Bulmash}}\ and\ \bibinfo {author} {\bibfnamefont {M.}~\bibnamefont {Barkeshli}},\ }\href {\doibase 10.1103/PhysRevResearch.2.043033} {\bibfield  {journal} {\bibinfo  {journal} {Phys. Rev. Res.}\ }\textbf {\bibinfo {volume} {2}},\ \bibinfo {pages} {043033} (\bibinfo {year} {2020})}\BibitemShut {NoStop}%
\bibitem [{\citenamefont {Bulmash}\ and\ \citenamefont {Barkeshli}(2022{\natexlab{b}})}]{Bulmash2021frac}%
  \BibitemOpen
  \bibfield  {author} {\bibinfo {author} {\bibfnamefont {D.}~\bibnamefont {Bulmash}}\ and\ \bibinfo {author} {\bibfnamefont {M.}~\bibnamefont {Barkeshli}},\ }\href {\doibase 10.1103/PhysRevB.105.125114} {\bibfield  {journal} {\bibinfo  {journal} {Phys. Rev. B}\ }\textbf {\bibinfo {volume} {105}},\ \bibinfo {pages} {125114} (\bibinfo {year} {2022}{\natexlab{b}})}\BibitemShut {NoStop}%
\bibitem [{\citenamefont {Lan}\ \emph {et~al.}(2024)\citenamefont {Lan}, \citenamefont {Yue},\ and\ \citenamefont {Wang}}]{Lan:2023uuq}%
  \BibitemOpen
  \bibfield  {author} {\bibinfo {author} {\bibfnamefont {T.}~\bibnamefont {Lan}}, \bibinfo {author} {\bibfnamefont {G.}~\bibnamefont {Yue}}, \ and\ \bibinfo {author} {\bibfnamefont {L.}~\bibnamefont {Wang}},\ }\href {\doibase 10.1007/JHEP11(2024)111} {\bibfield  {journal} {\bibinfo  {journal} {JHEP}\ }\textbf {\bibinfo {volume} {11}},\ \bibinfo {pages} {111} (\bibinfo {year} {2024})}\BibitemShut {NoStop}%
\bibitem [{\citenamefont {Zhou}\ and\ \citenamefont {Gu}(2024)}]{ZG2024}%
  \BibitemOpen
  \bibfield  {author} {\bibinfo {author} {\bibfnamefont {J.-R.}\ \bibnamefont {Zhou}}\ and\ \bibinfo {author} {\bibfnamefont {Z.-C.}\ \bibnamefont {Gu}},\ }\href {https://arxiv.org/abs/2410.19126} {\  (\bibinfo {year} {2024})},\ \Eprint {http://arxiv.org/abs/2410.19126} {arXiv:2410.19126 [cond-mat.str-el]} \BibitemShut {NoStop}%
\bibitem [{\citenamefont {{Wang}}\ \emph {et~al.}(2018)\citenamefont {{Wang}}, \citenamefont {{Wen}},\ and\ \citenamefont {{Witten}}}]{WWW2017}%
  \BibitemOpen
  \bibfield  {author} {\bibinfo {author} {\bibfnamefont {J.}~\bibnamefont {{Wang}}}, \bibinfo {author} {\bibfnamefont {X.-G.}\ \bibnamefont {{Wen}}}, \ and\ \bibinfo {author} {\bibfnamefont {E.}~\bibnamefont {{Witten}}},\ }\href {\doibase 10.1103/PhysRevX.8.031048} {\bibfield  {journal} {\bibinfo  {journal} {Physical Review X}\ }\textbf {\bibinfo {volume} {8}},\ \bibinfo {eid} {031048} (\bibinfo {year} {2018})}\BibitemShut {NoStop}%
\bibitem [{\citenamefont {Prakash}\ \emph {et~al.}(2018)\citenamefont {Prakash}, \citenamefont {Wang},\ and\ \citenamefont {Wei}}]{PWW2018}%
  \BibitemOpen
  \bibfield  {author} {\bibinfo {author} {\bibfnamefont {A.}~\bibnamefont {Prakash}}, \bibinfo {author} {\bibfnamefont {J.}~\bibnamefont {Wang}}, \ and\ \bibinfo {author} {\bibfnamefont {T.-C.}\ \bibnamefont {Wei}},\ }\href {\doibase 10.1103/PhysRevB.98.125108} {\bibfield  {journal} {\bibinfo  {journal} {Phys. Rev. B}\ }\textbf {\bibinfo {volume} {98}},\ \bibinfo {pages} {125108} (\bibinfo {year} {2018})}\BibitemShut {NoStop}%
\bibitem [{\citenamefont {Haldane}(1983{\natexlab{b}})}]{haldanechain}%
  \BibitemOpen
  \bibfield  {author} {\bibinfo {author} {\bibfnamefont {F.~D.~M.}\ \bibnamefont {Haldane}},\ }\href {\doibase 10.1103/PhysRevLett.50.1153} {\bibfield  {journal} {\bibinfo  {journal} {Phys. Rev. Lett.}\ }\textbf {\bibinfo {volume} {50}},\ \bibinfo {pages} {1153} (\bibinfo {year} {1983}{\natexlab{b}})}\BibitemShut {NoStop}%
\bibitem [{\citenamefont {Chen}\ \emph {et~al.}(2015)\citenamefont {Chen}, \citenamefont {Burnell}, \citenamefont {Vishwanath},\ and\ \citenamefont {Fidkowski}}]{chen2014symmetry}%
  \BibitemOpen
  \bibfield  {author} {\bibinfo {author} {\bibfnamefont {X.}~\bibnamefont {Chen}}, \bibinfo {author} {\bibfnamefont {F.~J.}\ \bibnamefont {Burnell}}, \bibinfo {author} {\bibfnamefont {A.}~\bibnamefont {Vishwanath}}, \ and\ \bibinfo {author} {\bibfnamefont {L.}~\bibnamefont {Fidkowski}},\ }\href {\doibase 10.1103/PhysRevX.5.041013} {\bibfield  {journal} {\bibinfo  {journal} {Phys. Rev. X}\ }\textbf {\bibinfo {volume} {5}},\ \bibinfo {pages} {041013} (\bibinfo {year} {2015})}\BibitemShut {NoStop}%
\bibitem [{Note1()}]{Note1}%
  \BibitemOpen
  \bibinfo {note} {The manifold should be spin with a chosen spin structure, as this is necessary to define fermions on it. However, since we are constructing gapped boundaries for a state on a region with the topology of an open disk, the issue of spin structure does not arise.}\BibitemShut {Stop}%
\bibitem [{\citenamefont {Chen}(2017)}]{Chen_2017}%
  \BibitemOpen
  \bibfield  {author} {\bibinfo {author} {\bibfnamefont {X.}~\bibnamefont {Chen}},\ }\href {\doibase 10.1016/j.revip.2017.02.002} {\bibfield  {journal} {\bibinfo  {journal} {Reviews in Physics}\ }\textbf {\bibinfo {volume} {2}},\ \bibinfo {pages} {3–18} (\bibinfo {year} {2017})}\BibitemShut {NoStop}%
\bibitem [{\citenamefont {Steenrod}(1947)}]{steenrod1947}%
  \BibitemOpen
  \bibfield  {author} {\bibinfo {author} {\bibfnamefont {N.~E.}\ \bibnamefont {Steenrod}},\ }\href {https://api.semanticscholar.org/CorpusID:124236740} {\bibfield  {journal} {\bibinfo  {journal} {Annals of Mathematics}\ }\textbf {\bibinfo {volume} {48}},\ \bibinfo {pages} {290} (\bibinfo {year} {1947})}\BibitemShut {NoStop}%
\bibitem [{\citenamefont {Lan}\ \emph {et~al.}(2019)\citenamefont {Lan}, \citenamefont {Zhu},\ and\ \citenamefont {Wen}}]{wen2019fspt}%
  \BibitemOpen
  \bibfield  {author} {\bibinfo {author} {\bibfnamefont {T.}~\bibnamefont {Lan}}, \bibinfo {author} {\bibfnamefont {C.}~\bibnamefont {Zhu}}, \ and\ \bibinfo {author} {\bibfnamefont {X.-G.}\ \bibnamefont {Wen}},\ }\href {\doibase 10.1103/PhysRevB.100.235141} {\bibfield  {journal} {\bibinfo  {journal} {Phys. Rev. B}\ }\textbf {\bibinfo {volume} {100}},\ \bibinfo {pages} {235141} (\bibinfo {year} {2019})}\BibitemShut {NoStop}%
\bibitem [{\citenamefont {Ren}\ \emph {et~al.}(2024)\citenamefont {Ren}, \citenamefont {Ning}, \citenamefont {Qi}, \citenamefont {Wang},\ and\ \citenamefont {Gu}}]{RNQWG2023}%
  \BibitemOpen
  \bibfield  {author} {\bibinfo {author} {\bibfnamefont {X.-Y.}\ \bibnamefont {Ren}}, \bibinfo {author} {\bibfnamefont {S.-Q.}\ \bibnamefont {Ning}}, \bibinfo {author} {\bibfnamefont {Y.}~\bibnamefont {Qi}}, \bibinfo {author} {\bibfnamefont {Q.-R.}\ \bibnamefont {Wang}}, \ and\ \bibinfo {author} {\bibfnamefont {Z.-C.}\ \bibnamefont {Gu}},\ }\href {\doibase 10.1103/PhysRevB.110.235117} {\bibfield  {journal} {\bibinfo  {journal} {Phys. Rev. B}\ }\textbf {\bibinfo {volume} {110}},\ \bibinfo {pages} {235117} (\bibinfo {year} {2024})}\BibitemShut {NoStop}%
\bibitem [{\citenamefont {Besche}\ \emph {et~al.}(2024)\citenamefont {Besche}, \citenamefont {Eick}, \citenamefont {O'Brien},\ and\ \citenamefont {Horn}}]{SmallGrp}%
  \BibitemOpen
  \bibfield  {author} {\bibinfo {author} {\bibfnamefont {H.~U.}\ \bibnamefont {Besche}}, \bibinfo {author} {\bibfnamefont {B.}~\bibnamefont {Eick}}, \bibinfo {author} {\bibfnamefont {E.}~\bibnamefont {O'Brien}}, \ and\ \bibinfo {author} {\bibfnamefont {M.}~\bibnamefont {Horn}},\ }\href@noop {} {\enquote {\bibinfo {title} {{SmallGrp, The GAP Small Groups Library, Version} 1.5.4},}\ }\bibinfo {howpublished} {\href{https://gap-packages.github.io/smallgrp/} {\texttt{https://gap-packages.github.io/smallgrp/}}} (\bibinfo {year} {2024}),\ \bibinfo {note} {{GAP package}}\BibitemShut {NoStop}%
\bibitem [{\citenamefont {Dokchitser}()}]{matydGroupNames}%
  \BibitemOpen
  \bibfield  {author} {\bibinfo {author} {\bibfnamefont {T.}~\bibnamefont {Dokchitser}},\ }\href@noop {} {\enquote {\bibinfo {title} {{GroupNames}},}\ }\bibinfo {howpublished} {\href{https://people.maths.bris.ac.uk/~matyd/GroupNames/61/C2^3.9D4.html}{\texttt{https://people.maths.bris.ac.uk/\symbol{126}matyd/GroupNames/ 61/C2\symbol{94}3.9D4.html}}}\BibitemShut {NoStop}%
\bibitem [{\citenamefont {Chen}\ \emph {et~al.}(2021)\citenamefont {Chen}, \citenamefont {Ellison},\ and\ \citenamefont {Tantivasadakarn}}]{chen2020}%
  \BibitemOpen
  \bibfield  {author} {\bibinfo {author} {\bibfnamefont {Y.-A.}\ \bibnamefont {Chen}}, \bibinfo {author} {\bibfnamefont {T.~D.}\ \bibnamefont {Ellison}}, \ and\ \bibinfo {author} {\bibfnamefont {N.}~\bibnamefont {Tantivasadakarn}},\ }\href {\doibase 10.1103/PhysRevResearch.3.013056} {\bibfield  {journal} {\bibinfo  {journal} {Phys. Rev. Res.}\ }\textbf {\bibinfo {volume} {3}},\ \bibinfo {pages} {013056} (\bibinfo {year} {2021})}\BibitemShut {NoStop}%
\bibitem [{\citenamefont {{Kobayashi}}\ \emph {et~al.}(2019)\citenamefont {{Kobayashi}}, \citenamefont {{Ohmori}},\ and\ \citenamefont {{Tachikawa}}}]{KOT2019}%
  \BibitemOpen
  \bibfield  {author} {\bibinfo {author} {\bibfnamefont {R.}~\bibnamefont {{Kobayashi}}}, \bibinfo {author} {\bibfnamefont {K.}~\bibnamefont {{Ohmori}}}, \ and\ \bibinfo {author} {\bibfnamefont {Y.}~\bibnamefont {{Tachikawa}}},\ }\href {\doibase 10.1007/JHEP11(2019)131} {\bibfield  {journal} {\bibinfo  {journal} {Journal of High Energy Physics}\ }\textbf {\bibinfo {volume} {2019}},\ \bibinfo {eid} {131} (\bibinfo {year} {2019})}\BibitemShut {NoStop}%
\bibitem [{\citenamefont {Else}\ and\ \citenamefont {Thorngren}(2019)}]{PhysRevB.99.115116}%
  \BibitemOpen
  \bibfield  {author} {\bibinfo {author} {\bibfnamefont {D.~V.}\ \bibnamefont {Else}}\ and\ \bibinfo {author} {\bibfnamefont {R.}~\bibnamefont {Thorngren}},\ }\href {\doibase 10.1103/PhysRevB.99.115116} {\bibfield  {journal} {\bibinfo  {journal} {Phys. Rev. B}\ }\textbf {\bibinfo {volume} {99}},\ \bibinfo {pages} {115116} (\bibinfo {year} {2019})}\BibitemShut {NoStop}%
\bibitem [{\citenamefont {Zhang}\ \emph {et~al.}(2020)\citenamefont {Zhang}, \citenamefont {Wang}, \citenamefont {Yang}, \citenamefont {Qi},\ and\ \citenamefont {Gu}}]{pointgroup}%
  \BibitemOpen
  \bibfield  {author} {\bibinfo {author} {\bibfnamefont {J.-H.}\ \bibnamefont {Zhang}}, \bibinfo {author} {\bibfnamefont {Q.-R.}\ \bibnamefont {Wang}}, \bibinfo {author} {\bibfnamefont {S.}~\bibnamefont {Yang}}, \bibinfo {author} {\bibfnamefont {Y.}~\bibnamefont {Qi}}, \ and\ \bibinfo {author} {\bibfnamefont {Z.-C.}\ \bibnamefont {Gu}},\ }\href {\doibase 10.1103/PhysRevB.101.100501} {\bibfield  {journal} {\bibinfo  {journal} {Phys. Rev. B}\ }\textbf {\bibinfo {volume} {101}},\ \bibinfo {pages} {100501} (\bibinfo {year} {2020})}\BibitemShut {NoStop}%
\bibitem [{\citenamefont {Zhang}\ \emph {et~al.}(2022{\natexlab{b}})\citenamefont {Zhang}, \citenamefont {Yang}, \citenamefont {Qi},\ and\ \citenamefont {Gu}}]{PhysRevResearch.4.033081}%
  \BibitemOpen
  \bibfield  {author} {\bibinfo {author} {\bibfnamefont {J.-H.}\ \bibnamefont {Zhang}}, \bibinfo {author} {\bibfnamefont {S.}~\bibnamefont {Yang}}, \bibinfo {author} {\bibfnamefont {Y.}~\bibnamefont {Qi}}, \ and\ \bibinfo {author} {\bibfnamefont {Z.-C.}\ \bibnamefont {Gu}},\ }\href {\doibase 10.1103/PhysRevResearch.4.033081} {\bibfield  {journal} {\bibinfo  {journal} {Phys. Rev. Res.}\ }\textbf {\bibinfo {volume} {4}},\ \bibinfo {pages} {033081} (\bibinfo {year} {2022}{\natexlab{b}})}\BibitemShut {NoStop}%
\bibitem [{\citenamefont {Song}\ \emph {et~al.}(2020)\citenamefont {Song}, \citenamefont {Xiong},\ and\ \citenamefont {Huang}}]{PhysRevB.101.165129}%
  \BibitemOpen
  \bibfield  {author} {\bibinfo {author} {\bibfnamefont {H.}~\bibnamefont {Song}}, \bibinfo {author} {\bibfnamefont {C.~Z.}\ \bibnamefont {Xiong}}, \ and\ \bibinfo {author} {\bibfnamefont {S.-J.}\ \bibnamefont {Huang}},\ }\href {\doibase 10.1103/PhysRevB.101.165129} {\bibfield  {journal} {\bibinfo  {journal} {Phys. Rev. B}\ }\textbf {\bibinfo {volume} {101}},\ \bibinfo {pages} {165129} (\bibinfo {year} {2020})}\BibitemShut {NoStop}%
\bibitem [{\citenamefont {Wang}\ and\ \citenamefont {Cheng}(2022)}]{U1SET}%
  \BibitemOpen
  \bibfield  {author} {\bibinfo {author} {\bibfnamefont {Q.-R.}\ \bibnamefont {Wang}}\ and\ \bibinfo {author} {\bibfnamefont {M.}~\bibnamefont {Cheng}},\ }\href {\doibase 10.1103/PhysRevB.106.115104} {\bibfield  {journal} {\bibinfo  {journal} {Phys. Rev. B}\ }\textbf {\bibinfo {volume} {106}},\ \bibinfo {pages} {115104} (\bibinfo {year} {2022})}\BibitemShut {NoStop}%
\bibitem [{\citenamefont {Pace}\ and\ \citenamefont {Wen}(2023)}]{PhysRevB.107.075112}%
  \BibitemOpen
  \bibfield  {author} {\bibinfo {author} {\bibfnamefont {S.~D.}\ \bibnamefont {Pace}}\ and\ \bibinfo {author} {\bibfnamefont {X.-G.}\ \bibnamefont {Wen}},\ }\href {\doibase 10.1103/PhysRevB.107.075112} {\bibfield  {journal} {\bibinfo  {journal} {Phys. Rev. B}\ }\textbf {\bibinfo {volume} {107}},\ \bibinfo {pages} {075112} (\bibinfo {year} {2023})}\BibitemShut {NoStop}%
\bibitem [{\citenamefont {Wang}\ \emph {et~al.}(2023)\citenamefont {Wang}, \citenamefont {Qi}, \citenamefont {Fang}, \citenamefont {Cheng},\ and\ \citenamefont {Gu}}]{IEI}%
  \BibitemOpen
  \bibfield  {author} {\bibinfo {author} {\bibfnamefont {Q.-R.}\ \bibnamefont {Wang}}, \bibinfo {author} {\bibfnamefont {Y.}~\bibnamefont {Qi}}, \bibinfo {author} {\bibfnamefont {C.}~\bibnamefont {Fang}}, \bibinfo {author} {\bibfnamefont {M.}~\bibnamefont {Cheng}}, \ and\ \bibinfo {author} {\bibfnamefont {Z.-C.}\ \bibnamefont {Gu}},\ }\href {\doibase 10.1103/PhysRevB.108.L121104} {\bibfield  {journal} {\bibinfo  {journal} {Phys. Rev. B}\ }\textbf {\bibinfo {volume} {108}},\ \bibinfo {pages} {L121104} (\bibinfo {year} {2023})}\BibitemShut {NoStop}%
\bibitem [{\citenamefont {Barkeshli}\ \emph {et~al.}(2024)\citenamefont {Barkeshli}, \citenamefont {Chen}, \citenamefont {Hsin},\ and\ \citenamefont {Kobayashi}}]{Barkeshli:2022edm}%
  \BibitemOpen
  \bibfield  {author} {\bibinfo {author} {\bibfnamefont {M.}~\bibnamefont {Barkeshli}}, \bibinfo {author} {\bibfnamefont {Y.-A.}\ \bibnamefont {Chen}}, \bibinfo {author} {\bibfnamefont {P.-S.}\ \bibnamefont {Hsin}}, \ and\ \bibinfo {author} {\bibfnamefont {R.}~\bibnamefont {Kobayashi}},\ }\href {\doibase 10.21468/SciPostPhys.16.4.089} {\bibfield  {journal} {\bibinfo  {journal} {SciPost Phys.}\ }\textbf {\bibinfo {volume} {16}},\ \bibinfo {pages} {089} (\bibinfo {year} {2024})}\BibitemShut {NoStop}%
\bibitem [{Note2()}]{Note2}%
  \BibitemOpen
  \bibinfo {note} {The simplified form of $\protect \mathcal O_5$ agrees with the result derived in Appendix C of Ref.~\protect \rev@citealpnum {Barkeshli:2022edm}.}\BibitemShut {Stop}%
\end{thebibliography}

\begin{widetext}

\appendix

\section{On gapped boundaries of (1+1)D $(\mathbb Z_2^f \times_{\om_2} G)$-FSPT states}
\label{sec:1Dbdy}

In this section, we attempt to perform pullback trivialization for the (1+1)D FSPT state with $\mathbb Z_2^f \times_{\om_2} G$ symmetry.
The defining decoration data is $(\om_2;n_1,\nu_2)$ with consistency conditions \cite{wang2018construction}
\begin{align}
\dd \om_2 &\={2} 0,\\\label{dn1om1d}
\dd n_1 &\={2} 0,\\\label{dnu2om1d}
\dd \nu_2 &\={1} \frac12\om_2\smile n_1.
\end{align}

Similar to (2+1)D and (3+1)D cases, we will employ the following sequence of pullback trivializations,
\begin{align}\nonumber
(\om_2;n_1,\nu_2)
&\,\overset{\text{  pullback  }}{\sim} (\om'_2;n'_1,\nu_2') = (\dd \tau'_1;n'_1,\nu_2')\\\nonumber
&\overset{\text{coboundary}}{\sim} (0;N'_1,\mathcal V'_2)
= (0;\dd m'_0,\mathcal V_2')\\\nonumber
&\overset{\text{coboundary}}{\sim} (0;0,\mathscr V_2')\\\nonumber
&\,\overset{\text{  pullback  }}{\sim} (0;0,\mathscr V''_2) = (0;0,\dd \mu''_1)\\
&\overset{\text{coboundary}}{\sim} (0;0,0).
\end{align}

We note that, in principle, only the trivialization of $\omega_2$ and $\nu_2$ can be achieved via pullback. The degree of the complex fermionic decoration data, $n_1$, is too low to be trivialized in an extended group if it is nontrivial. The same issue is discussed in the main text for the Majorana chain decoration in (2+1)D. We include the $n_1$ data here solely for completeness. This means that, initially, we assume $n_1$ (and hence its pullback) to be a 1-coboundary.

\subsection{Pullback trivialization of $\om_2$}
Let us first trivialize the extension 2-cocycle $\om_2$ by considering the symmetry extension
\begin{align}
1 \rightarrow A \rightarrow G' \overset{\pi}{\rightarrow} G \rightarrow 1.
\end{align}
Pulling back to $G'$, the extension 2-cocycle of $G'$ is denoted by $\om'_2 :=\pi^\ast (\om_2)$, which can be written as a coboundary $\om'_2 \={2} \dd \tau'_1$. 
The twisted cocycle equation now becomes
\begin{align}
    \dd \nu_2' \={1} \frac12\dd\tau_1'\smile n_1'
    \={1} \dd\!\left(\frac12\tau_1'\smile n_1'\right).
\end{align}

By appropriate transformations, 
we define a trivial extension 2-cocycle $\Omega_2'\={2}\om'_2 +\dd \tau'_1\={2}0$
and the shifted cocycles
\begin{align}
    N_1' &\={2}n_1',\\
    \mathcal V_2' &\={1} \nu_2'+\frac12\tau_1'\smile n_1',
\end{align}
such that their consistency equations are independent of $\om'_2$, where
\begin{align}
    \dd N_1' &\={2}0,\\
    \dd \mathcal V_2' &\={1}0.
\end{align}
The data $(0;N_1',\mathcal V_2')$ describes an FSPT phase with $\Z_2^f\times G'$ symmetry.

\subsection{Complex fermion decoration $n_1$}
Owing to the low degree of the cocycle $n_1$, we consider $N_1'\={2}\dd m_0'$ to be a 1-coboundary of $G'$, for an appropriate 0-cochain $m_0'$. Next, we perform an FSLU or ``gauge transformation'' on $N_1'\={2}\dd m_0'$ such that a BSPT phase is obtained at the cochain level.

Since the pullback 2-cocycle satisfies the condition
\begin{align}\nonumber
\dd \mathcal V'_2 &\={1} 0,
\end{align}
we can define a $\R/\Z$-valued 2-cocycle of $G'$,
\begin{align}
\mathscr V'_2 :\={1} \mathcal V'_2\={1} \nu'_2 + \frac12\tau'_1 \smile n'_1.
\end{align}
The data $(0;0, \mathscr V_2')$ now represents a (1+1)D $G'$-BSPT phase. 

\subsection{Bosonic SPT $\nu_2$}
Extending further to a larger symmetry $G''$,
\begin{align}
    1 \rightarrow A' \rightarrow G'' \overset{\pi'}{\rightarrow} G' \rightarrow 1,
\end{align}
we obtain a trivial 2-cocycle of $G''$ of the form
\begin{align}
    \mathscr V''_2 :\={1} \dd\mu_1'',
\end{align}
and by shifting with the coboundary $\dd\mu_1''$, we obtain a trivial $G''$-BSPT phase connected adiabatically to the vacuum (i.e., trivial product state). Heuristically, compactifying the SPT phases corresponding to the enhanced symmetries $G'$ and $G''$ gives us a (0+1)D boundary SPT theory with symmetry $(G_f''=\Z_2^f\times_{\om_2}G'')$ for the (1+1)D $(G_f=\Z_2^f\times_{\om_2}G)$ SPT phase.

\section{Simplified expression of $\mathcal O_5$ with $\om_2\neq 0$}
\label{sec:O5om}

In this appendix, we aim to simplify the last consistency equation for (3+1)D FSPT states within the general supercohomology framework, where the symmetry group is given by $G_f = \Zf \times_{\omega_2} G$ with nontrivial $\om_2$ extension. To begin with, the twisted cocycle equation of $\nu_4$ is given by \cite{wang2018construction}
\begin{align}
\dd \nu_4 
&\={1}\mathcal O_5[\om_2;n_2,n_3]\\\nonumber
&\={1} \frac12 [\om_2\smile n_3+\Sq^2(n_3)]+\frac14 [\dd n_3(01245)\dd n_3(01234)\ (\text{mod}\ 2)]\\\nonumber
&\quad-\frac14 \{[\dd n_3(12345)+\dd n_3(02345)+\dd n_3(01345)]\dd n_3(01235)\ (\text{mod}\ 2)\}\\\label{dv4om}
&\quad+\frac12\dd n_3(02345)\dd n_3(01235)+\frac12 \om_2(013)\dd n_3(12345)
+\frac12 \om_2(023)[\dd n_3(01245)+\dd n_3(01235)+\dd n_3(01234)].
\end{align}

With the equation $\dd n_3 \={2} (\om_2+n_2)\smile n_2$, one can simplify the expression of obstruction function $\mathcal O_5[\om_2;n_2,n_3]$. The last line of \eq{dv4om} can be written as
\begin{align}
    \frac12(\om_2+n_2)(012)(\om_2+n_2)(023)n_2(235)n_2(345)+\frac12 \{[\om_2\smile_1(\om_2+n_2)]\smile n_2\}(012345).
\end{align}
For terms involving a $\frac14$ coefficient, we have
\begin{align}
\frac14 [(\om_2+n_2)(012)n_2(245)n_2(234) \ (\text{mod}\ 2)]
-\frac14 [(\om_2+n_2)(012) n_2(235)n_2(345)\ (\text{mod}\ 2)].
\end{align}
The square parentheses with (mod 2) indicate that the expression inside should only take value in $\Z_2=\{0,1\}$. By considering all possible values of $(\om_2+n_2)(012)$, one notices that $(\om_2+n_2)(012)=2$ when both $\om_2(012)$ and $n_2(012)$ are 1. By subtracting $2\om_2(012)n_2(012)$ from $(\om_2+n_2)(012)$, we have discarded the outlier and arrived at a refined expression: $[(\om_2+n_2)(012)\ (\mathrm{mod}\ 2)] = (\om_2+n_2)(012) - 2\om_2(012)n_2(012)$. Therefore, the $\frac14$-term takes this form without (mod 2):
\begin{align}
&\quad\frac14 [(\om_2+n_2)(012) - 2\om_2(012)n_2(012)]n_2(245)n_2(234)
-\frac14 [(\om_2+n_2)(012) -2\om_2(012)n_2(012)]n_2(235)n_2(345)\\
&\={1}
-\frac14 [(\om_2+n_2)\smile\beta_2  n_2](012345) +\frac12 [(\om_2\smile_2 n_2)\smile(n_2\smile_1 n_2)](012345),
\end{align}
where $\beta_2  n_2:=\frac12 \dd n_2$ is the Bockstein homomorphism, which is a map $H^2(G,\Z_2)\rightarrow H^3(G,\Z)$. Given that $\dd n_2\={2} 0$, one can check that $\beta_2  n_2(ijkl)\={\Z}n_2(jkl)n_2(ijl)-n_2(ikl)n_2(ijk)$ as enumerated in Table \ref{tab:bockstein}. Taking mod 2 after Bockstein homomorphism, we also have the relation $\beta_2 n_2 \={2} n_2\smile_1 n_2$.
\begin{table}[h]
\centering
\begin{tabular}{|c c c c|c|c|}
\hline
$n_2(ijk)$ & $n_2(ijl)$ & $n_2(ikl)$ & $n_2(jkl)$ & $\beta_2  n_2(ijkl)$ & $n_2(jkl)n_2(ijl)-n_2(ikl)n_2(ijk)$ \\
\hline
\hline
0 & 0 & 0 & 0 & 0 & 0\\
\hline
0 & 0 & 1 & 1 & 0 & 0\\
\hline
0 & 1 & 0 & 1 & 1 & 1\\
\hline
0 & 1 & 1 & 0 & 0 & 0\\
\hline
1 & 0 & 0 & 1 & 0 & 0\\
\hline
1 & 0 & 1 & 0 & -1 & -1\\
\hline
1 & 1 & 0 & 0 & 0 & 0\\
\hline
1 & 1 & 1 & 1 & 0 & 0\\
\hline
\end{tabular}
\caption{Relation between $\beta_2  n_2(ijkl):=\frac12 (\dd n_2)(ijkl)$ and $n_2(jkl)n_2(ijl)-n_2(ikl)n_2(ijk)$ for $\dd n_2=0\ (\text{mod}\ 2)$.}
\label{tab:bockstein}
\end{table}

Putting all the terms together, we have the twisted cocycle equation\footnote{The simplified form of $\mathcal O_5$ agrees with the result derived in Appendix C of \refn{Barkeshli:2022edm}.} 
\begin{align}\label{dnu4omapp}\nonumber
\dd \nu_4
&\={1}\frac12\left[\om_2\smile n_3+\Sq^2(n_3)\right]
-\frac14(\om_2+n_2)\smile\beta_2  n_2
+\frac12 [\om_2\smile_1(\om_2+n_2)]\smile n_2\\
&\quad 
+\frac12 (\om_2\smile_2 n_2)\smile(n_2\smile_1 n_2)
+\frac12(\om_2+n_2)(012)(\om_2+n_2)(023)n_2(235)n_2(345)
\end{align}
which reduces to \eq{dnu4_} when $\om_2=0$.

\section{Trivialization equation of $\Z$-valued 2-cocycle $n_2$}\label{sec:triveqz}

Upon performing the pullback to an extended group, the trivial $\Z_2$-valued 2-cocycle $n_2$ satisfies the trivialization equation
\begin{align}\label{n2triv}
    n_2(ijk)\={2}\dd m_1(ijk).
\end{align}
However, this is a mod 2 equation, and the left-hand and right-hand sides may differ by multiples of 2. Since the expression for the obstruction function typically restricts $n_2$ to $\{0,1\}$, it would be preferable to find a trivialization equation for $n_2$ that holds in integers without mod 2.

Using the definition of the differential,
\begin{align}
    \dd m_1(ijk) :\={\Z}m_1(jk)-m_1(ik)+m_1(ij),
\end{align}
let us calculate all possible values of $\dd m_1(ijk)$ for the simplex $\langle ijk\rangle$ in Table \ref{tab:dm1}. Notice that we omit the superscripts on cochains and cocycles denoting the symmetry in this appendix for simplicity.
\begin{table}[h]
\centering
\begin{tabular}{|c c c|c|}
\hline
$m_1(jk)$ & $m_1(ik)$ & $m_1(ij)$ & $\dd m_1(ijk)$ \\
\hline
\hline
0 & 0 & 0 & 0\\
\hline
0 & 0 & 1 & 1\\
\hline
0 & 1 & 0 & -1\\
\hline
0 & 1 & 1 & 0\\
\hline
1 & 0 & 0 & 1\\
\hline
1 & 0 & 1 & 2\\
\hline
1 & 1 & 0 & 0\\
\hline
1 & 1 & 1 & 1\\
\hline
\end{tabular}
\caption{Possible values of $\Z$-valued coboundary $\dd m_1(ijk)$.
}
\label{tab:dm1}
\end{table}

We anticipate $\dd m_1$ to be a 2-coboundary which takes value in $\{0,1\}$. Nevertheless, Table \ref{tab:dm1} shows that $\dd m_1$ can also take values $-1$ and $+2$ other than the anticipated values. 
This implies that we have to refine the values of $\dd m_1$ by removing the outliers. Hence we have
\begin{align}
    n_2(ijk) &\={\Z} \dd m_1(ijk)+2[1-m_1(jk)]m_1(ik)[1-m_1(ij)]-2m_1(jk)[1-m_1(ik)]m_1(ij)\\\nonumber
    &\={\Z} \dd m_1(ijk)+2m_1(ik)-2m_1(ik)m_1(ij)-2m_1(ik)m_1(jk)\\ &\quad+4m_1(ik)m_1(jk)m_1(ij)-2m_1(ij)m_1(jk)\\
    &\={\Z} \dd m_1(ijk)
    -2m_1(ik)[-m_1(ik)+m_1(jk)+m_1(ij)-2m_1(jk)m_1(ij)]-2(m_1\smile m_1)(ijk)\\
    &\={\Z} [\dd m_1 - 2 m_1 \smile_1 (\dd m_1 - 2 m_1\smile m_1)- 2 m_1\smile m_1](ijk).
\end{align}
The trivialization equation of $\Z$-valued $n_2$ is then
\begin{align}
n_2 &\={\Z} \dd m_1 - 2 m_1^2 - 2 m_1 \smile_1 \dd m_1 +4 m_1 \smile_1 m_1^2\\
&\={\Z} \dd m_1 - 2\Sq^1(m_1) + 4 m_1 \smile_1 m_1^2.
\end{align}
If we view $n_2$ as a $\Z_4$-valued cocycle with values 0 and 1, $n_2$ satisfies the mod-4 trivialization equation
\begin{align}\label{App_dm1}
n_2 \={4} \dd m_1 - 2\Sq^1(m_1).
\end{align}

\section{Pullback trivialization of $\mathcal O_5$ with $\om_2=0$}\label{sec:O5pb}
The pullback from $G$ to $G'$ leads to a trivial 2-cocycle $n'_2$, which is defined as $n'_2 \={\Z} \dd m'_1 - 2\Sq^1(m'_1)+ 4 m'_1 \smile_1 (m_1')^2$ taking values $\{0,1\}$ in $\Z$. The obstruction function $\mathcal O_5$ for the twisted cocycle condition in \eq{dnu4_} depends nontrivially on $n_2'$. We can, however, remove such a dependence by a gauge transformation.

The pulled-back obstruction function takes the form
\begin{align}
    \mathcal O_5[n'_2,n'_3] 
    &\={1} \frac12 \Sq^2(n'_3)
    -\frac14(n'_2\smile \beta_2 n'_2)+\frac12(n'_2)^2(02345)(n'_2)^2(01235).
\end{align}
Expressing $n'_2$ in terms of $m_1'$, we have the $\frac14$-term:
\begin{align}
    -\frac14(n_2'\smile \beta_2 n_2')
    &\={1} \frac14 [\dd m_1' (\dd m_1'\smile_1 \dd m_1')] 
    + \frac12 \left[\Sq^1(m_1') (\dd m_1'\smile_1 \dd m_1') \right]
    + \frac12 \dd m_1'\left[\dd m_1'\smile_1 m_1'^2+m_1'\smile_1 \dd (m_1'^2)\right]\\
    &\={1} \frac14 \dd [(\dd m_1') \Sq^1(m_1')]
    + \frac12\Sq^1(m_1') \dd\Sq^1(m_1')
    + \frac12 \dd[\dd m_1'\smile(m_1'\smile_1 m_1'^2)].
\end{align}
Hence, the obstruction function can be expressed as follows:
\begin{align}\label{O5wn2}\nonumber
    \mathcal O_5[m'_1,n'_3] 
    &\={1} \frac12 \Sq^2(n'_3)
    +\frac14 \dd [(\dd m_1') \Sq^1(m_1')]
    + \frac12\Sq^1(m_1') \dd\Sq^1(m_1')\\
    &\quad + \frac12 \dd[\dd m_1'\smile(m_1'\smile_1 m_1'^2)]
    +\frac12\dd m'_1(012)\dd m'_1(023)\dd m'_1(235)\dd m'_1(345).
\end{align}

Our task is to remove $n'_2$ dependence in $\mathcal O_5$. This means, with the 3-cochain $N'_3 :\={2} n'_3 + m'_1 \smile \dd m'_1$, we should have an obstruction function of the data $(0,N'_3)$:
\begin{align}\label{O5won2}
    \Tilde{\mathcal O}_5[0,N'_3] 
    & \={1} \frac12 \Sq^2(n'_3)+\frac12 \Sq^2(m_1' \dd m_1')
    +\frac12 \dd\!\sqb{(n'_3+m_1' \dd m_1')\smile_2 (m_1' \dd m_1')}.
\end{align}
Here we have applied the action of $\Sq^2$ to the sum $n'_3+m_1' \dd m_1'$, which we refer to in \eq{sqk2c}, along with the trivialization equation $\dd n'_3\={2} \dd(m_1'\smile\dd m_1')$.
The difference between $\mathcal O_5[m'_1,n'_3]$ and 
$\Tilde{\mathcal O}_5[0,N'_3]$ yields a coboundary,
\begin{align}\label{df4}\nonumber
    \dd f_4 &\={1} 
    \dd\!\sqb{\frac14 (\dd m_1') \Sq^1(m_1')
    +\frac12\dd m_1'(m_1'\smile_1 m_1'^2)
    +\frac12 (n'_3+m_1' \dd m_1')\smile_2 (m_1' \dd m_1')}\\
    &\quad +\frac12 \Sq^2(m_1' \dd m_1')
    + \frac12\Sq^1(m_1')(\dd m_1'\smile_1 \dd m_1')
    +\frac12\dd m'_1(012)\dd m'_1(023)\dd m'_1(235)\dd m'_1(345).
\end{align}
By numerical verification, the last line of \eq{df4} also forms a coboundary. This leads to
\begin{align}\nonumber
    \dd f_4 &\={1} 
    \dd\!\sqb{\frac14 (\dd m_1') \Sq^1(m_1')
    +\frac12\dd m_1'(m_1'\smile_1 m_1'^2)
    +\frac12 (n'_3+m_1' \dd m_1')\smile_2 (m_1' \dd m_1')}\\
    &\quad +\dd\!\sqb{
    \frac12(\dd m_1')(m_1'^2\smile_2 \dd m_1'+m_1'\smile_1 m_1'^2)
    +\frac12 m_1'^2 (\dd m_1'\smile_1 m_1')
    +\frac12m_1'(01)m_1'(12)m_1'(14)(\dd m_1')(234)
    }.
\end{align}
Hence, from $\dd f_4$, we have found a 4-cochain,
\begin{align}\nonumber
    f_4[m_1',n_3']
    &\={1}\frac14 (\dd m_1') \Sq^1(m_1')
    +\frac12 (n'_3+m_1' \dd m_1')\smile_2 (m_1' \dd m_1')\\
    &\quad+\frac12(\dd m_1')(m_1'^2\smile_2 \dd m_1')
    +\frac12 m_1'^2 (\dd m_1'\smile_1 m_1')
    +\frac12m_1'(01)m_1'(12)m_1'(14)\dd m_1'(234),
\end{align}
which is used to define $\mathcal V'_4$ in \eq{N4nu4}.

Further, we consider the pullback from symmetry $G'$ to an even larger symmetry $G''$. This gives us a trivial 3-cocycle $N''_3$, which can be expressed as a coboundary,
\begin{align}
N''_3 :\={2} \dd m''_2.
\end{align}
The pulled-back 4-cocycle is now obstructed by
\begin{align}
    \Tilde{\mathcal O}_5[0,N''_3]
    \={1} \frac12 \Sq^2(N''_3)
    \={1} \frac12 \Sq^2(\dd m''_2)
    \={1} \frac12 \dd\Sq^2(m''_2).
\end{align}
Performing an FSLU or ``gauge'' transformation on $\Tilde{\mathcal O}_5[0,N''_3]$ simply removes its dependence on $N''_3$,
\begin{align}
\Tilde{\mathcal O}_5[0,N''_3]\rightarrow\dtilde{\mathcal O}_5[0,0] :\={1} \Tilde{\mathcal O}_5[0,N''_3]-\frac12 \dd\Sq^2(m''_2).
\end{align}
Defining the pulled-back cochains $m_1'':=\pi'(m_1')$ and $n_3'':=\pi'(n_3')$ via the central extension 
\begin{align}
    1 \rightarrow A' \rightarrow G'' \overset{\pi'}{\rightarrow} G' \rightarrow 1,
\end{align}
one can show that $\dtilde{\mathcal O}_5[0,0]-\mathcal O_5[m_1'',n_3'':\={2}\dd m_2''+m_1''\dd m_1'']$ is a coboundary $\dd \mathcal F_4$, where $\mathcal F_4$ is the nontrivial 4-cochain in \eq{f4maj},
\begin{align}\nonumber
\mathcal F_4[m_1''',m_2''']
&\={1} 
-\frac14(\dd m_1''')\Sq^1(m_1''')
+\frac12 [\Sq^2(m_2''') + \dd m_2'''\smile_2 (m_1''' \dd m_1''')]\\
&\quad+\frac12(\dd m_1''')\sqb{(m_1''')^2\smile_2 \dd m_1'''}
+\frac12 (m_1''')^2 (\dd m_1'''\smile_1 m_1''')
+\frac12m_1'''(01)m_1'''(12)m_1'''(14)\dd m_1'''(234).
\end{align}

\end{widetext}

\end{document}